\documentclass[a4paper,fleqn]{cas-sc}

\usepackage[authoryear,longnamesfirst]{natbib}

\usepackage{float}
\usepackage{subcaption}
\usepackage{bm}

\usepackage[authoryear,longnamesfirst]{natbib}

\def\tsc#1{\csdef{#1}{\textsc{\lowercase{#1}}\xspace}}
\tsc{WGM}
\tsc{QE}

\begin{document}
\let\WriteBookmarks\relax
\def\floatpagepagefraction{1}
\def\textpagefraction{.001}

\shorttitle{Akhmediev breathers at the interface of two fluid half-spaces}

\shortauthors{O. Avramenko et~al.}

\title [mode = title]{Parameter mapping and physical reconstruction of Akhmediev breathers at the interface of two fluid half-spaces}

%

\author[1,2]{O. Avramenko}[
                        orcid=https:0000-0002-7960-1436]

\ead{o.avaramenko@ukma.edu.ua}


\credit{Conceptualization  of this study, Methodology, Software, Visualization, Writing – original draft preparation}

\affiliation[1]{organization={National University of Kyiv-Mohyla Academy},
            addressline={Skovorody 2},
            city={Kyiv},
            postcode={04070},
            country={Ukraine}}

\author[3]{V. Naradovyi}[
                        orcid=https:0000-0001-5187-8831]

\ead{v.v.naradovyi@cuspu.edu.ua}

\ead[url]{}


\credit{Formal analysis, Validation}

\affiliation[2]{organization={Vytautas Magnus University},
            addressline={K. Donelai\v{c}io g. 58},
            city={Kaunas},
            postcode={44248},
            country={Lithuania}}

\affiliation[3]{organization={Volodymyr Vynnychenko Central Ukraine State University},
            addressline={Shevchenka 1},
            city={Kropyvnytskyi},
            postcode={25006},
            country={Ukraine}}

\cortext[cor1]{Corresponding author}


\begin{abstract}
A methodology combining parameter mapping and physical
reconstruction of Akhmediev breathers at the interface
between two fluid half-spaces is developed on the basis
of the Nayfeh model. Parameter maps are constructed to
characterize the breather modulation period, the
modulational instability growth rate, and the relative
contribution of the bound second harmonic to the
reconstructed interfacial profile. Their combined
analysis provides a physically meaningful classification
of breather regimes beyond the conventional focusing
condition of the nonlinear Schrödinger equation.
Reconstruction of the physical interfacial profile
establishes a quantitative relation between nonlinear
wave deformation and the contribution of the bound
second harmonic, making it possible to identify the
range of applicability of the weakly nonlinear
approximation. Representative regimes from different
modulational instability regions are analyzed to
demonstrate the influence of resonance and focusing
boundaries on breather characteristics and interfacial
wave profiles. The proposed approach provides a direct
link between the mathematical description of Akhmediev
breathers and their physical interpretation and can be
extended to other localized solutions of the nonlinear
Schrödinger equation and to a broad class of stratified
hydrodynamic systems.

\end{abstract}





\maketitle
\thispagestyle{empty}
\pagestyle{empty}

\section*{Introduction}

A general theoretical framework for describing the slow
modulation of weakly nonlinear wave packets was established by
\citet{Whitham1965}, laying the foundation for the modern
theory of nonlinear dispersive waves. Within this framework,
modulational instability (MI) emerged as a fundamental
mechanism governing the evolution of weakly nonlinear wave
packets.

MI was first identified for deep-water gravity waves by
\citet{Benjamin1967}, and is therefore often referred to as
the Benjamin--Feir instability in the context of water waves.
Its mathematical description was
formulated within the nonlinear Schrödinger equation (NLS),
derived by \citet{Zakharov1968} and further developed by
\citet{Hasimoto1972}. The NLS is now the standard envelope
equation for weakly nonlinear dispersive waves.

The universality of MI was soon recognized beyond
hydrodynamics. In plasma physics,
\citet{TaniutiWashimi1968} investigated the self-trapping and
instability of magnetohydrodynamic waves, while
\citet{Hasegawa1970} demonstrated analogous nonlinear
processes for cyclotron waves. These studies established that
MI is a generic feature of nonlinear
dispersive media.


The Akhmediev breather is an exact solution of the focusing
nonlinear Schrödinger equation (NLS) describing the nonlinear stage
of MI on a finite background. It is
spatially periodic and temporally
localized~\citep{Akhmediev1986}. A unified analytical
framework linking periodic, soliton, and localized solutions
of the focusing NLS was established by
\citet{Akhmediev1987}, with the Peregrine breather recovered
as the limiting case corresponding to an infinitely long
modulation period~\citep{Peregrine1983}.

Akhmediev breathers provide an effective description of the
nonlinear evolution of MI in a wide
range of dispersive media. Their role has been demonstrated
most clearly in nonlinear optics through the dynamics of
unstable continuous waves and supercontinuum
generation~\citep{Dudley2009}. More generally, breather
solutions are regarded as universal models of extreme wave
events, although their physical realizability depends on the
properties of the medium and on the validity of the NLS
approximation~\citep{Dudley2014}. A comprehensive review of
these developments was presented by
\citet{ZakharovOstrovsky2009}.

Interest in breather solutions has been further stimulated by
the problem of rogue-wave formation in the ocean.
\citet{Onorato2001} showed that MI can
generate extreme waves in random sea states, while
\citet{Janssen2003} demonstrated the importance of nonlinear
four-wave interactions in this process. Direct experimental
evidence for NLS breathers in hydrodynamics was obtained by
\citet{Chabchoub2011,Chabchoub2012}, who observed Peregrine
and higher-order breather solutions in deep-water wave-tank
experiments, confirming that localized NLS structures can be
physically realized in hydrodynamic wave systems. These
experiments, however, were performed for classical deep-water
gravity waves, whereas the corresponding problem for
interfacial waves is considerably more complex because the
dispersive and nonlinear properties of the wave envelope
depend strongly on the density contrast, surface tension, and
carrier wavenumber.


The universality of breather dynamics has led to their
application in a wide variety of nonlinear dispersive
systems. In plasma physics, NLS breather solutions have been
used to describe localized extreme wave structures and the
nonlinear evolution of MI~\citep{ElTantawy2017}, including the formation of
Akhmediev breathers in electronegative plasmas and the
influence of plasma parameters on their
characteristics~\citep{KHAN2025108967}. Experimental evidence
for the stability of fundamental breather modes has recently
been reported in hydrodynamics~\citep{HE2026135098}, while
related manifestations of MI have also
been investigated for waves propagating in viscous
fluid-filled elastic tubes~\citep{Ding2024}.


The NLS governing weakly nonlinear wave packets propagating
along the interface between two fluid half-spaces was first
derived by \citet{Nayfeh1978}. The dependence of its
dispersive and nonlinear coefficients on the density
contrast, surface tension, and carrier wavenumber leads to a
considerably richer parameter space than in the classical
deep-water wave problem.

The theory was subsequently extended by
\citet{Grimshaw1985}, who carried out a systematic analysis
of MI of finite-amplitude interfacial
waves. Their study revealed the existence of several
disconnected MI regions determined by
the resonant properties of the dispersion relation. For
finite-depth two-layer fluids,
\citet{Christodoulides1995} investigated MI of interfacial capillary--gravity waves using the
method of multiple scales together with
Davey--Stewartson-type equations.


Subsequent studies of MI in
two-layer fluids focused on the influence of the physical
properties of the medium on the NLS coefficients and the
structure of instability regions. In particular,
\citet{Selezov2003} demonstrated that surface tension may
substantially modify the instability diagram and give rise
to additional focusing regimes. The dispersive and weakly
nonlinear properties of wave packets in finite-depth
two-layer fluids with a free surface were analyzed by
\citet{Avramenko2015,Avramenko2016}, who examined the
combined influence of the density contrast, layer geometry,
and surface tension on both surface and interfacial waves.

More recently, NLS models have been extended to increasingly
realistic configurations of stratified fluids.
\citet{Li2020} derived an NLS equation for interfacial waves
in two-layer fluids of arbitrary finite depth and
constructed MI diagrams, demonstrating
that the instability boundaries depend strongly on the
density ratio and layer depths. This approach was
subsequently generalized to include background shear currents
in both fluid layers~\citep{Li2021,Li2022}. Similar effects
of background shear on the instability boundaries and growth
rates were also reported by
\citet{CHOWDHURY2023103186}.


Breather dynamics in stratified fluids has been investigated
within a variety of theoretical models, including
weakly nonlinear long-wave
models~\citep{Grimshaw1999,Pelinovsky2000,Grimshaw2001},
NLS-based descriptions of internal
waves~\citep{Chan2018,Chow2019,Talipova2020}, and fully
nonlinear Euler
models~\citep{Lamb2007}. Experimental and theoretical aspects
of rogue-wave formation were comprehensively reviewed by
\citet{Kharif2009Observation}. However, these studies
consider physical configurations that differ substantially
from interfacial waves propagating between two fluid
half-spaces considered in the present work.


Recent studies of interfacial waves in two-layer fluids have
revealed the topology of MI diagrams,
including localized instability regions, stability
corridors, and resonant structures in the system parameter
space~\citep{Avramenko_Naradovyi_2025,Avramenko_Naradovyi_2026}.
These results identify the parameter domains in which
Akhmediev breathers may exist. However, they provide little
information about the physical characteristics of these
breathers, including their spatial scales, the growth rate
of MI, the importance of nonlinear
corrections, and the resulting shape of the reconstructed
interfacial surface.


The present work provides a systematic characterization of
Akhmediev breather regimes for waves propagating along the
interface between two fluid half-spaces. Parameter maps are
constructed to quantify their principal physical
properties, and representative regimes are then selected for
the reconstruction of the corresponding interfacial wave
profiles. This approach establishes a direct connection
between the mathematical properties of NLS breathers and the
physical structure of the interfacial surface.

\section{Mathematical model}
\label{sec:model}

\subsection{Preliminaries}
\label{subsec:nls}

We consider the propagation of weakly nonlinear wave packets
along the interface separating two incompressible,
inviscid fluids with densities \(\rho_1\) and \(\rho_2\),
occupying the infinitely deep half-spaces
\(\Omega_1=\{(x,z):|x|<\infty,\; z<0\}\) and
\(\Omega_2=\{(x,z):|x|<\infty,\; z>0\}\),
respectively. Surface tension with coefficient \(T\) is
taken into account at the interface.

The governing problem for the velocity potentials
\(\phi_j(x,z,t)\;(j=1,2)\) and the interface elevation
\(z=\eta(x,t)\) is formulated in dimensionless variables
using characteristic scales based on the gravitational
acceleration \(g\), the density of the lower fluid
\(\rho_1\), and a reference value of the surface tension
coefficient \(T_0\):
\[
L=\sqrt{T_0/(\rho_1 g)},
\qquad
t_0=\sqrt{L/g},
\qquad
m_0=\rho_1L^3.
\]

The dimensionless variables (the superscript ``*'' is
omitted hereafter) are introduced as
\[
(x^*,z^*)=\frac{(x,z)}{L},
\quad
(\rho_1^*,\rho_2^*)
=
\frac{(\rho_1,\rho_2)}{\rho_1},
\quad
t^*=\frac{t}{t_0},
\quad
T^*=\frac{T}{T_0},
\quad
\eta^*=\frac{\eta}{\alpha L},
\quad
(\phi_1^*,\phi_2^*)
=
\frac{(\phi_1,\phi_2)}
{\alpha L^2/t_0},
\]
where \(\alpha=a_c/l\) is the small wave-steepness parameter,
\(a_c\) is the characteristic amplitude of the interface
deformation, and \(l\) is the characteristic wavelength, and
\(
\rho=\rho_2/\rho_1
\)
is the density ratio.

Unlike the formulation of \cite{Nayfeh1978}, in which
the characteristic length scale is defined using the actual
value of the surface tension coefficient, the present study
employs a fixed reference value \(T_0\). This choice allows
the dimensionless surface tension coefficient \(T\) to be
treated as an independent parameter of the problem. Although
\(T\) is retained explicitly in the governing equations, only
the case \(T=1\) is considered in the present work.

In dimensionless variables, the governing equations take the
form
\begin{align}
& \quad	\Delta \phi_{j}  = 0 \quad \text{in} \quad \Omega_{j} ,  \label{statmentOfPr}\\
&	 \eta_{,t} -  \phi_{j,z} = - \alpha \eta_{,x} \phi_{j,x} \quad \text{at} \quad z=\alpha \eta(x,t),
\nonumber\\
&	  \phi_{1,t} \!-\!   \rho \phi_{2,t}\! + \!(1 \! - \! \rho) \eta\! +\! 0.5 \alpha \left( \nabla \phi_{1} \right)^{2} \! - \! 0.5 \alpha \rho \left( \nabla \phi_{2} \right)^{2} - T \left( 1 + \left( \alpha \eta_{,x} \right)^{2} \right)^{-3/2} \! \eta_{,xx} = 0 \,\,\,\, \text{at} \, z= \alpha \eta(x,t),
\nonumber\\
&  \left| \nabla \phi_{j} \right|  \rightarrow 0 \quad \text{at} \quad z \rightarrow \pm\infty.
\nonumber
\end{align}

Application of the method of multiple scales yields the
asymptotic expansions
\[
(\eta,\phi_j)
=
\sum_{n=1}^{3}
\alpha^{n-1}
(\eta_n,\phi_{jn})
+
O(\alpha^3),
\]
where the slow variables are introduced as
\[
x_n=\alpha^n x,
\qquad
t_n=\alpha^n t,
\qquad
n=0,1,2,
\]
and the functions \(\eta_n\) and \(\phi_{jn}\) represent
successive approximations to the solution.

In what follows, we present only those results
from~\cite{Nayfeh1978} that are required for the subsequent
analysis. The first-order dispersion relation implies that
linear waves exist only if
\begin{equation}
k>k_{\mathrm{cr}},
\qquad
k_{\mathrm{cr}}
=
\sqrt{\frac{\rho-1}{T}},
\label{eq:linear_condition}
\end{equation}
for \(\rho>1\), whereas for \(\rho\leq1\) this condition is
satisfied for all \(k>0\).

The complex envelope \(A\) is defined as the amplitude of the
fundamental harmonic of the interface elevation,
\begin{equation}
\eta_1=
Ae^{i\theta}
+\overline{A}e^{-i\theta},
\label{eq:eta_1}
\end{equation}
where
\(
\theta=kx-\omega t
\)
is the fast phase variable.

The second- and third-order approximations account for weakly
nonlinear effects, while the solvability condition of the
third-order problem yields the evolution equation
\begin{equation}
 iA_t
+\frac12\omega'' A_{xx}
+4\alpha^2\omega^{-1}J |A|^2A=0,
\label{eq:Nayfeh}
\end{equation}
where
\begin{equation}
J=
-\frac{k}{16(1+\rho)}
\left[
4(1-\rho)\omega^2\Lambda
+
4(1+\rho)\omega^2 k
-
3Tk^4
\right],
\label{eq:J}
\end{equation}
\begin{equation}
\Lambda=
\frac{\omega^2(1-\rho)}
{1-\rho-2Tk^2},
\label{eq:Lambda}
\end{equation}
and
\[
\omega''=\frac{d^2\omega}{dk^2}.
\]

A standard stability analysis of the plane-wave solution of
Eq.~(\ref{eq:Nayfeh}) yields the MI condition
\begin{equation}
\omega''J>0.
\label{eq:focusing_J}
\end{equation}

The corresponding regions in the parameter plane
\((\rho,k)\) are bounded by the curves
\(\omega''=0\) and \(J=0\), together with the resonance curve
\begin{equation}
1-\rho-2Tk^2=0,
\label{eq:second_harmonic_resonance}
\end{equation}
at which the coefficients \(\Lambda\) and \(J\) become
unbounded.

For the subsequent analysis, the evolution
equation~(\ref{eq:Nayfeh}) is written in the standard form of
the NLS,
\begin{equation}
iA_t+PA_{xx}+Q|A|^2A=0,
\label{eq:NLS}
\end{equation}
where
\begin{equation}
P=\frac12\omega'',
\qquad
Q=4\alpha^2\omega^{-1}J.
\label{eq:P_and_Q}
\end{equation}
The focusing NLS regime is determined by the condition
\begin{equation}
PQ>0.
\label{eq:focusing}
\end{equation}
Substituting~(\ref{eq:P_and_Q}) gives
\[
PQ
=
2\alpha^2\omega^{-1}\omega''J,
\]
so that, for \(\omega>0\), conditions
(\ref{eq:focusing}) and~(\ref{eq:focusing_J}) are
equivalent. Consequently, the MI diagrams obtained by \cite{Nayfeh1978} for interfacial waves directly determine the parameter domains in which Akhmediev breathers exist.

\subsection{Akhmediev breather characteristics}
\label{subsec:Akhmediev}

In the focusing regime defined by
condition~(\ref{eq:focusing}), Eq.~(\ref{eq:NLS})
admits the Akhmediev breather
solution~\cite{Akhmediev1986,Akhmediev1987}. We introduce the
scaled variables

\begin{equation}
A=A_0\psi,
\qquad
\tau=|Q| A_0^2 t,
\qquad
\chi=
A_0
\sqrt{
\left|
\frac{Q}{2P}
\right|
}\,x ,
\label{eq:A_tau_Xi}
\end{equation}
where \(A_0\) is the amplitude of the uniform background wave
from which the MI develops. In the
variables \((\chi,\tau)\), the Akhmediev breather is given by

\begin{equation}
\psi(\chi,\tau)=
\left[
\frac{
(1-4a)\cosh(b\tau)
+\sqrt{2a}\cos(\Omega\chi)
+i b\sinh(b\tau)}
{\cosh(b\tau)-\sqrt{2a}\cos(\Omega\chi)}
\right]e^{i\tau}.
\label{eq:AB}
\end{equation}

The breather parameters are defined as

\begin{equation}
\Omega=2\sqrt{1-2a},
\qquad
b=\sqrt{8a(1-2a)},
\qquad
0<a<\frac12 .
\end{equation}

The parameter \(\Omega\) determines the spatial modulation of
the breather in the normalized coordinate \(\chi\), whereas
\(b\), which appears in the hyperbolic functions of
Eq.~(\ref{eq:AB}), determines the growth rate of
MI.

Since the spatial dependence of the Akhmediev breather
solution~(\ref{eq:AB}) is contained in the factor
\(\cos(\Omega\chi)\), substituting the scaled coordinate
(\ref{eq:A_tau_Xi}) and returning to the physical coordinate
\(x\) yields
\[
\cos(\Omega\chi)=\cos(K_Bx),
\]
where
\begin{equation}
K_B=
\Omega A_0
\sqrt{
\left|
\frac{Q}{2P}
\right|
}.
\label{eq:KB}
\end{equation}
Here \(K_B\) is the wavenumber of the envelope modulation,
and the corresponding spatial period of the breather is
defined as
\begin{equation}
L_B=
\frac{2\pi}{K_B}.
\label{eq:LB}
\end{equation}

On the other hand, the temporal localization of the breather
in the normalized variables is governed by the parameter
\(b\), which appears in the factors
\(\cosh(b\tau)\) and \(\sinh(b\tau)\) of
Eq.~(\ref{eq:AB}). Substituting the time scaling
\(\tau=|Q|A_0^2t\) from Eq.~(\ref{eq:A_tau_Xi}) gives the
arguments of the hyperbolic functions in the form
\(\Gamma t\), where
\begin{equation}
\Gamma
=
|Q|A_0^2 b
=
|Q|A_0^2
\sqrt{8a(1-2a)} .
\label{eq:Gamma}
\end{equation}

Thus, combining the Akhmediev breather parameters with the
NLS coefficients determined by Nayfeh's model yields two
principal physical characteristics of the breather regime:
the spatial scale \(L_B\) defined by Eq.~(\ref{eq:LB}) and
the growth rate \(\Gamma\) given by
Eq.~(\ref{eq:Gamma}). As in previous studies of breather
structures in stratified fluids and mKdV-type models, these
quantities establish a direct connection between the
parameters of the normalized solution and the physical
scales of the wave field~\citep{Lamb2007}.

\subsection{Reconstruction of the physical interfacial profile}
\label{subsec:reconstruction}
To reconstruct the interface elevation, the Akhmediev
breather solution~(\ref{eq:AB}) is written in the form
\(
\psi=(U+iV)e^{i\tau}.
\)
In what follows, the reconstructed interface elevation is
interpreted as the physical interfacial profile.

With the amplitude scaling introduced above, the complex
envelope becomes

\begin{equation}
A(x,t)=
A_0\left[U(x,t)+iV(x,t)\right]
\exp\left(iQA_0^2t\right),
\label{eq:A_physical}
\end{equation}
where \(Q=4\alpha^2\omega^{-1}J\), and the functions
\(U\) and \(V\) are determined by the Akhmediev breather
solution.

In physical coordinates, the functions
\(U\) and \(V\) depend on the combination
\(x-\omega't\), where
\(
\omega'=\frac{d\omega}{dk}
\)
is the group velocity of the wave packet, and are given by

\begin{equation}
U(x,t)=
\frac{
(1-4a)\cosh(\Gamma t)
+
\sqrt{2a}\,
\cos\left[
K_B\left(x-\omega' t\right)
\right]
}
{D(x,t)} ,
\label{eq:U_AB}
\end{equation}

\begin{equation}
V(x,t)
=
\frac{
b\sinh(\Gamma t)
}
{D(x,t)},
\label{eq:V_AB}
\end{equation}

where

\begin{equation}
D(x,t)
=
\cosh(\Gamma t)
-
\sqrt{2a}
\cos
\left[
K_B(x-\omega' t)
\right].
\label{eq:D_AB}
\end{equation}

To second-order accuracy, the interface elevation is
represented as
\begin{equation}
\eta(x,t)=\eta_1+\alpha\eta_2 .
\label{eq:eta_sum}
\end{equation}

The first-order approximation corresponds to the fundamental
harmonic given by Eq.~(\ref{eq:eta_1}), whereas the
second-order approximation describes the bound second
harmonic generated by the quadratic nonlinearity,
\begin{equation}
\eta_2=
\Lambda A^2 e^{2i\theta}
+
\Lambda \overline{A}^{\,2}e^{-2i\theta},
\label{eq:eta2}
\end{equation}
where the coefficient \(\Lambda\), defined by
Eq.~(\ref{eq:Lambda}), characterizes the strength of
second-harmonic generation and depends on the parameters
of the carrier wave \((\rho,k,T)\).

Substituting Eqs.~(\ref{eq:A_physical}) and
(\ref{eq:eta2}) into Eq.~(\ref{eq:eta_sum}) yields

\begin{equation}
\eta(x,t)
=
2A_0
\left[
U\cos\Theta
-
V\sin\Theta
\right]
+
2\alpha A_0^2\Lambda
\left[
(U^2-V^2)\cos 2\Theta
-
2UV\sin 2\Theta
\right],
\label{eq:eta_physical}
\end{equation}
where

\begin{equation}
\Theta=
kx-\omega t+QA_0^2t .
\label{eq:Theta}
\end{equation}

The first term in Eq.~(\ref{eq:eta_physical})
represents the fundamental harmonic with the characteristic
amplitude \(A_0\), whereas the second term represents the
nonlinearly generated bound second harmonic with the characteristic
amplitude \(\alpha A_0^2\Lambda\).
To quantify the relative contribution of the second harmonic,
we introduce the dimensionless parameter
\begin{equation}
R_{20}
=
\frac{\alpha A_0^2|\Lambda|}{A_0}
=
\alpha A_0|\Lambda|.
\label{eq:R20}
\end{equation}
The subscript ``2'' refers to the second harmonic, whereas
the subscript ``0'' indicates normalization by the
characteristic amplitude \(A_0\) of the fundamental harmonic.

The parameter \(R_{20}\) characterizes the relative
contribution of the second harmonic to the reconstructed
interface elevation and, together with \(L_B\) and
\(\Gamma\), is used to construct maps of the breather
regimes.

\section{Breather property maps}
\label{sec:maps}

\subsection{Construction of breather property maps}
\label{subsec:maps_construction}

Based on the nonlinear Schrödinger
equation~(\ref{eq:NLS}) and the Akhmediev breather
solution~(\ref{eq:AB}), parameter maps were constructed to
characterize the properties of breathers at the interface
between two fluid half-spaces.

Representative breather regimes were selected using maps of
the breather modulation period
\(L_B\)~(\ref{eq:LB}),
the MI growth rate
\(\Gamma\)~(\ref{eq:Gamma}),
and the relative contribution of the second harmonic
\(R_{20}\)~(\ref{eq:R20}),
presented in this section.

All maps show the boundaries of the MI regions
derived in Section~\ref{sec:model}: the curve
\(J=0\) (red), the curve
\(\omega''=0\) (green), and the resonance curve
\(J=\infty\) (blue), defined by
Eq.~(\ref{eq:second_harmonic_resonance}).
The gray region below the black curve
\(
k_c=\sqrt{(\rho-1)/T},
\)
corresponds to the domain of linear instability of the
interface. The maps of the parameter \(R_{20}\) also include
the contour lines \(R_{20}=0.1\) and \(R_{20}=1\), which
separate regions with weak and significant contributions of
the second harmonic.

In the upper MI region, the curve \(J=0\) has two
vertical asymptotes,
\[
\rho=3-2\sqrt2\approx0.1716,
\qquad
\rho=3+2\sqrt2\approx5.8284,
\]
which it approaches as the wavenumber \(k\) increases.
Near the first asymptote, the shape of the upper MI
region and the distribution of the breather parameters
change markedly. For this reason, enlarged views of the
region
\(0.12<\rho<0.18\),
\(0<k<12\)
are provided for several maps, allowing a more detailed
examination of the breather characteristics in the vicinity
of
\(\rho\approx0.1716\).

Since all characteristics considered here are determined by
the same NLS coefficients, the MI regions have the same
geometry in all parameter maps. The maps differ only in the
distribution of the corresponding characteristic within these
regions.

\subsection{Maps of the breather spatial scale}
\label{subsec:LB}

\begin{figure}
\centering

\begin{minipage}{0.4\textwidth}
\centering
\includegraphics[width=\linewidth]{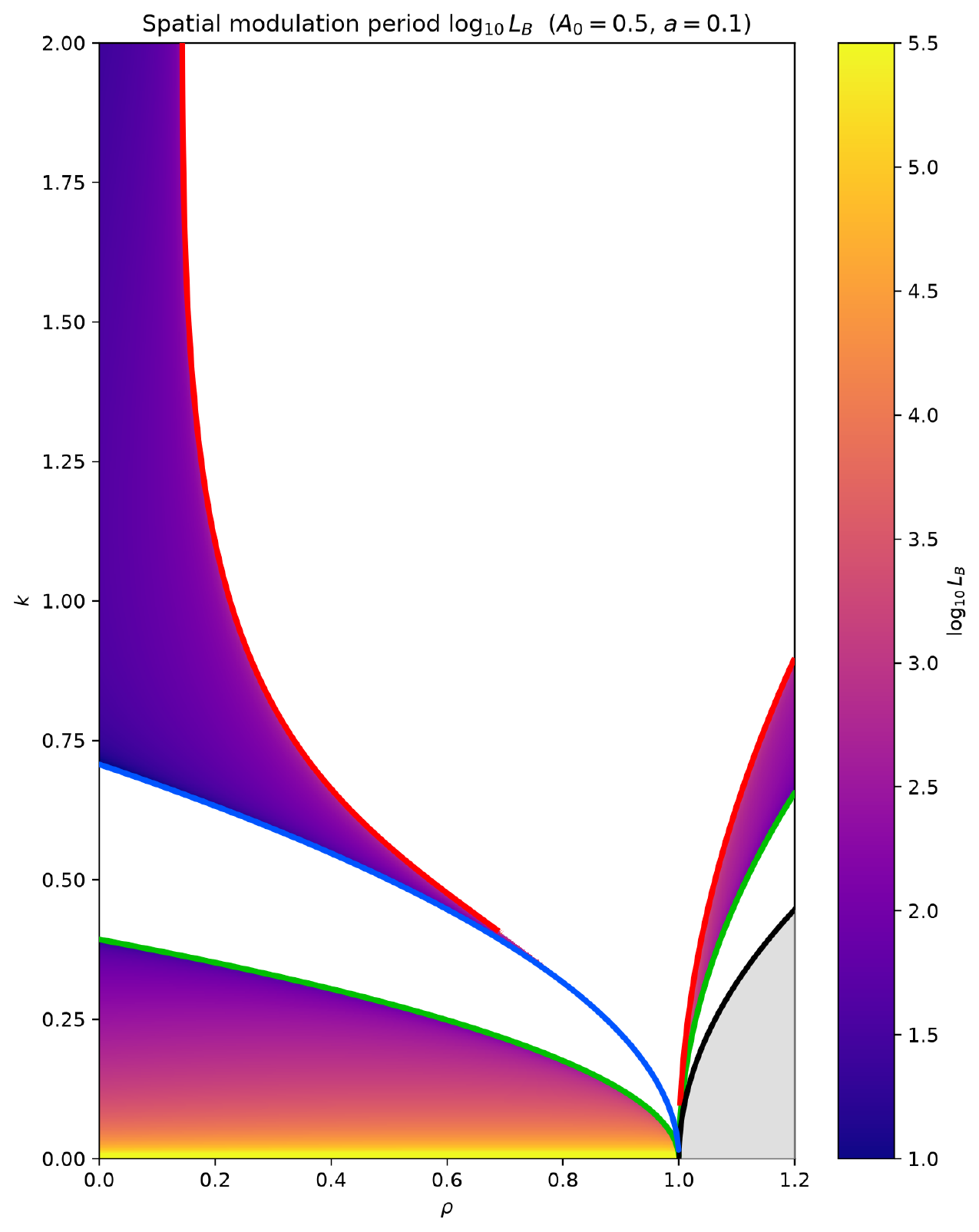}
\end{minipage}
\hspace{0.4cm}
\begin{minipage}{0.28\textwidth}
\centering
\includegraphics[width=\linewidth]{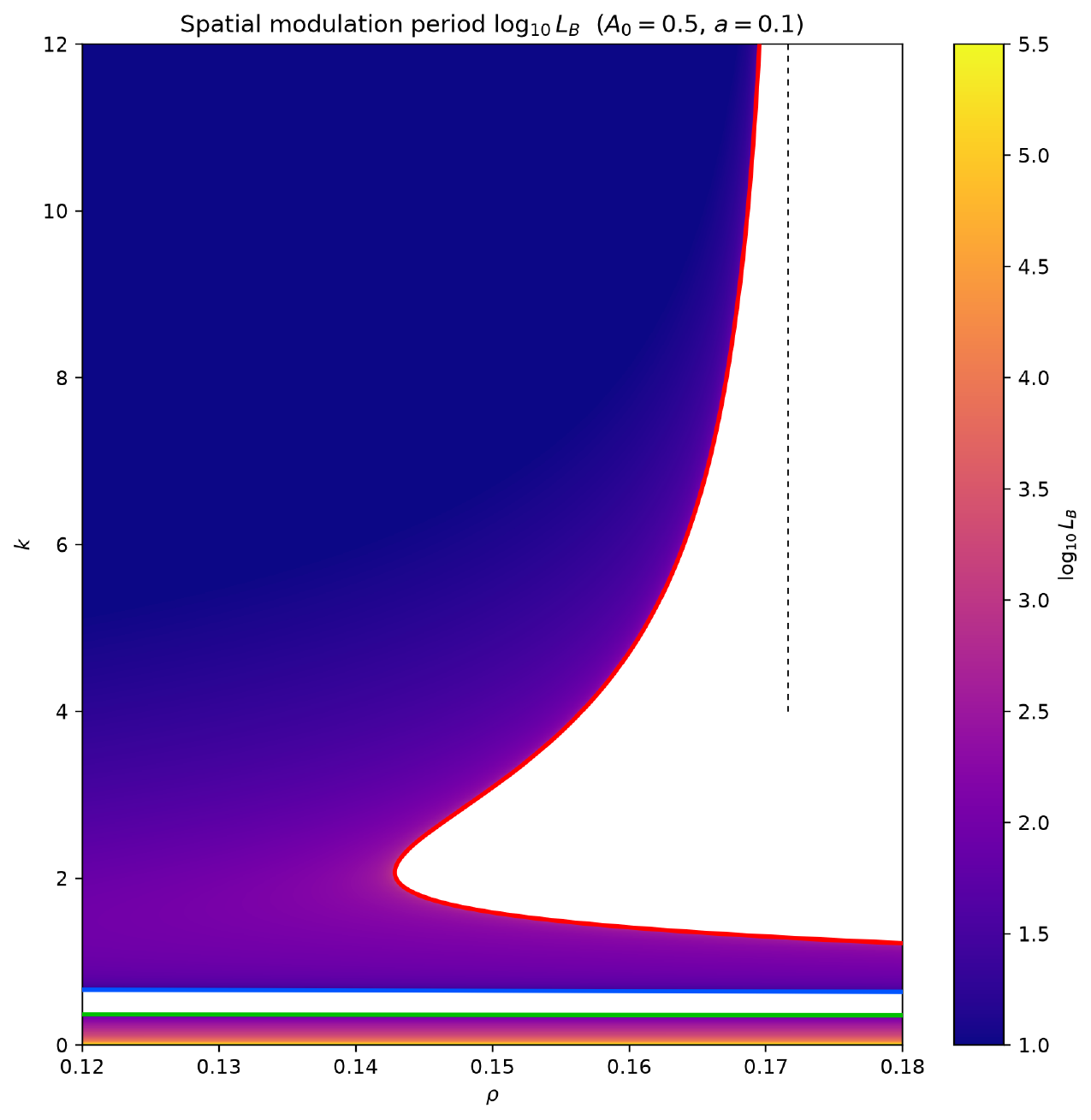}
\end{minipage}

\caption{
Distribution of the breather modulation period
\(\log_{10}L_B\)
for \(A_0=0.5\) and \(a=0.1\):
the full parameter domain (left) and an enlarged view
of the region near the asymptote
\(\rho\approx0.1716\) (right).
}
\label{fig:LBLog}
\end{figure}

The spatial scale of the Akhmediev breather is characterized
by the modulation period \(L_B\) defined by
Eq.~(\ref{eq:LB}).
Since the values of \(L_B\) span several orders of magnitude,
Fig.~\ref{fig:LBLog} shows the distribution of
\(\log_{10}L_B\).
Small values correspond to compact breather structures,
whereas large values indicate long-wavelength spatial
modulations.

For fixed values of
\((\rho,k)\), the modulation period scales as
\(L_B\propto (A_0\Omega(a))^{-1}\).
Therefore, varying the parameters \(A_0\) and \(a\)
primarily changes the overall magnitude of \(L_B\)
without significantly altering its distribution over the
\((\rho,k)\) parameter plane.
Accordingly, the following analysis is carried out for the
representative parameter values
\(A_0=0.5\) and \(a=0.1\).

In the lower MI region,
\(\log_{10}L_B\) spans nearly nine orders of magnitude,
whereas in the upper MI region it is generally
restricted to
\(\log_{10}L_B\lesssim3\).
A sharp increase in the modulation period is observed only
in the vicinity of the red boundary \(J=0\).

The spatial period exhibits fundamentally different
behavior in the lower and upper MI regions.
In the lower MI region, for a fixed value of
\(\rho\), the modulation period \(L_B\) decreases
almost monotonically with increasing wavenumber \(k\),
from the axis \(k=0\) to the green boundary
\(\omega''=0\).
The variation occurs primarily along the \(k\)-direction,
whereas the dependence on \(\rho\) is considerably weaker.

In the upper MI region, the smallest values of
\(L_B\) are found near the blue boundary
\(J=\infty\).
As the upper red boundary \(J=0\) is approached, the
modulation period increases gradually, while the
\(\log_{10}L_B\) contours become increasingly aligned with
this boundary.
Most of the upper MI region is characterized by
\(\log_{10}L_B\lesssim2\), whereas only a narrow band
adjacent to the \(J=0\) boundary exhibits a rapid increase
in \(L_B\), reaching
\(\log_{10}L_B>4\)--\(5\) and diverging as the boundary is
approached. This behavior is consistent with the scaling
\(L_B\propto |Q|^{-1/2}\) together with the limiting relation
\(Q\propto J\to0\).

An additional analysis over the extended parameter domain
\((0<\rho<7.2,\;0<k<12)\)
shows that the observed trends persist for larger values of
\(\rho\) and \(k\).
In particular, the \(\log_{10}L_B\) contours become
progressively aligned with the \(J=0\) boundary, an effect
that is especially pronounced in the upper MI region
at large values of \(\rho\) and \(k\).

\subsection{Maps of the MI growth rate}
\label{subsec:Gamma}

\begin{figure}
\centering

\begin{minipage}{0.4\textwidth}
\centering
\includegraphics[width=\linewidth]{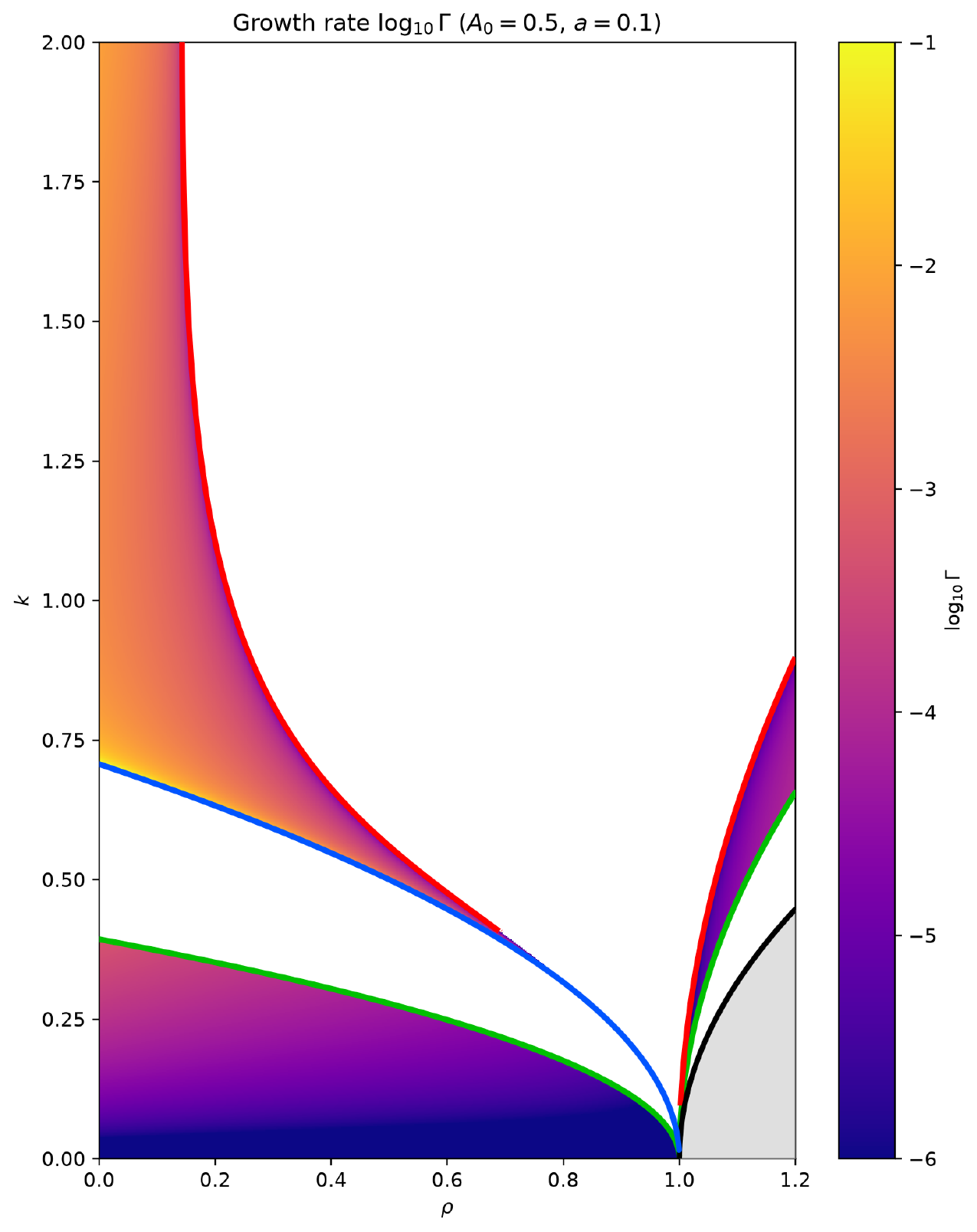}
\end{minipage}
\hspace{0.4cm}
\begin{minipage}{0.28\textwidth}
\centering
\includegraphics[width=\linewidth]{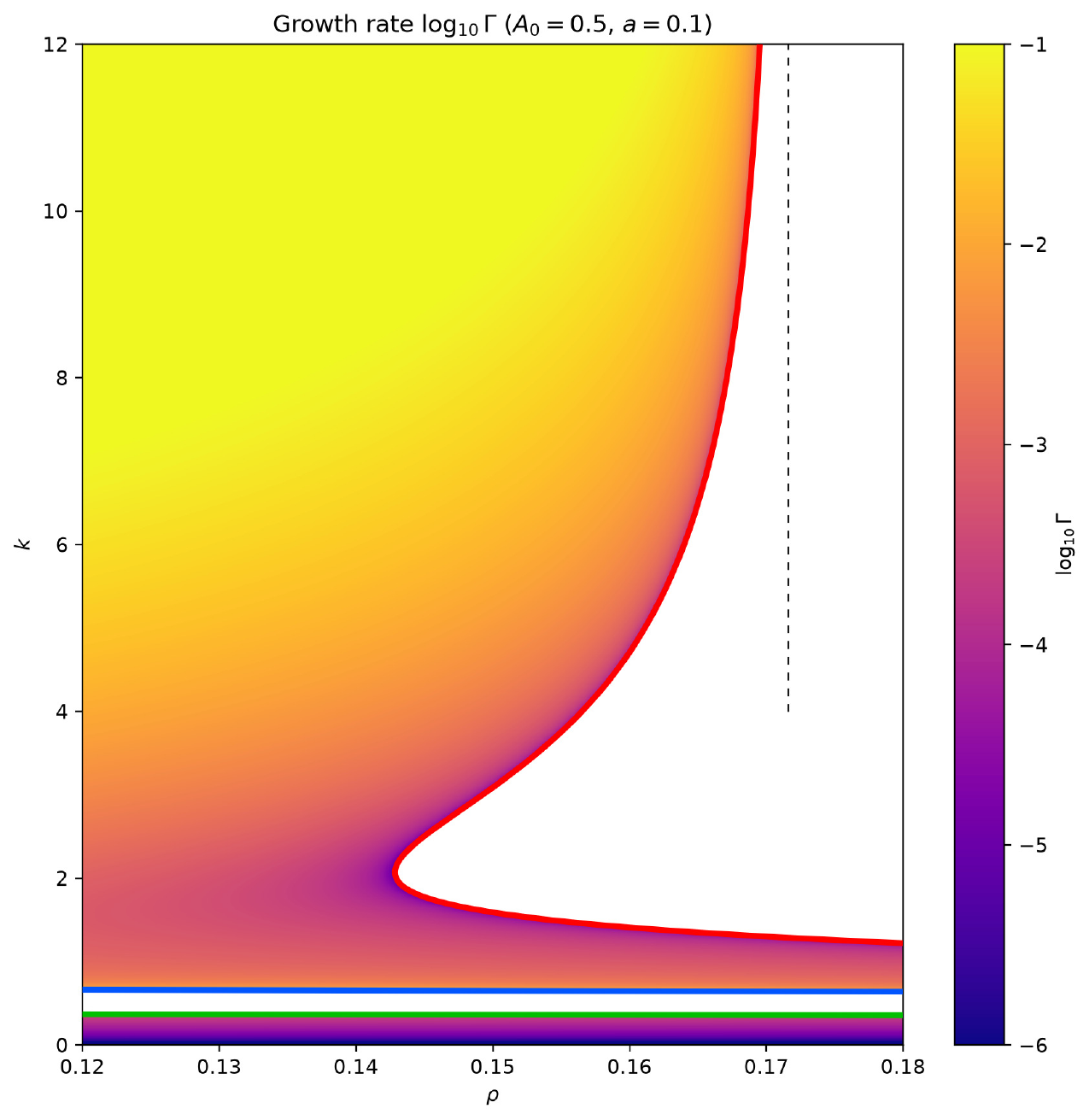}
\end{minipage}

\caption{
Distribution of the MI growth rate
\(\log_{10}\Gamma\)
for \(A_0=0.5\) and \(a=0.1\):
the full parameter domain (left) and an enlarged view
of the region near the vertical asymptote
\(\rho\approx0.1716\) (right).
}
\label{fig:GammaLog}
\end{figure}

The rate of Akhmediev breather formation is characterized by
the MI growth rate \(\Gamma\),
defined by Eq.~(\ref{eq:Gamma}).
Large values of \(\Gamma\) correspond to the rapid
development of MI, whereas small
values indicate slow formation of the breather structure.
Because \(\Gamma\) spans several orders of magnitude,
Fig.~\ref{fig:GammaLog} presents the distribution of
\(\log_{10}\Gamma\).

According to Eq.~(\ref{eq:Gamma}),
the MI growth rate is proportional to the square of the
background-wave amplitude,
\(\Gamma\propto A_0^2\).
Consequently, increasing \(A_0\) produces a uniform increase
in the instability growth rate throughout the entire
MI region.
The dependence on the breather parameter \(a\) is governed by
the factor
\(b=\sqrt{8a(1-2a)}\),
which attains its maximum at
\(a=0.25\) and takes the same value for
\(a=0.1\) and \(a=0.4\).
The following analysis is therefore performed for the
representative parameter values
\(A_0=0.5\) and \(a=0.1\).

In contrast to the modulation period \(L_B\), the largest
values of \(\Gamma\) are located well away from the
\(J=0\) boundary. As this boundary is approached, the
growth rate decreases monotonically and tends to zero.
Consequently, the regions with the largest values of
\(L_B\) are simultaneously characterized by the slowest
development of MI.

In the lower MI region, the growth rate
\(\Gamma\) increases primarily with increasing
wavenumber \(k\).
At the same time, the
\(\log_{10}\Gamma\) contours exhibit a noticeable tilt
toward increasing values of \(\rho\), indicating an
additional, although weaker, influence of the density
ratio on the instability growth rate.

In the upper MI region, the largest values of
\(\Gamma\) are found near the blue boundary
\(J=\infty\).
Toward the upper red boundary \(J=0\), the growth rate
decreases gradually, while the
\(\log_{10}\Gamma\) contours become increasingly aligned
with this boundary.
As the \(J=0\) boundary is approached,
\(\Gamma\to0\), indicating that exponential growth of the modulation is no longer possible.
 This behavior is consistent with the scaling
\(\Gamma\propto |Q|\) together with the limiting relation
\(Q\propto J\to0\).

As for the modulation period \(L_B\), the growth rate
was also analyzed over the extended parameter domain
\((0<\rho<7.2,\;0<k<12)\).
The results show that the observed trends persist for
large values of \(\rho\) and \(k\).
In particular, the
\(\log_{10}\Gamma\) contours become progressively aligned
with the \(J=0\) boundary, whereas the rapid decrease in
the growth rate is confined to its immediate vicinity.

\subsection{Maps of the relative contribution of the second harmonic}
\label{subsec:R20}

To assess the range of validity of the weakly nonlinear
approximation, we employ the dimensionless parameter
\(R_{20}\), defined by Eq.~(\ref{eq:R20}), which
characterizes the relative contribution of the bound second
harmonic to the reconstructed interface elevation.
Small values of \(R_{20}\) correspond to regimes in which
the fundamental harmonic dominates, whereas
\(R_{20}=O(1)\) indicates that the contribution of the
second harmonic becomes comparable to that of the
fundamental harmonic.

\begin{figure}
\centering

\begin{minipage}{0.32\textwidth}
\centering
\includegraphics[width=\linewidth]{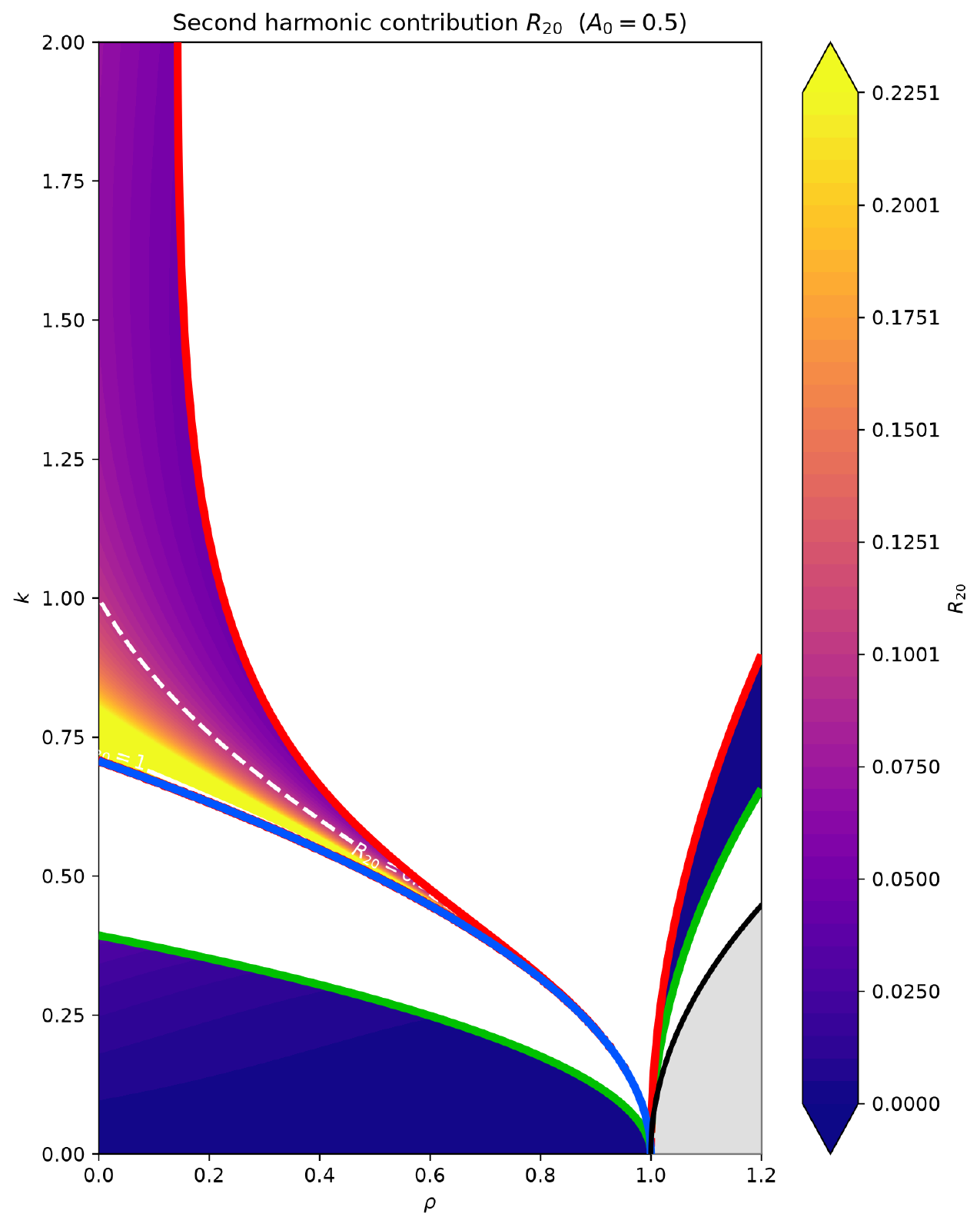}

\vspace{2mm}

\includegraphics[width=0.6\linewidth]{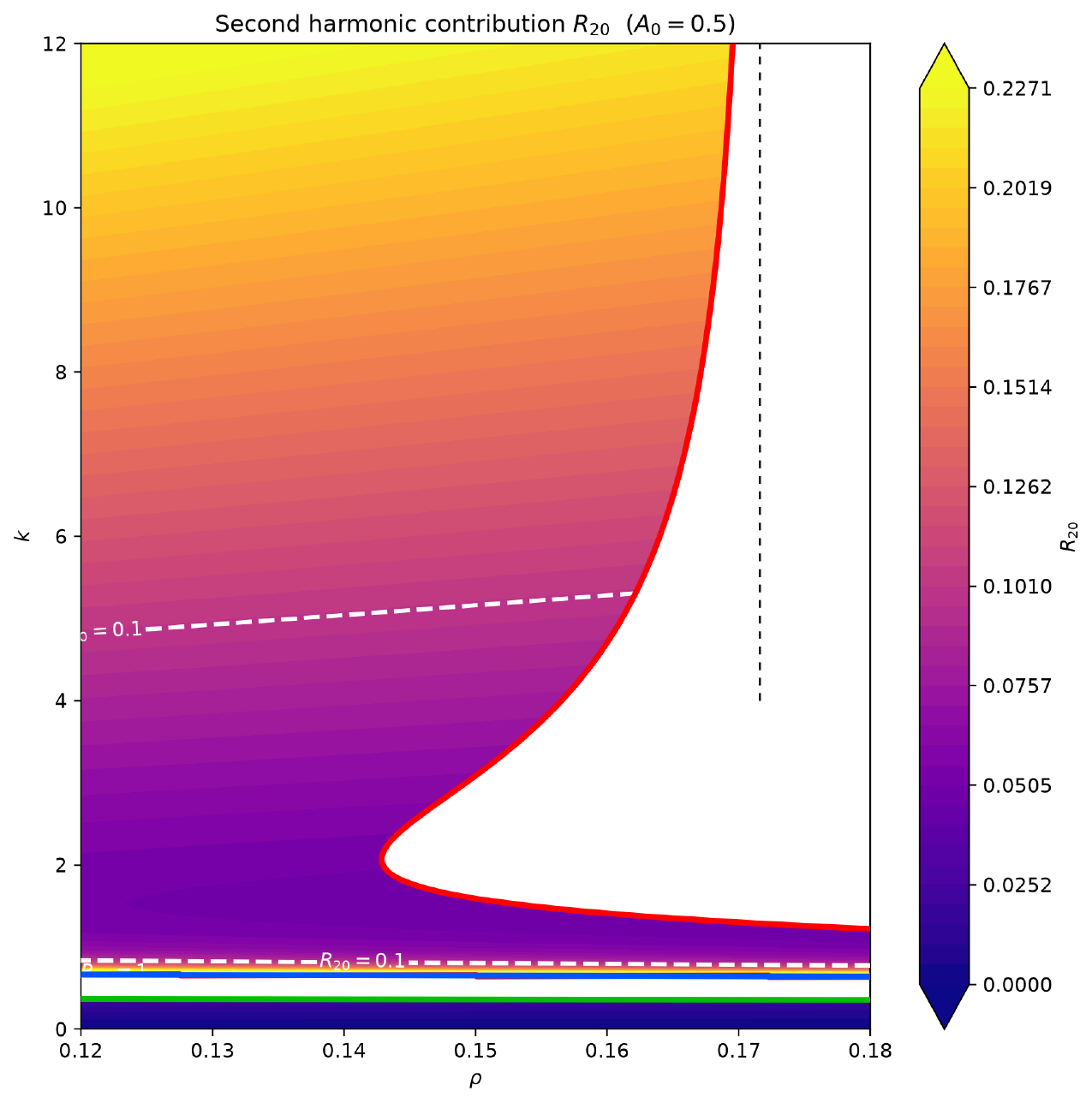}

\vspace{1mm}

\textbf{(a)} $A_0=0.5$
\end{minipage}
\hfill
\begin{minipage}{0.32\textwidth}
\centering
\includegraphics[width=\linewidth]{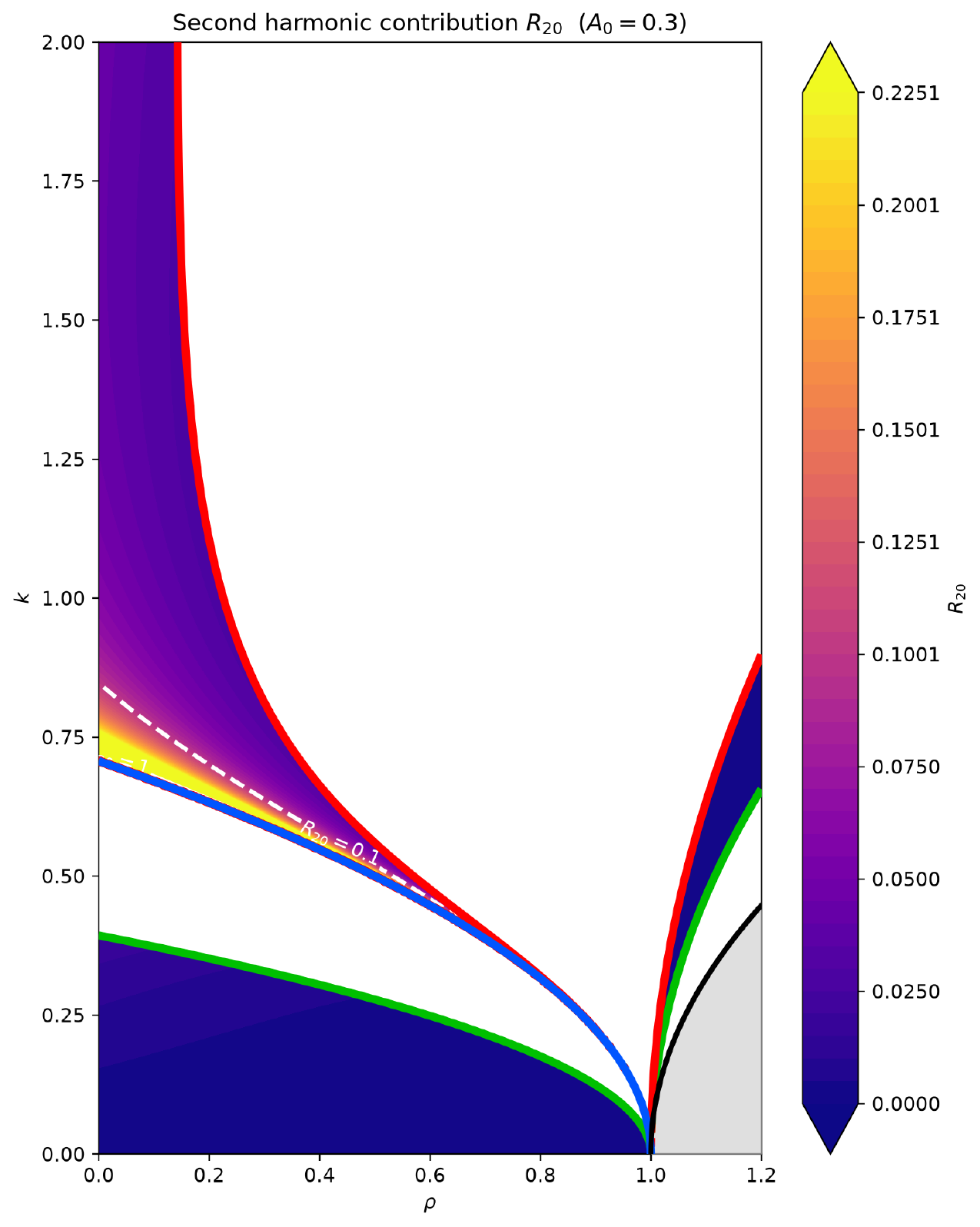}

\vspace{2mm}

\includegraphics[width=0.6\linewidth]{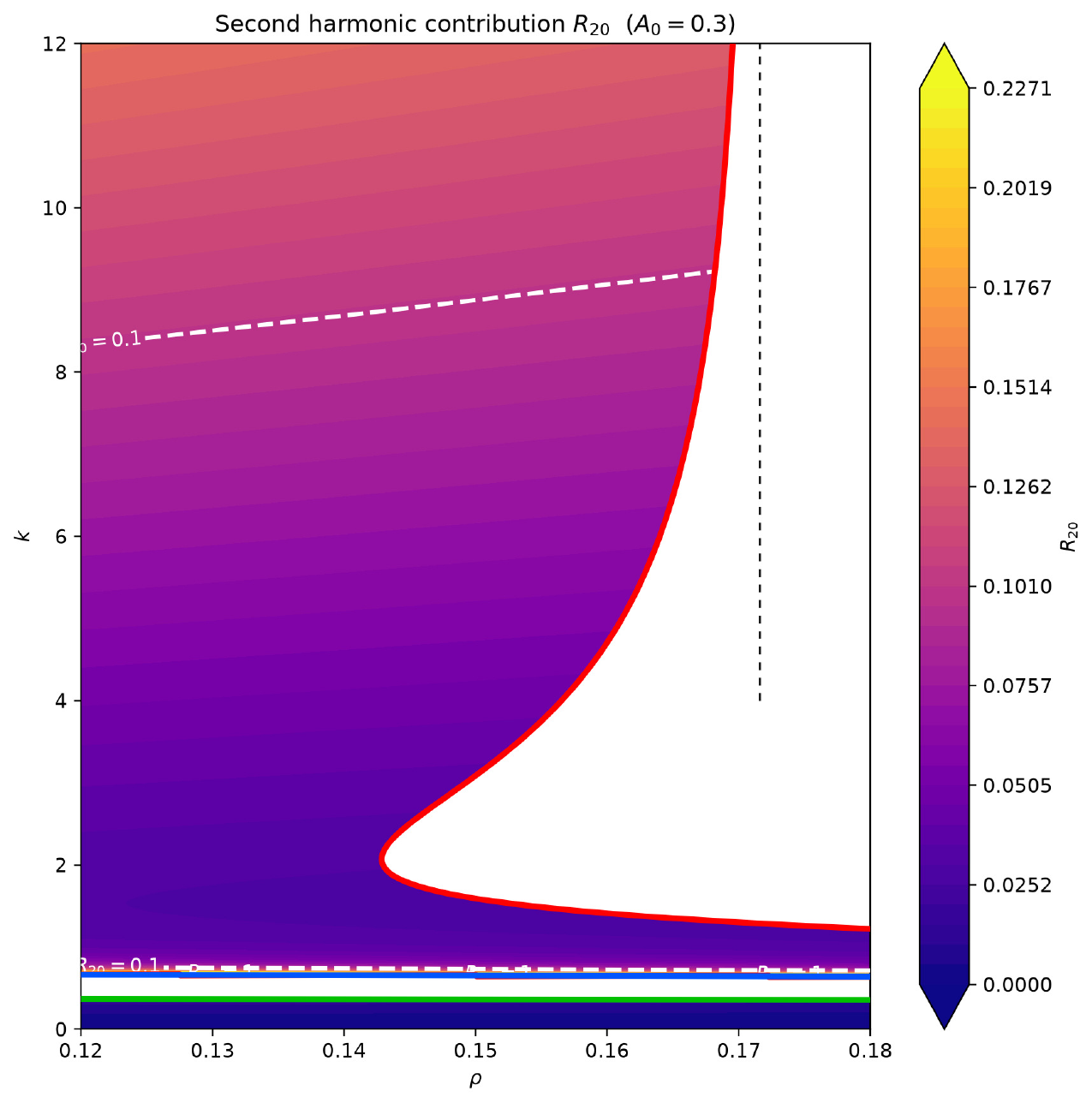}

\vspace{1mm}

\textbf{(b)} $A_0=0.3$
\end{minipage}
\hfill
\begin{minipage}{0.32\textwidth}
\centering
\includegraphics[width=\linewidth]{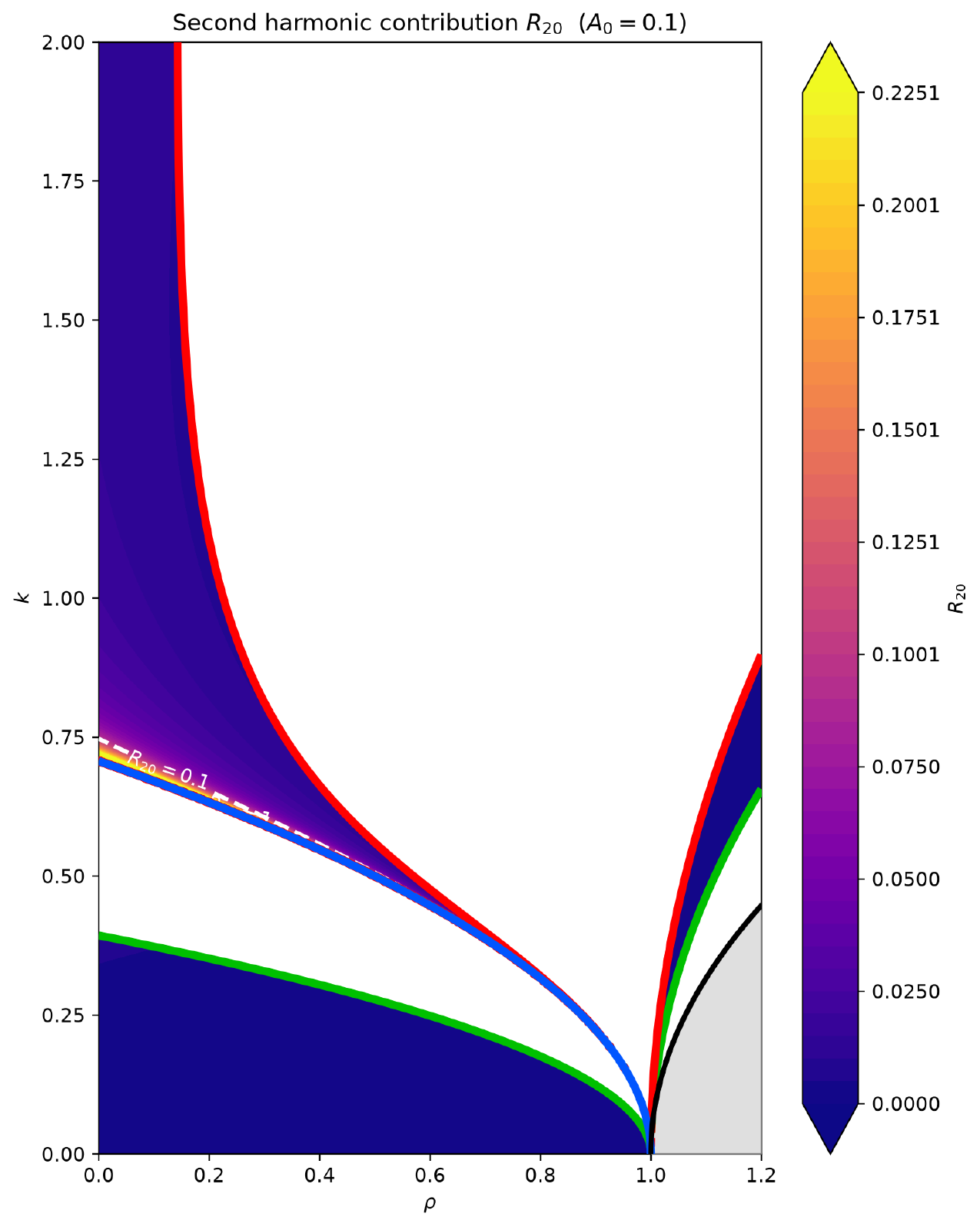}

\vspace{2mm}

\includegraphics[width=0.6\linewidth]{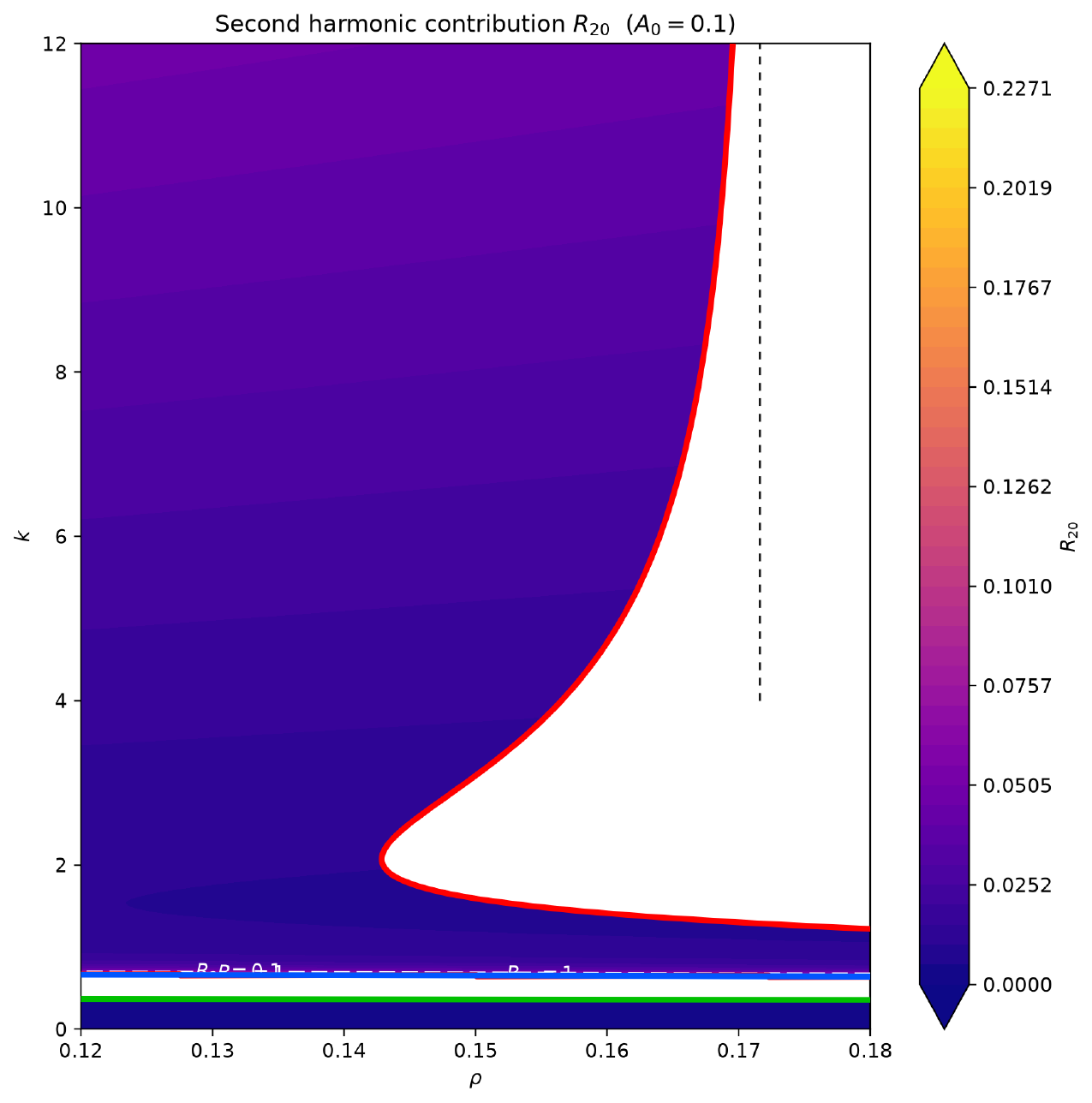}

\vspace{1mm}

\textbf{(c)} $A_0=0.1$
\end{minipage}

\caption{
Distribution of the relative contribution of the second
harmonic \(R_{20}\)
(top row) and enlarged views of the region near the
vertical asymptote
\(\rho\approx0.1716\)
(bottom row) for three values of the background-wave
amplitude:
(a) \(A_0=0.5\),
(b) \(A_0=0.3\),
and (c) \(A_0=0.1\).
}
\label{fig:R20_A0}
\end{figure}

\begin{figure}
\centering

\begin{minipage}{0.32\textwidth}
\centering
\includegraphics[width=\linewidth]{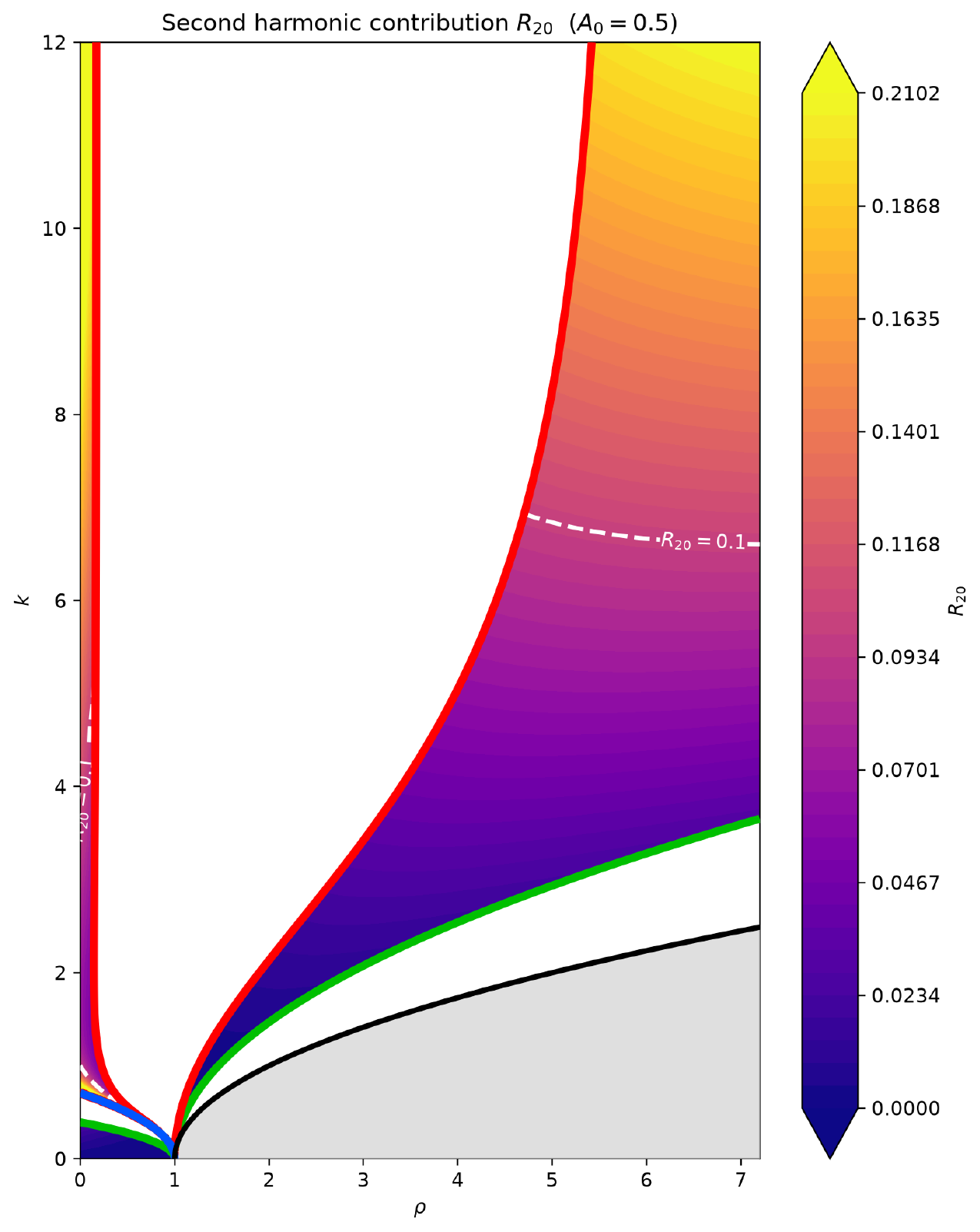}\\[-1mm]
\textbf{(a)} $A_0=0.5$
\end{minipage}
\hfill
\begin{minipage}{0.32\textwidth}
\centering
\includegraphics[width=\linewidth]{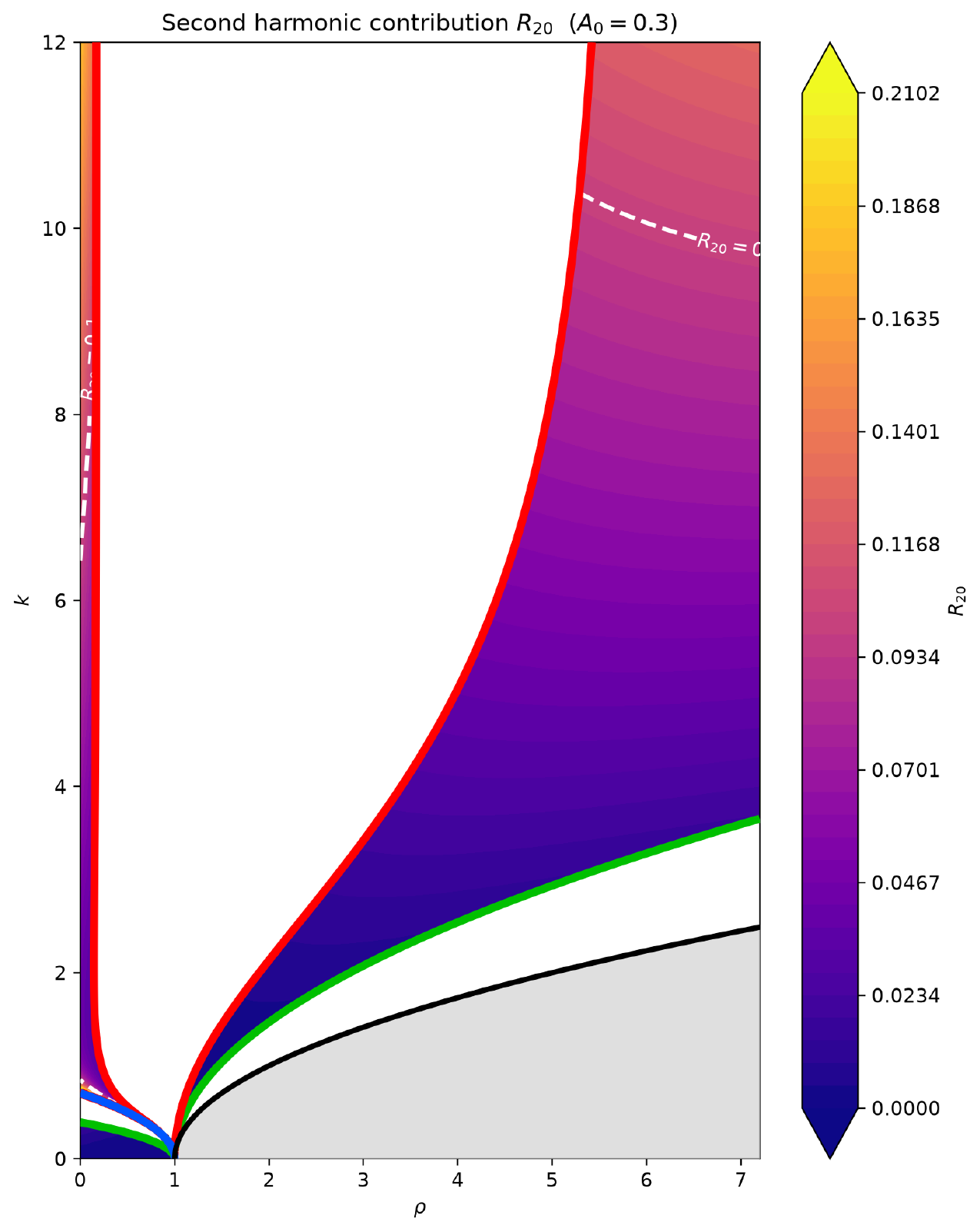}\\[-1mm]
\textbf{(b)} $A_0=0.3$
\end{minipage}
\hfill
\begin{minipage}{0.32\textwidth}
\centering
\includegraphics[width=\linewidth]{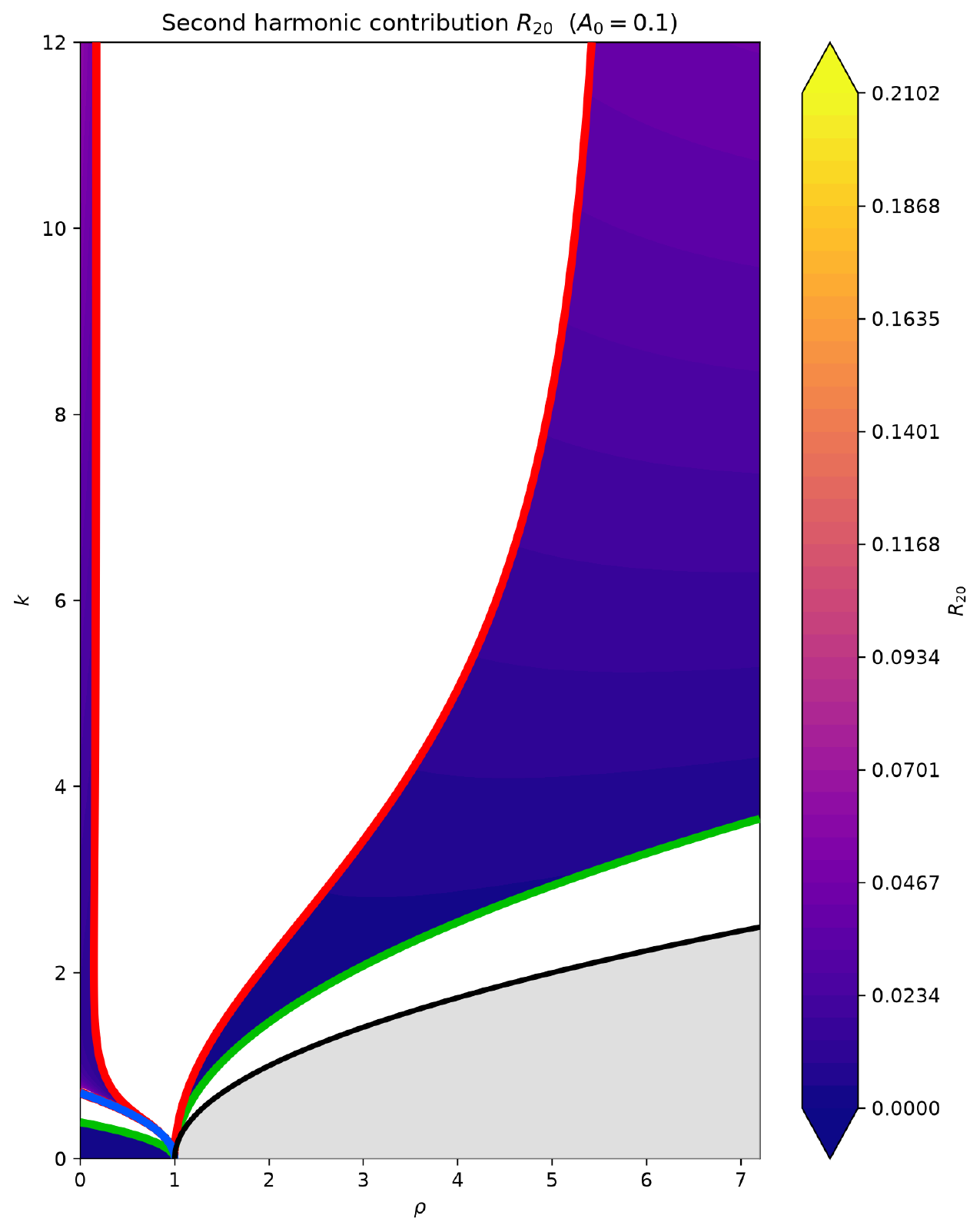}\\[-1mm]
\textbf{(c)} $A_0=0.1$
\end{minipage}

\caption{
Distribution of the relative contribution of the second
harmonic \(R_{20}\)
over the extended parameter domain
for three values of the background-wave amplitude:
(a) \(A_0=0.5\),
(b) \(A_0=0.3\),
and (c) \(A_0=0.1\).
}
\label{fig:R20_A0_large}
\end{figure}

Unlike the parameters \(L_B\) and \(\Gamma\),
\(R_{20}\) is independent of the breather parameter \(a\).
According to Eq.~(\ref{eq:R20}),
\(R_{20}\propto A_0\); therefore, increasing the
background-wave amplitude produces a uniform increase in
\(R_{20}\) throughout the parameter plane, while leaving
the overall structure of the maps essentially unchanged.
Accordingly, Figs.~\ref{fig:R20_A0} and
\ref{fig:R20_A0_large} present the distributions of
\(R_{20}\) for three values of the amplitude
\(A_0\) over the small and extended parameter domains,
respectively.

As shown in Figs.~\ref{fig:R20_A0} and
\ref{fig:R20_A0_large}, increasing the amplitude
\(A_0\) shifts the contours
\(R_{20}=0.1\) and \(R_{20}=1\) outward,
so that they encompass progressively larger portions of
the MI region.

As shown in Fig.~\ref{fig:R20_A0}, in the lower MI
region the values of \(R_{20}\) increase primarily with
increasing wavenumber \(k\), whereas their dependence on
the density ratio \(\rho\) is considerably weaker.
Unlike the parameters \(L_B\) and \(\Gamma\),
\(R_{20}\) is determined directly by the second-harmonic
coefficient \(\Lambda\).
Accordingly, the \(R_{20}\) contours are nearly parallel
to the \(\rho\)-axis, indicating that the wavenumber
\(k\) is the dominant factor controlling the relative
contribution of the second harmonic.

In the upper MI region shown in
Fig.~\ref{fig:R20_A0}, the values of \(R_{20}\)
increase monotonically from the blue boundary
\(J=\infty\) toward the upper red boundary
\(J=0\), while remaining relatively smooth throughout
the MI region.
As \(A_0\) increases, the overall level of
\(R_{20}\) rises uniformly, causing the region where
\(R_{20}>0.1\) to expand and shifting the
\(R_{20}=0.1\) and \(R_{20}=1\) contours toward
smaller wavenumbers.
At the same time, the \(R_{20}\) contours become
progressively aligned with the upper \(J=0\) boundary.

The extended parameter domain
(Fig.~\ref{fig:R20_A0_large})
shows that the trends identified in the small-domain maps
persist for large values of both \(\rho\) and \(k\).
At the same time, the effect of the background-wave
amplitude \(A_0\) becomes even more pronounced.
For small amplitudes, a substantial portion of the
MI region corresponds to regimes in which the
contribution of the second harmonic remains small or
moderate.
As \(A_0\) increases, the region where
\(R_{20}>0.1\) expands considerably, particularly in the
upper MI region and at large values of
\(\rho\) and \(k\).
Consequently, regimes with
\(R_{20}<0.1\) remain confined to only a limited portion
of the MI region.

Even over the extended parameter domain,
\(R_{20}\) varies relatively smoothly with increasing
\(\rho\).
Therefore, the relative contribution of the second
harmonic is more sensitive to variations in the
wavenumber \(k\) and the background-wave amplitude
\(A_0\) than to changes in the density ratio.

In contrast to the parameters \(L_B\) and \(\Gamma\),
which characterize the breather modulation period and the
growth rate of MI, respectively,
\(R_{20}\) quantifies the relative contribution of the
second harmonic and thus provides a measure of the range
of validity of the weakly nonlinear approximation.
The resulting maps demonstrate that the focusing condition
\(PQ>0\) is a necessary but not sufficient criterion for
its applicability.
Within the MI region, \(R_{20}\) may vary by
several orders of magnitude, indicating that the
contribution of the second harmonic depends strongly on
the choice of physical parameters.

The \(R_{20}\) maps therefore distinguish regimes in
which the contribution of the second harmonic remains
small from those in which nonlinear corrections become
significant.
Together with the \(L_B\) and \(\Gamma\) maps, they
demonstrate that the breather modulation period, the
growth rate of MI, and the relative
contribution of the second harmonic are governed by
different regions of the \((\rho,k)\) parameter plane.
Consequently, the selection of representative physical
regimes for reconstructing interfacial wave profiles
should be based on the combined analysis of all three
parameters.

\section{Physical breather profiles}
\label{sec:physical_profiles}

\subsection{Selection of representative regimes and reconstruction of interfacial profiles}
\label{subsec:selection}

To analyze the physical profile of the fluid interface,
this section employs the reconstruction given by
Eq.~(\ref{eq:eta_physical}), which incorporates the
contributions of both the fundamental and bound
second harmonics.
This reconstruction provides a direct link between the
parameter maps presented in
Section~\ref{sec:maps} and the corresponding physical
breather profiles.
The subsequent analysis focuses on representative
regimes selected from different MI regions,
which differ in the breather modulation period, the
MI growth rate, and the relative
contribution of the second harmonic.

\begin{figure}
\centering

\begin{minipage}{0.4\textwidth}
\centering
\includegraphics[width=\textwidth]{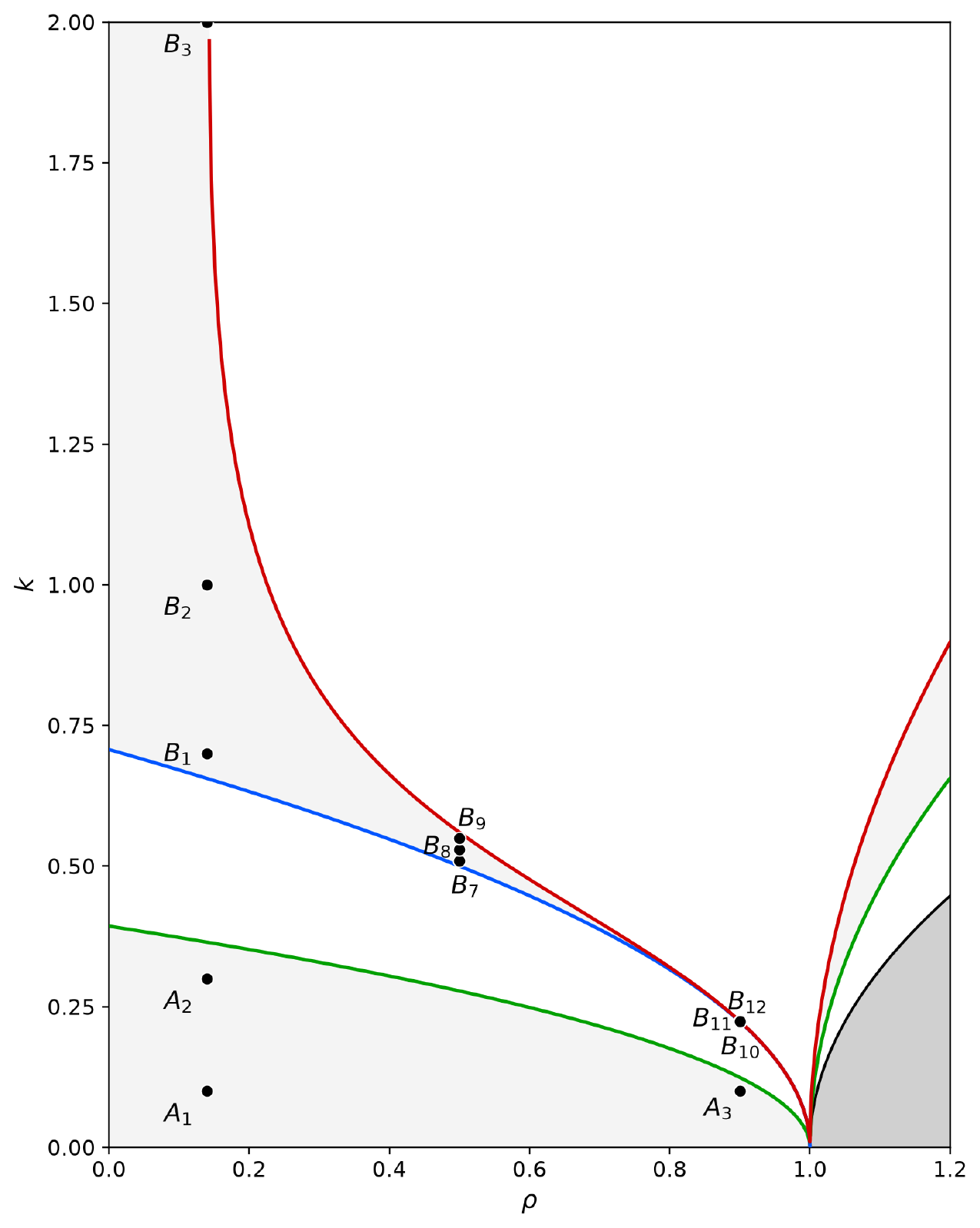}
\end{minipage}
\hspace{0.4cm}
\begin{minipage}{0.3\textwidth}
\centering
\includegraphics[width=\textwidth]{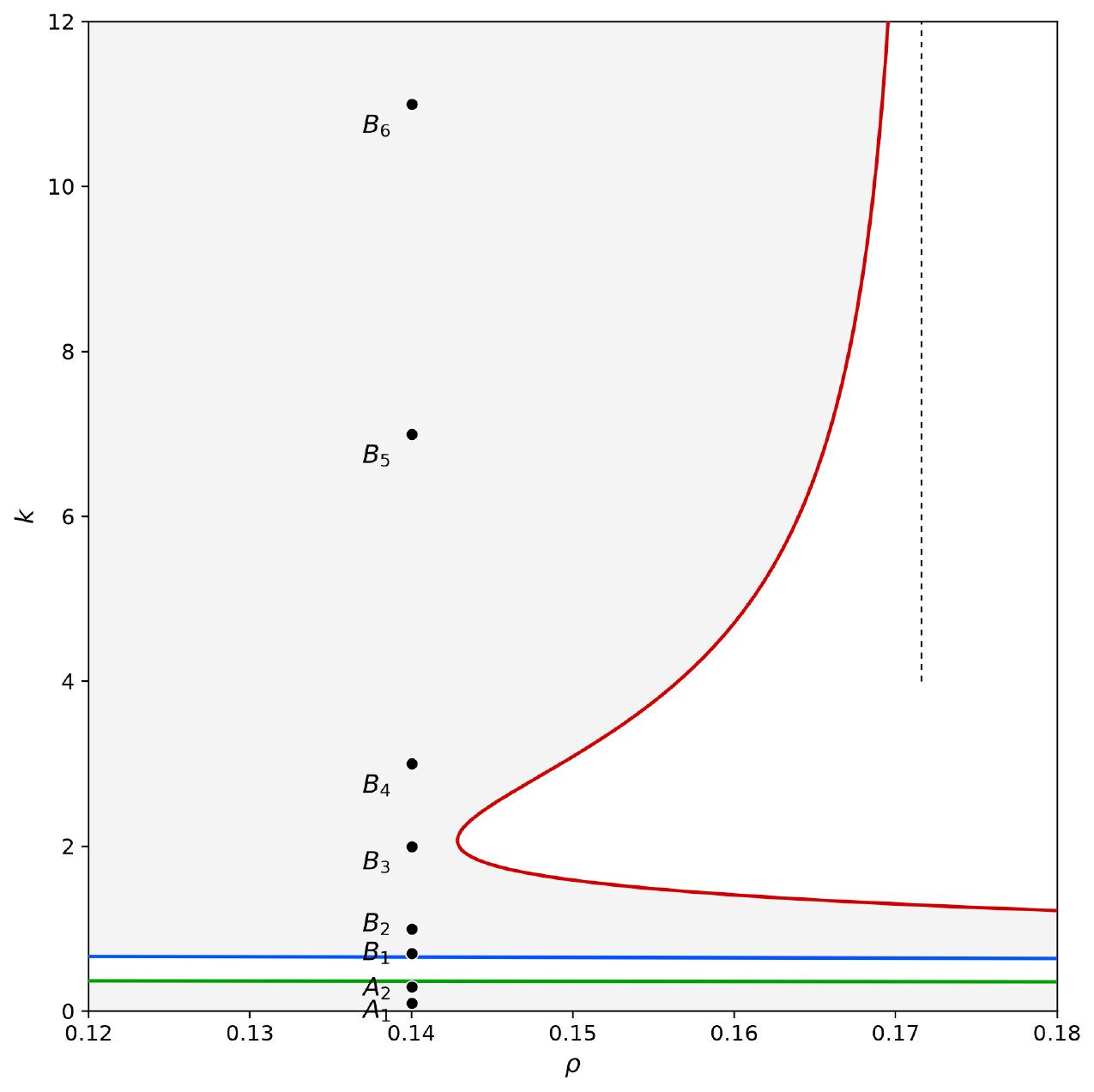}
\end{minipage}

\caption{
Representative regimes selected for the reconstruction
of physical breather profiles:
the \((\rho,k)\) parameter map (left) and an enlarged
view of the region near the vertical asymptote
\(\rho\approx0.1716\) (right).
}
\label{fig:selected_regimes}
\end{figure}

Representative regimes and their locations in the
\((\rho,k)\) parameter plane are shown in
Fig.~\ref{fig:selected_regimes}.
The points \(A_i\) (\(i=1,\ldots,3\))
belong to the lower MI region, whereas the
points \(B_i\) (\(i=1,\ldots,12\))
represent different regimes within the upper MI
region.
For the points \(B_1\)--\(B_6\), an enlarged view
of the map is also provided to clearly illustrate their
locations in the vicinity of the first vertical asymptote
of the \(J=0\) boundary.

All figures in this section follow the same layout.
The left panels show the reconstructed interfacial profile
\(\eta=\eta_1+\alpha\eta_2\), whereas the right
panels display separately the contributions of the
fundamental (blue curve) and bound second
(orange curve) harmonics.
The columns (from left to right) correspond to the
breather parameter values
\(a=0.10\), \(0.25\), and \(0.40\),
whereas the rows (from top to bottom) correspond to the
background-wave amplitudes
\(A_0=0.5\), \(0.3\), and \(0.1\).
For each regime, the corresponding values of
\(L_B\), \(\Gamma\), and \(R_{20}\)
are also indicated, allowing the features of the
reconstructed interfacial profile to be directly related
to the corresponding breather characteristics.

The remainder of this section is organized according to
the structure of the MI regions.
The discussion begins with representative regimes from
the lower MI region for small and
large density ratios and then proceeds to representative
regimes from the upper MI region.

\subsection{Lower MI region}
\label{subsec:lower_instability}

The lower MI region corresponds to
long-wave gravity-dominated regimes of MI.
Breather structures exist over a broad range of
wavenumbers and density ratios, while the contribution
of the bound second harmonic generally remains moderate.
This makes the lower MI region
particularly suitable for analyzing the influence of the
physical parameters on the reconstructed interfacial
profiles within the framework of the weakly nonlinear
approximation.

\paragraph{Small density ratios.}

\begin{figure}
\centering

\begin{subfigure}{\textwidth}
\centering
\includegraphics[width=0.49\linewidth]{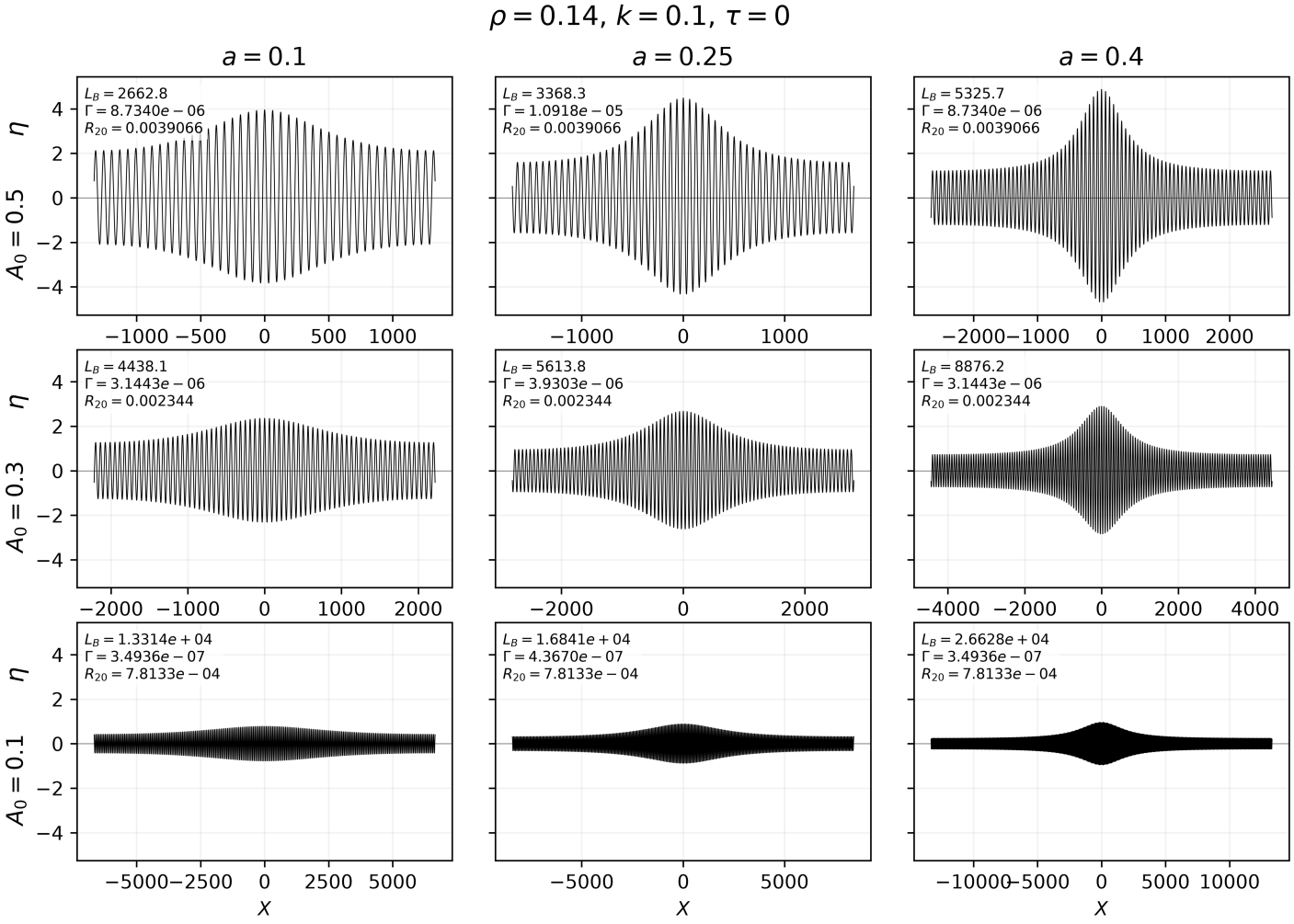}\hfill
\includegraphics[width=0.49\linewidth]{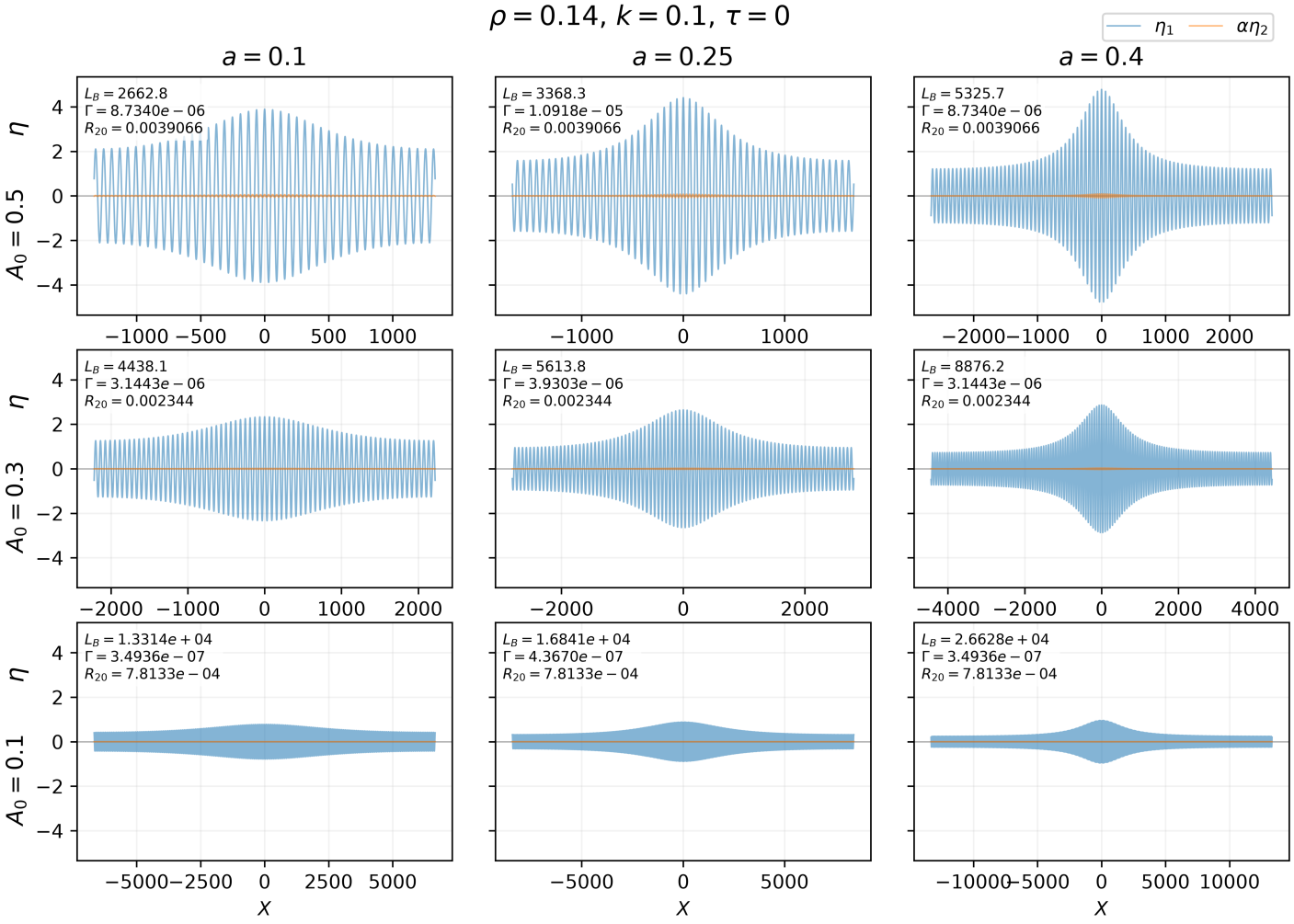}
\caption{$k=0.1$}
\label{fig:rho014_lower_branch:a}
\end{subfigure}

\vspace{2mm}

\begin{subfigure}{\textwidth}
\centering
\includegraphics[width=0.49\linewidth]{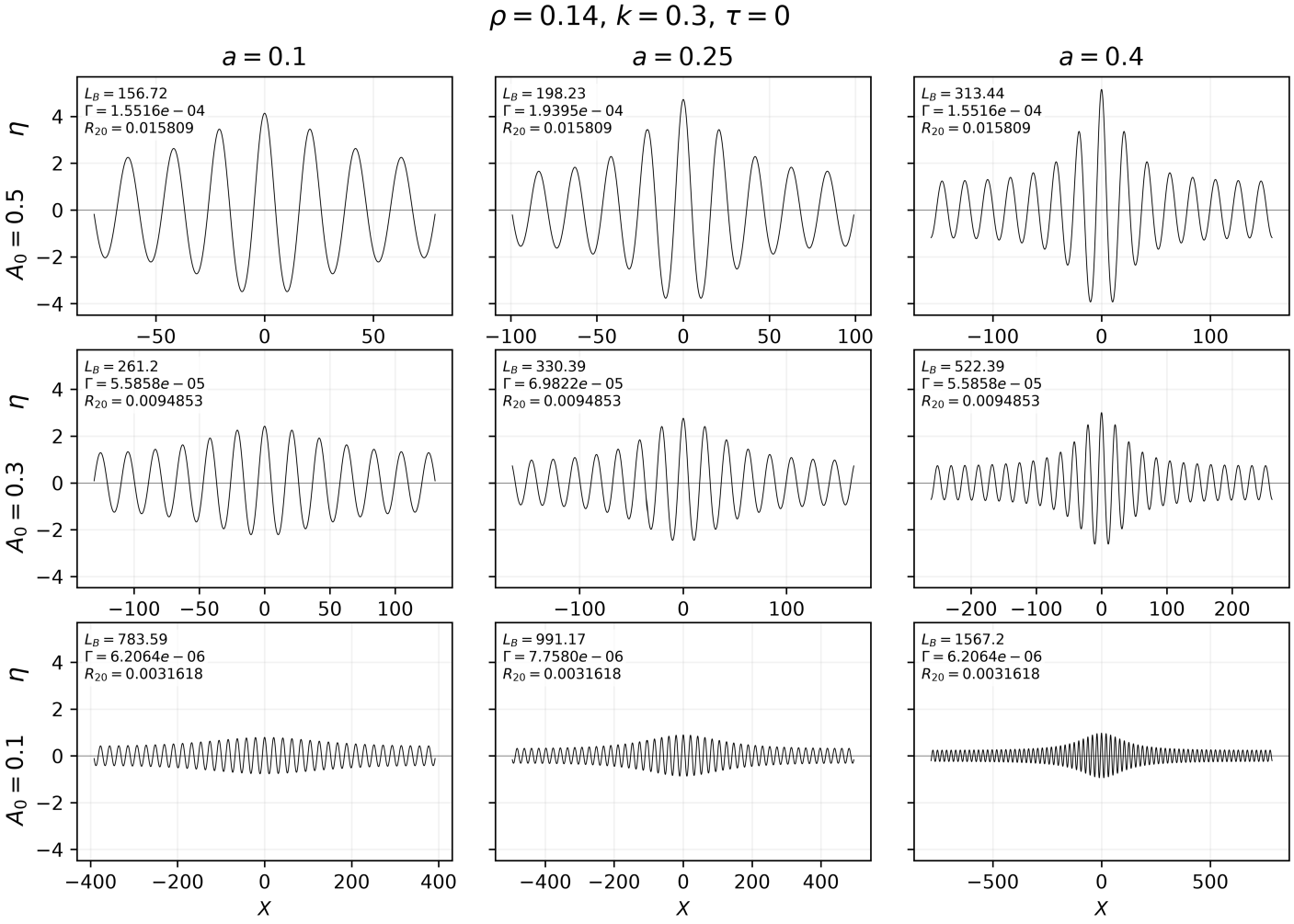}\hfill
\includegraphics[width=0.49\linewidth]{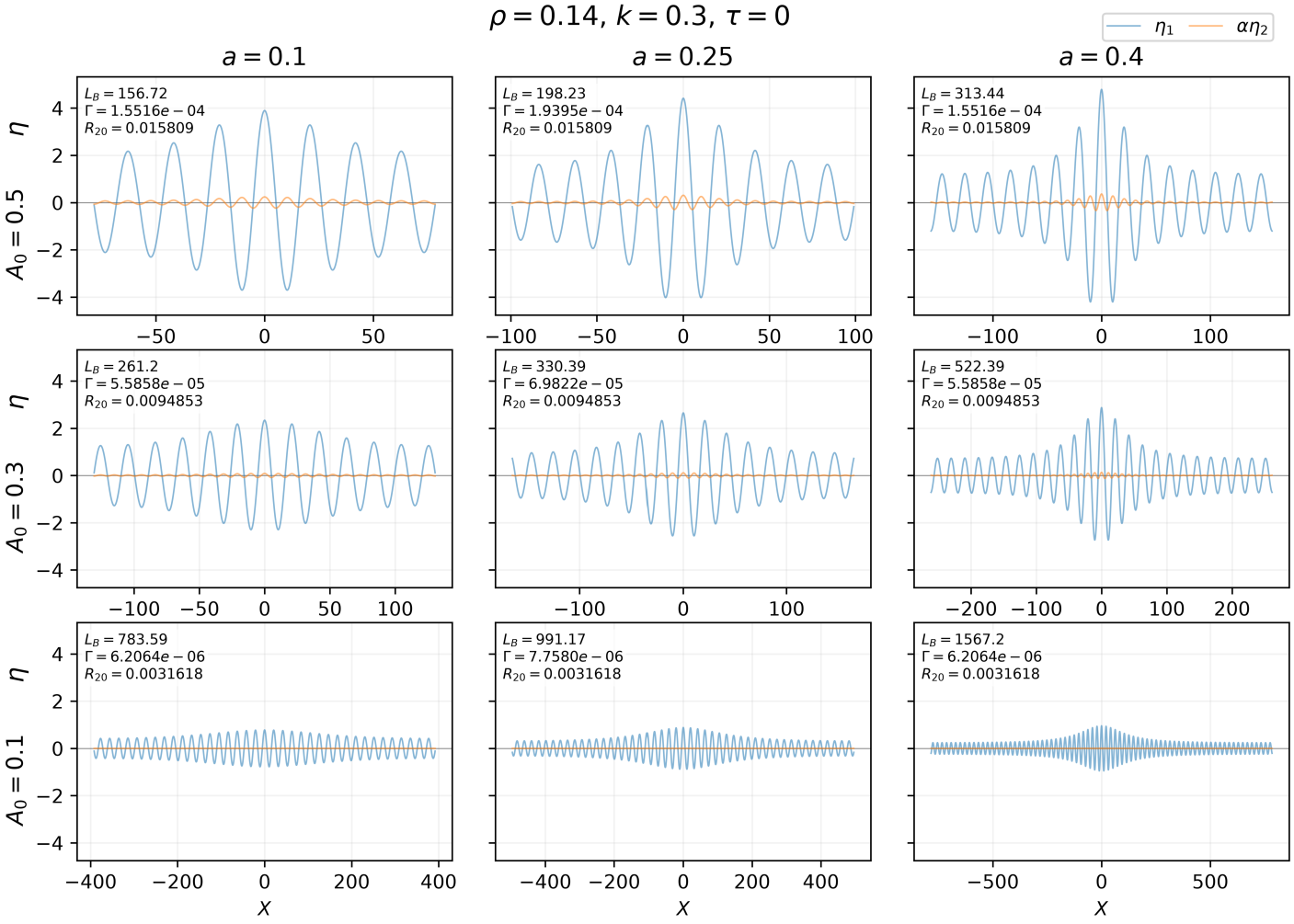}
\caption{$k=0.3$}
\label{fig:rho014_lower_branch:b}
\end{subfigure}

\caption{
Reconstructed breather profiles in the lower
MI region for \(\rho=0.14\).
Left: reconstructed interfacial profile
\(\eta=\eta_1+\alpha\eta_2\).
Right: contributions of the fundamental harmonic
\(\eta_1\) and the bound second harmonic
\(\alpha\eta_2\).
}
\label{fig:rho014_lower_branch}
\end{figure}

To examine the lower MI region at
small density ratios, the representative value
\(\rho=0.14\) was selected, corresponding to systems
with a pronounced density contrast.
Figure~\ref{fig:rho014_lower_branch} presents the
reconstructed breather profiles at the points
\(A_1\) and \(A_2\)
(Fig.~\ref{fig:selected_regimes}),
corresponding to the wavenumbers
\(k=0.1\) and \(k=0.3\), respectively.

For both values of \(k\), the reconstructed interfacial
profiles exhibit the characteristic shape of a localized
breather wave packet.
As the breather parameter \(a\) increases, the central
peak becomes narrower and its amplitude increases,
although the breather modulation period \(L_B\) also
increases.
The MI growth rate \(\Gamma\)
reaches its maximum at \(a=0.25\).
Since the parameter \(R_{20}\) is independent of \(a\),
the relative contribution of the bound second harmonic
remains unchanged within each row of the matrix.

As the background-wave amplitude \(A_0\) decreases, the
breather wave packet becomes more extended, its maximum
amplitude decreases, the breather modulation period
\(L_B\) increases, the MI growth
rate \(\Gamma\) decreases significantly, and the
relative contribution of the bound second harmonic is
also reduced.
Despite these quantitative changes, the shape of the
reconstructed interfacial profile remains nearly
unchanged, indicating the self-similar nature of the
breather solution.

A comparison of the cases \(k=0.1\)
(Fig.~\ref{fig:rho014_lower_branch:a}) and
\(k=0.3\)
(Fig.~\ref{fig:rho014_lower_branch:b})
demonstrates the pronounced influence of the wavenumber
on the spatial scale of the breather structure.
For all values of \(A_0\) and \(a\) considered here,
increasing \(k\) reduces the breather modulation period
\(L_B\) by more than an order of magnitude, while the
MI growth rate \(\Gamma\) and the
parameter \(R_{20}\) increase substantially.
As a result, for \(k=0.1\) the long-wave modulation
extends over many carrier-wave periods, whereas for
\(k=0.3\) a more compact breather wave packet with
stronger spatial localization is formed.

The right panels of
Figs.~\ref{fig:rho014_lower_branch:a} and
\ref{fig:rho014_lower_branch:b}
show that the reconstructed interfacial profile is
dominated by the fundamental harmonic, whereas the
bound second harmonic exhibits the same spatial
structure but with a much smaller amplitude.
Even when the wavenumber increases from \(k=0.1\) to
\(k=0.3\), resulting in an approximately fourfold
increase in the parameter \(R_{20}\), its contribution
remains small and does not produce any qualitative
change in the reconstructed profile.
Thus, for the lower MI region at
small density ratios, the weakly nonlinear
reconstruction remains well justified, and the physical
interfacial profile is determined predominantly by the
fundamental harmonic.

\paragraph{Large density ratios.}

To examine the lower MI region at
large density ratios, the representative value
\(\rho=0.9\) was selected.
This value is sufficiently close to the limiting case
\(\rho=1\), while avoiding the pronounced narrowing of
the MI region.

\begin{figure}
\centering

\includegraphics[width=0.49\textwidth]{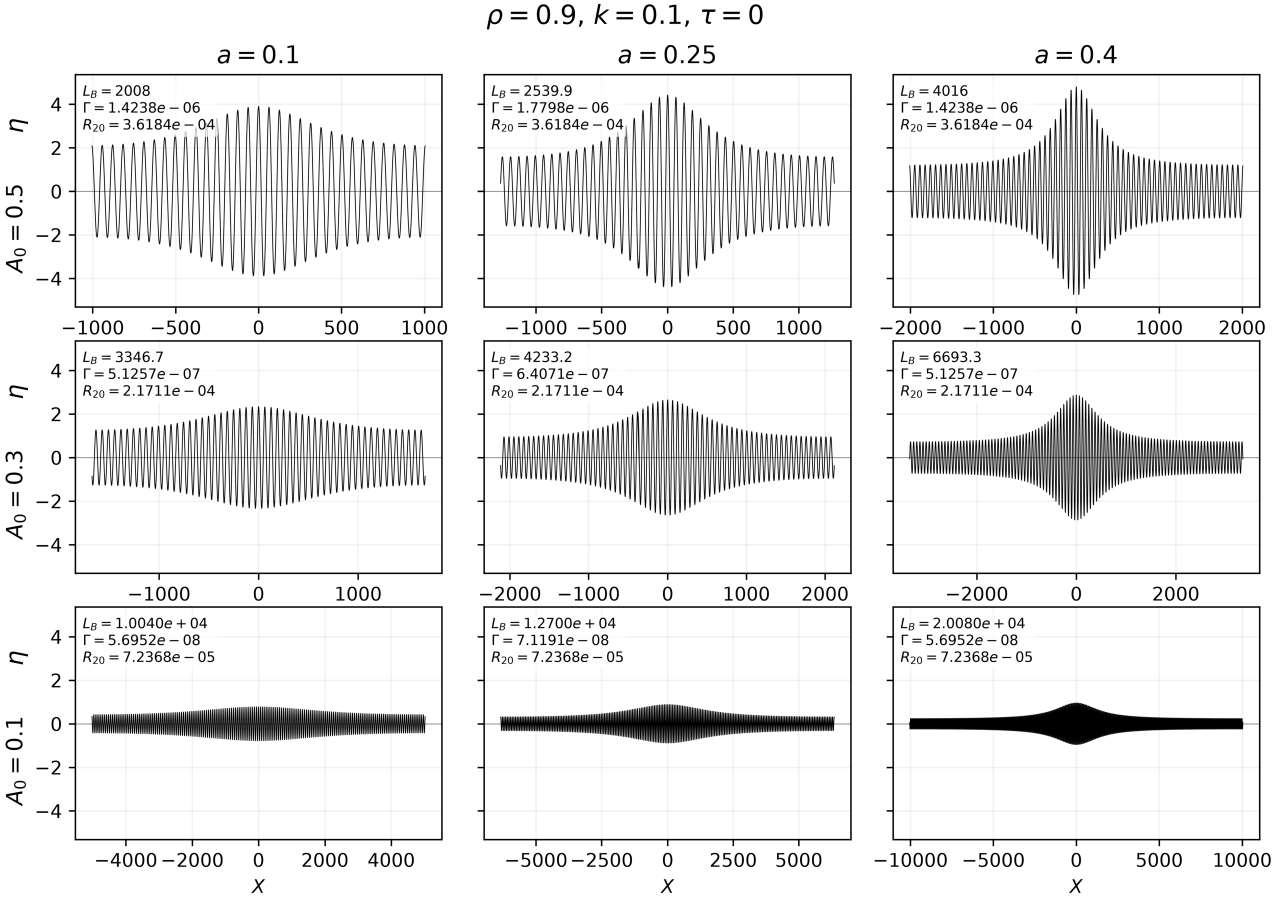}
\hfill
\includegraphics[width=0.49\textwidth]{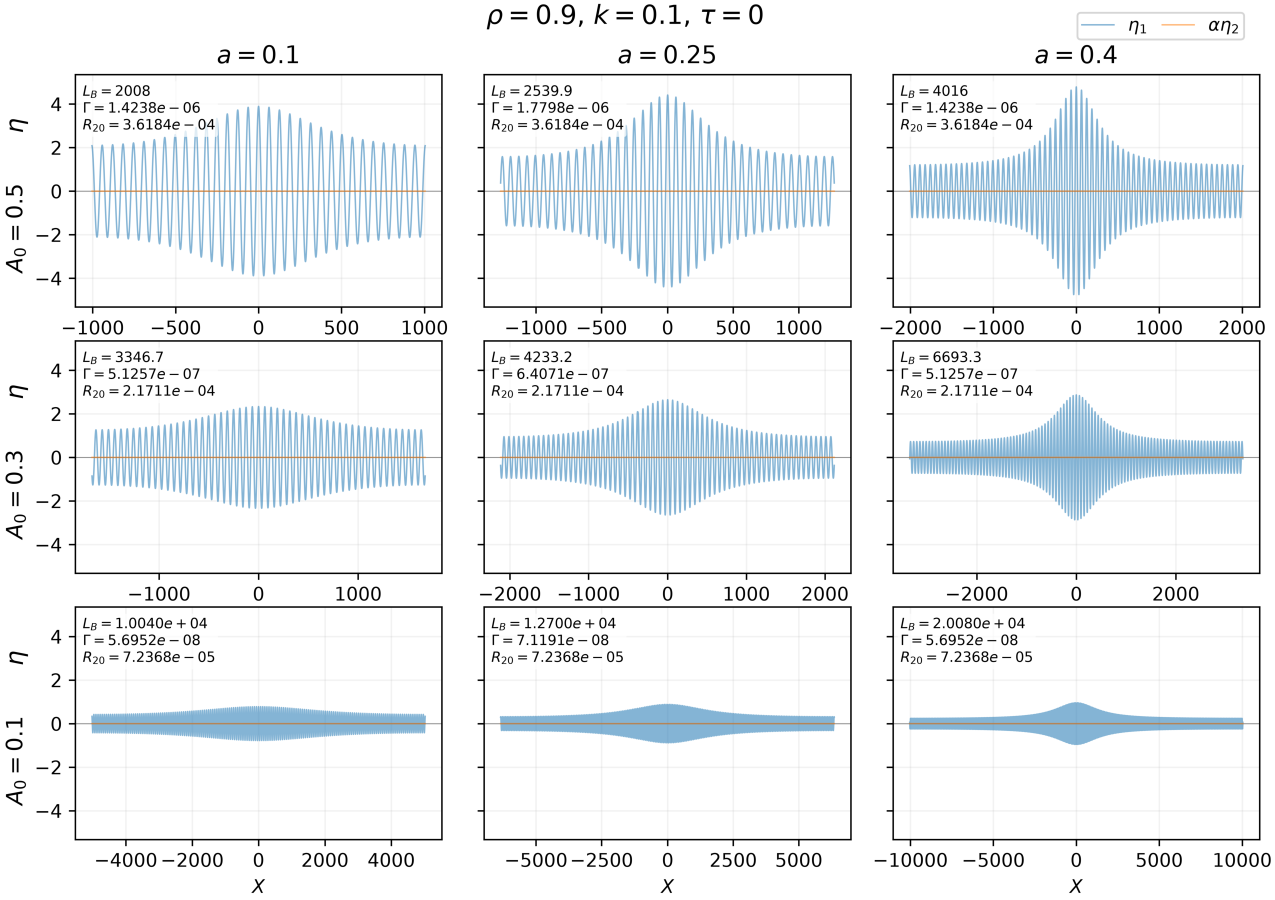}

\caption{
Reconstructed Akhmediev breather profiles for
\(\rho=0.9\), \(k=0.1\), and \(\tau=0\).
The left panel shows the reconstructed interfacial profile
\(\eta=\eta_1+\alpha\eta_2\).
The right panel shows separately the contributions of the
fundamental harmonic \(\eta_1\) and the bound second
harmonic \(\alpha\eta_2\).
Rows correspond to the background-wave amplitudes
\(A_0=0.5\), \(0.3\), and \(0.1\),
whereas columns correspond to the breather parameter
\(a=0.1\), \(0.25\), and \(0.4\).
}
\label{fig:rho09_lower_branch}
\end{figure}

Figure~\ref{fig:rho09_lower_branch} shows the
reconstructed breather profiles for the wavenumber
\(k=0.1\) (point \(A_3\) in
Fig.~\ref{fig:selected_regimes}).
Overall, their behavior remains similar to that observed
for small density ratios.
As the breather parameter \(a\) increases, the central
peak becomes more pronounced, whereas decreasing the
background-wave amplitude \(A_0\) causes the breather
wave packet to broaden and its maximum amplitude to
decrease.

Comparison with the case of small density ratios shows
that increasing the density ratio has little effect on
the qualitative structure of the breather.
The main differences are quantitative and are primarily
associated with its spatial scale and the relative
contribution of the bound second harmonic.
For the same value of \(k\), the breather modulation
period \(L_B\) becomes somewhat shorter, leading to
stronger spatial localization.
At the same time, the breather envelope contains fewer
carrier-wave periods, although the overall profile
remains essentially unchanged.

The right panels of
Fig.~\ref{fig:rho09_lower_branch}
show that the contribution of the bound second harmonic
is even smaller than that observed for small density
ratios.
For the same values of \(A_0\) and \(k\), the parameter
\(R_{20}\) decreases by nearly an order of magnitude,
so that the reconstructed interfacial profile is almost
entirely determined by the fundamental harmonic.
Thus, as the density ratio approaches
\(\rho\approx1\), the breather structure undergoes no
qualitative changes: its spatial scale becomes
somewhat smaller, while the contribution of the bound
second harmonic becomes even less significant.

Calculations performed for intermediate density ratios
show the same qualitative behavior of the reconstructed
interfacial profiles.
They are therefore omitted, since they reveal no new
qualitative features and simply represent a smooth
transition between the regimes corresponding to small
and large density ratios.

Overall, throughout the lower MI
region, the reconstructed profiles retain the
characteristic form of the Akhmediev breather.
Variations in the wavenumber, background-wave
amplitude, and density ratio primarily affect the
breather modulation period and the relative
contribution of the bound second harmonic, without
altering the qualitative structure of the wave field.
These results confirm the validity of the weakly
nonlinear reconstruction throughout the entire
parameter range considered.

\subsection{Upper MI region}
\label{subsec:upper_instability}

Unlike the lower MI region, the
contribution of the bound second harmonic can become
substantial in the upper MI
region, leading to pronounced modifications of the
reconstructed interfacial profile.
The properties of the breather structures are governed
both by the density ratio and by the location of the
selected regime within the MI
region.

The analysis begins with small density ratios, for
which the neighborhoods of the lower resonance
boundary, the turning point of the upper boundary, and
the capillary regime are examined in sequence.
The discussion then proceeds to moderate and large
density ratios, with particular attention devoted to
the central part of the upper MI
region and the narrow near-gravity subregion.

\subsubsection{Small density ratios}
\label{subsubsec:upper_smallrho}

\begin{figure}
\centering

\includegraphics[width=0.49\textwidth]{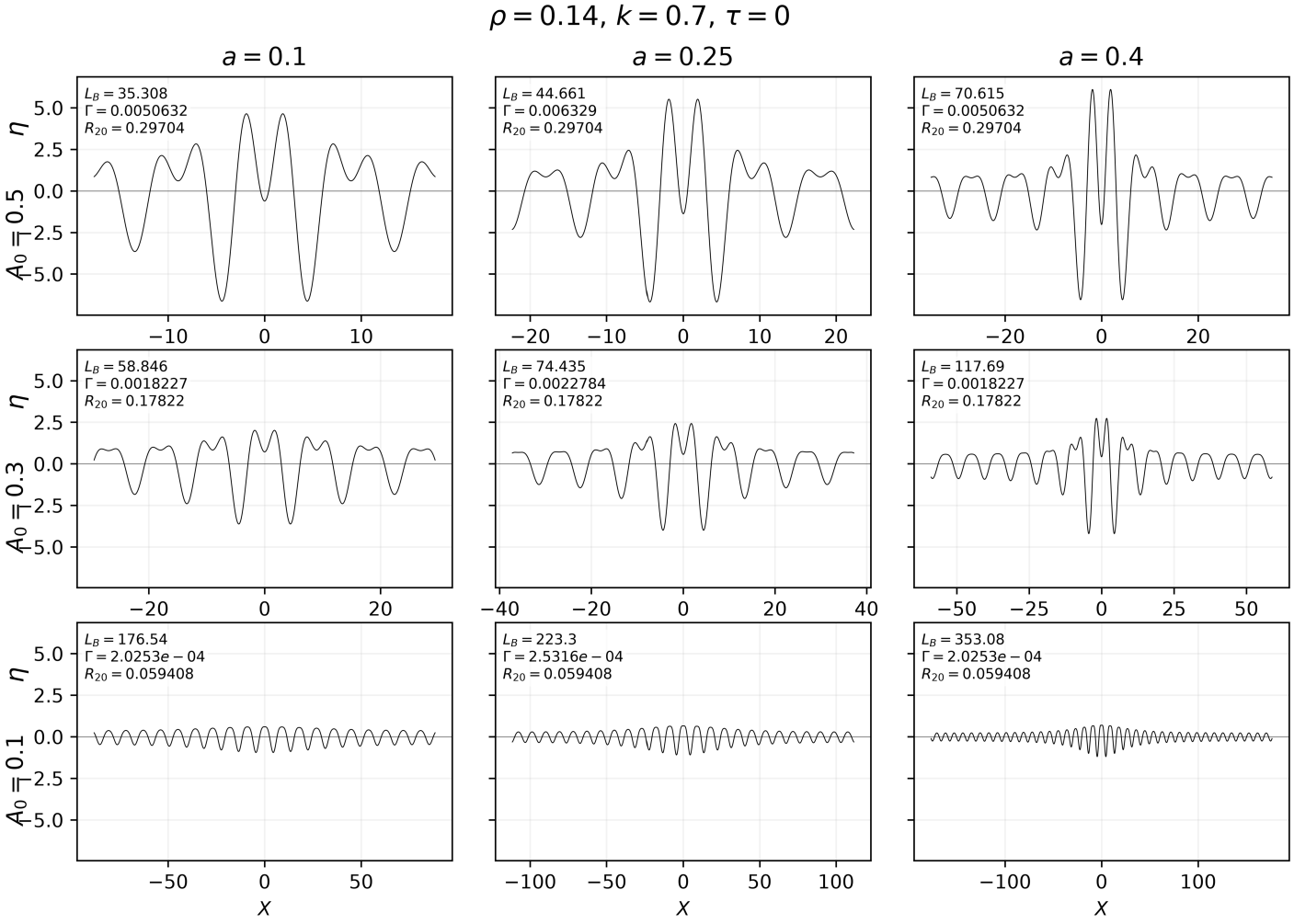}\hfill
\includegraphics[width=0.49\textwidth]{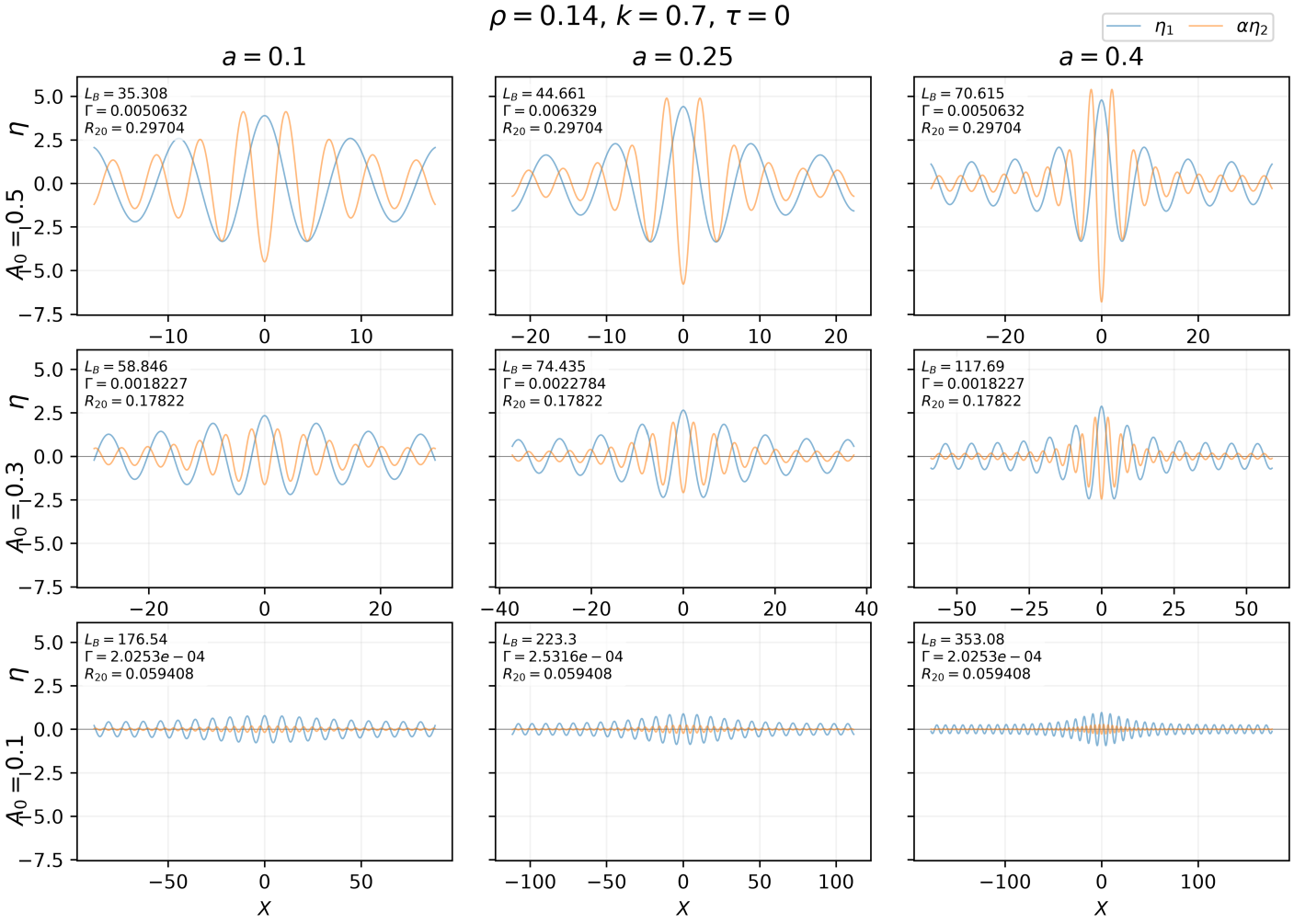}

\caption{
Reconstructed Akhmediev breather profiles for
\(\rho=0.14\), \(k=0.7\), and \(\tau=0\).
The left panel shows the reconstructed interfacial profile
\(\eta=\eta_1+\alpha\eta_2\).
The right panel shows separately the contributions of the
fundamental harmonic \(\eta_1\) and the bound second
harmonic \(\alpha\eta_2\).
Rows correspond to the background-wave amplitudes
\(A_0=0.5\), \(0.3\), and \(0.1\),
whereas columns correspond to the breather parameter
\(a=0.1\), \(0.25\), and \(0.4\).
}
\label{fig:k07_profiles}
\end{figure}

\textbf{Near the lower resonance boundary.}
The analysis begins with the vicinity of the lower
resonance boundary of the upper MI
region at small density ratios.
Figure~\ref{fig:k07_profiles} presents the reconstructed
interfacial profiles for \(\rho=0.14\) and \(k=0.7\),
corresponding to point \(B_1\) in
Fig.~\ref{fig:selected_regimes}.

The overall breather localization is preserved;
however, the reconstructed interfacial profile differs
substantially from those characteristic of the lower
MI region.
A complex multi-peak pattern develops near the center
of the wave packet and becomes progressively more
pronounced as the breather parameter \(a\) increases.
For \(a=0.1\), the envelope remains relatively smooth,
whereas for \(a=0.25\), and especially for \(a=0.4\),
the central region evolves into a narrow packet of
intense oscillations with a well-defined internal
structure.

This increased complexity of the reconstructed profile
is consistent with the parameter maps of \(R_{20}\)
presented in Section~\ref{sec:maps}.
Near the lower resonance boundary, the second-harmonic
coefficient \(\Lambda\) increases sharply, so that even
at small density ratios the contribution of the bound
second harmonic is no longer asymptotically small.
For \(A_0=0.5\), the parameter \(R_{20}\) reaches
approximately \(0.3\).
Thus, although the nonlinear corrections can no longer
be regarded as asymptotically small, the weakly
nonlinear reconstruction remains justified because
\(R_{20}<1\).

The right panels of
Fig.~\ref{fig:k07_profiles} support this conclusion.
The bound second harmonic is no longer a small
correction to the fundamental harmonic; instead, its
contribution becomes most pronounced in the central
part of the wave packet.
The superposition of the two harmonics gives rise to
the complex multi-peak structure of the reconstructed
interfacial profile.

As the background-wave amplitude \(A_0\) decreases, the
parameter \(R_{20}\) decreases rapidly, leading to a
corresponding reduction in the contribution of the
bound second harmonic.
Consequently, the reconstructed interfacial profile
gradually approaches the classical Akhmediev breather.
For \(A_0=0.1\), the bound second harmonic becomes
considerably weaker, although its influence on the
central part of the wave packet remains noticeable.

\clearpage

\begingroup
\renewcommand{\floatpagefraction}{0.95}
\renewcommand{\topfraction}{0.99}
\renewcommand{\textfraction}{0.01}

\begin{figure}[p]
\centering

\begin{subfigure}{\textwidth}
\centering
\includegraphics[width=0.49\linewidth]{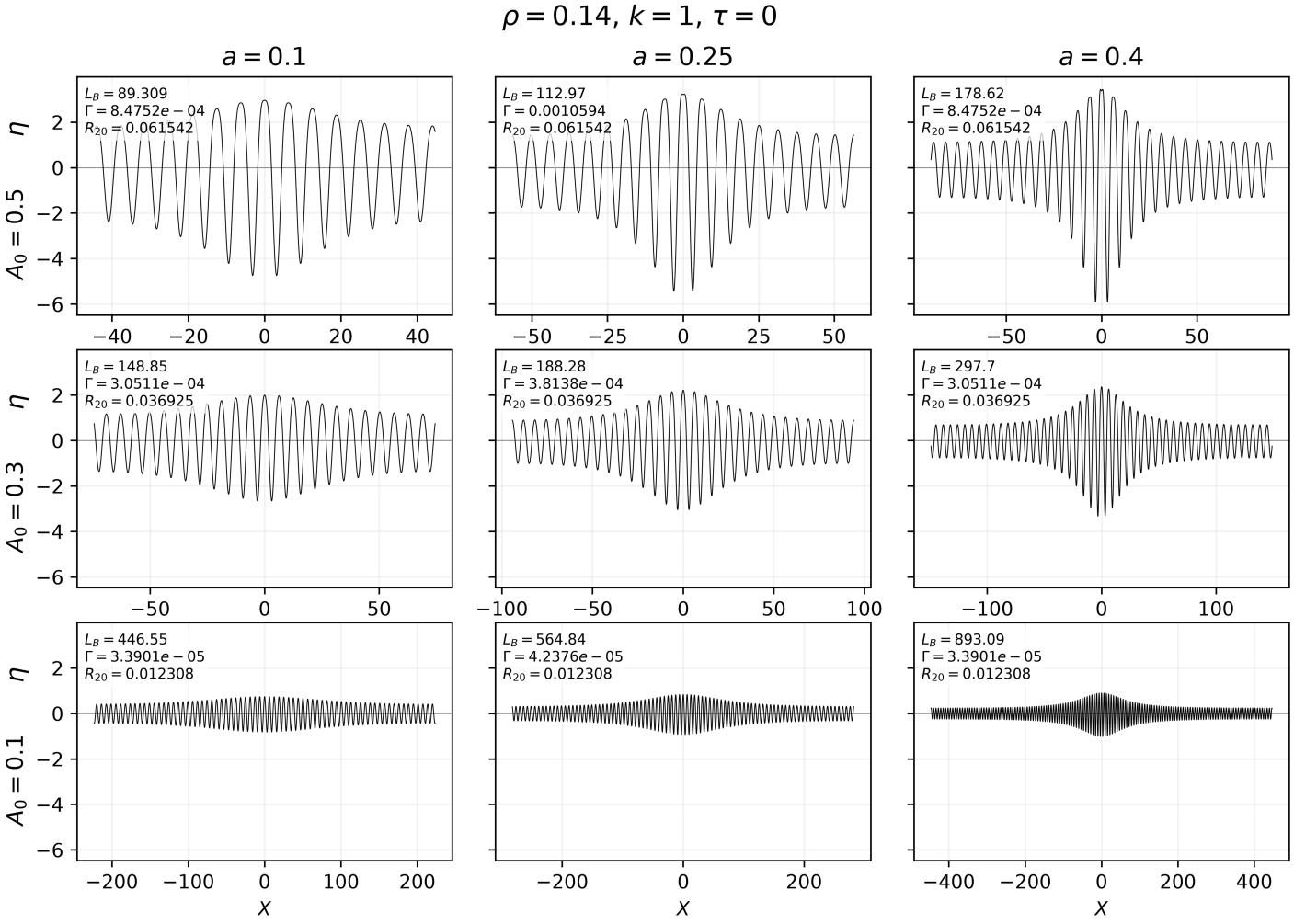}\hfill
\includegraphics[width=0.49\linewidth]{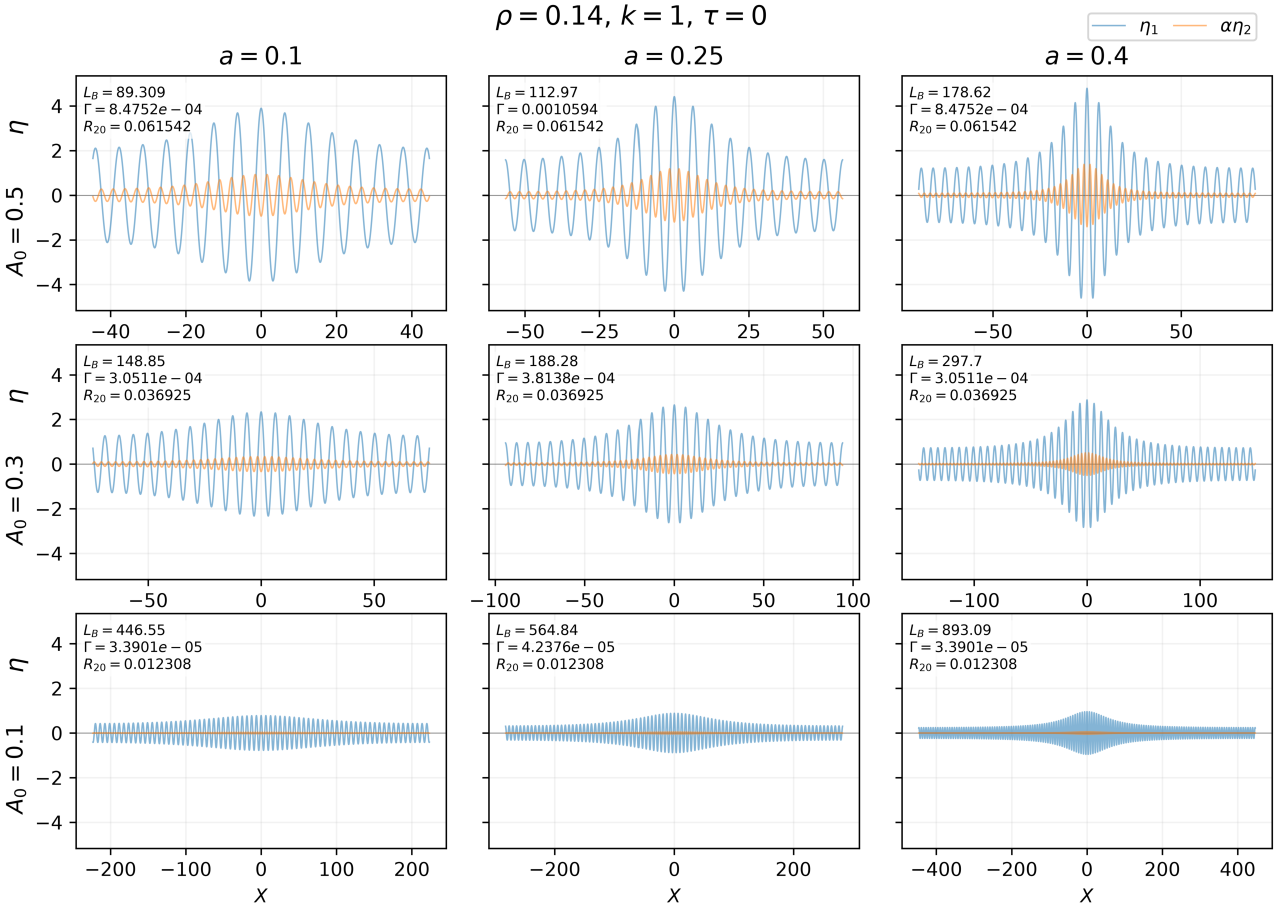}
\caption{\(k=1\)}
\label{fig:rho014_upper_branch_k1}
\end{subfigure}

\vspace{-2mm}

\begin{subfigure}{\textwidth}
\centering
\includegraphics[width=0.49\linewidth]{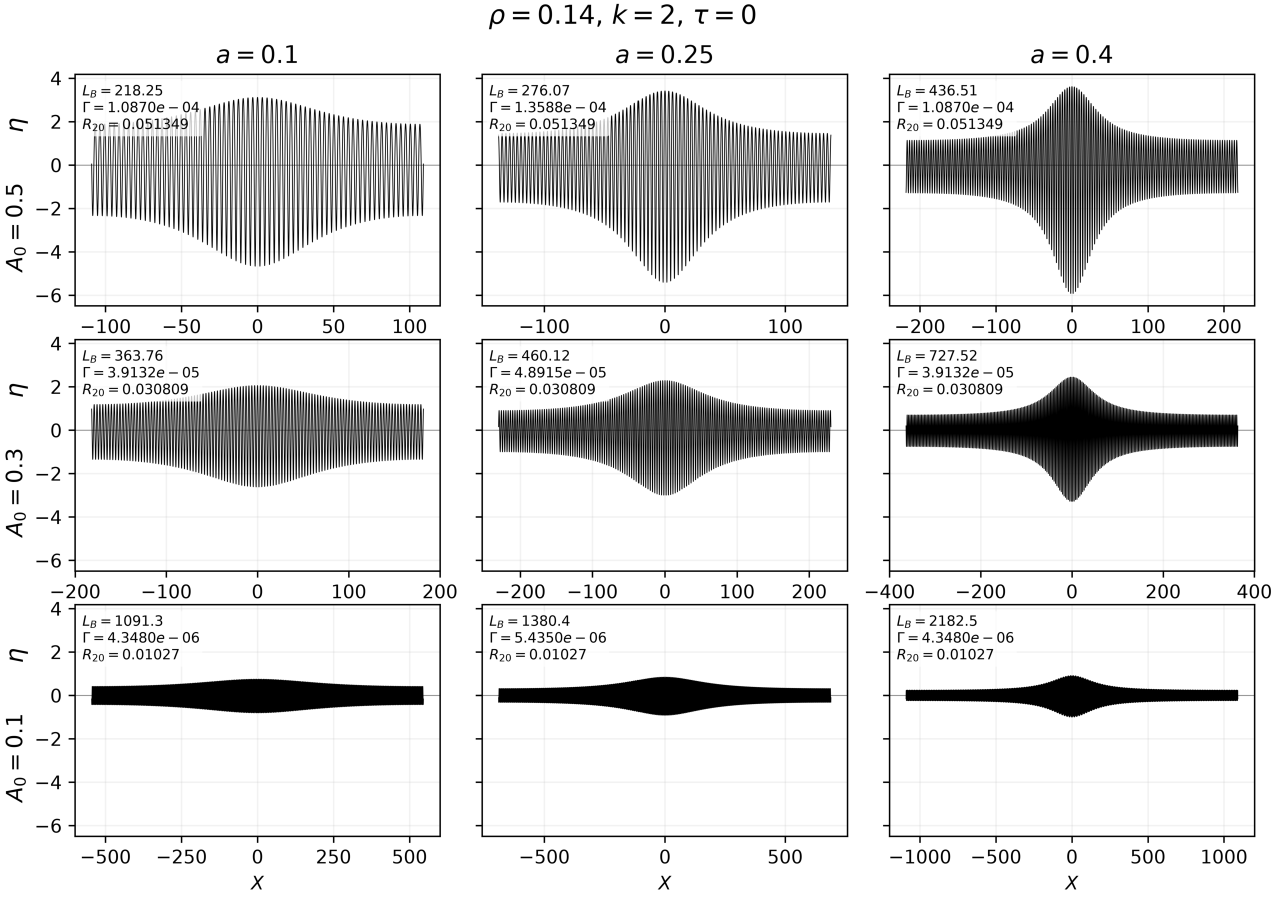}\hfill
\includegraphics[width=0.49\linewidth]{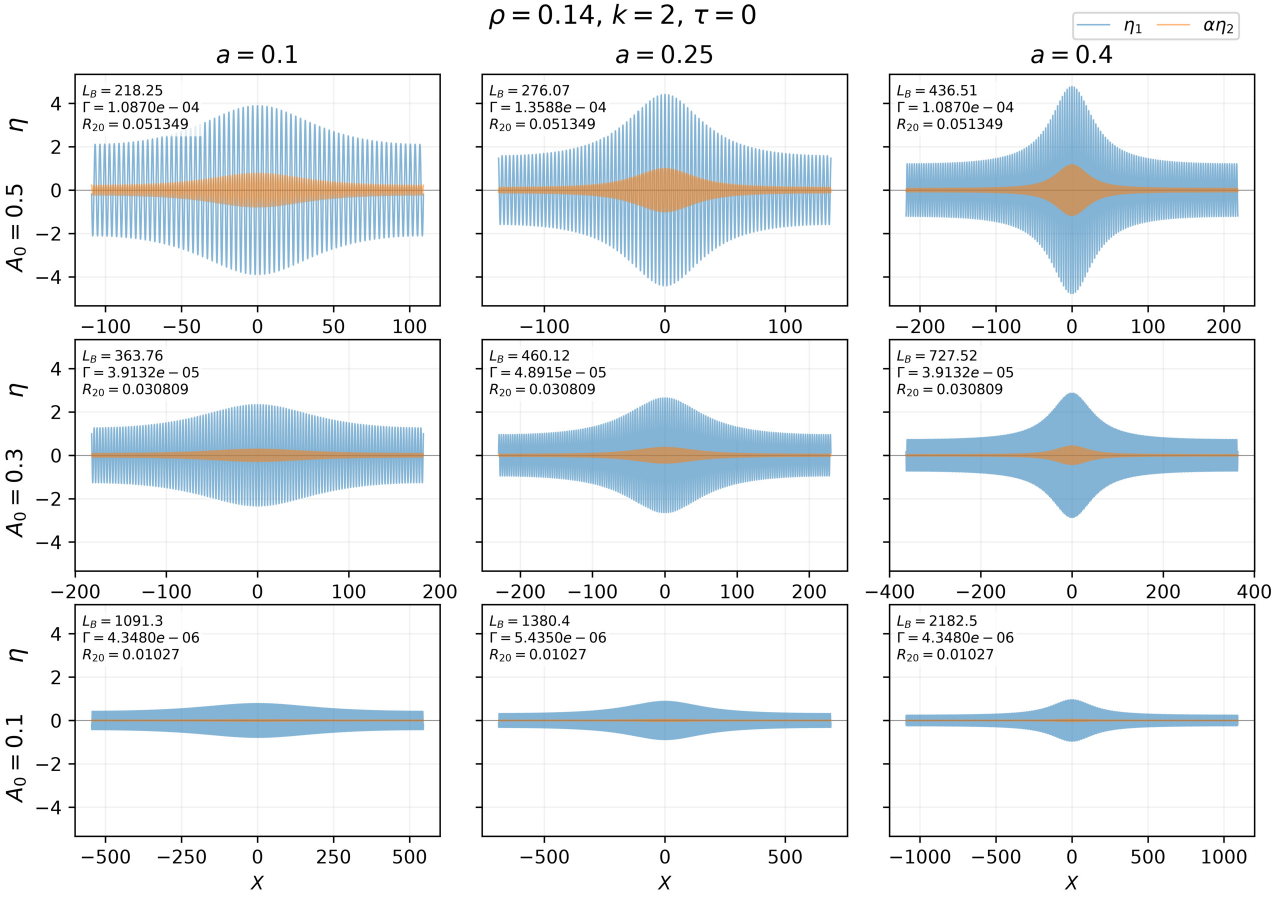}
\caption{\(k=2\)}
\label{fig:rho014_upper_branch_k2}
\end{subfigure}

\vspace{-2mm}

\begin{subfigure}{\textwidth}
\centering
\includegraphics[width=0.49\linewidth]{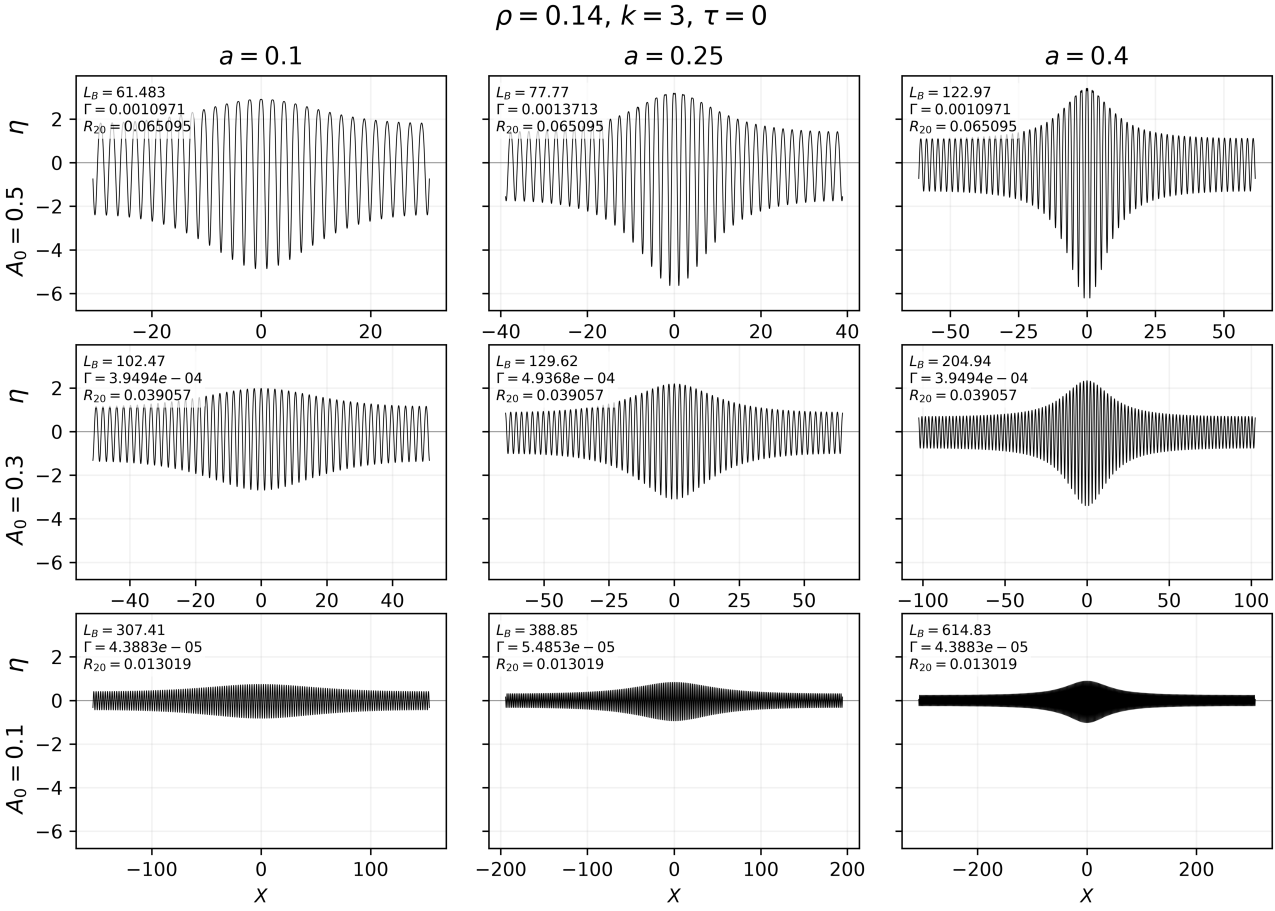}\hfill
\includegraphics[width=0.49\linewidth]{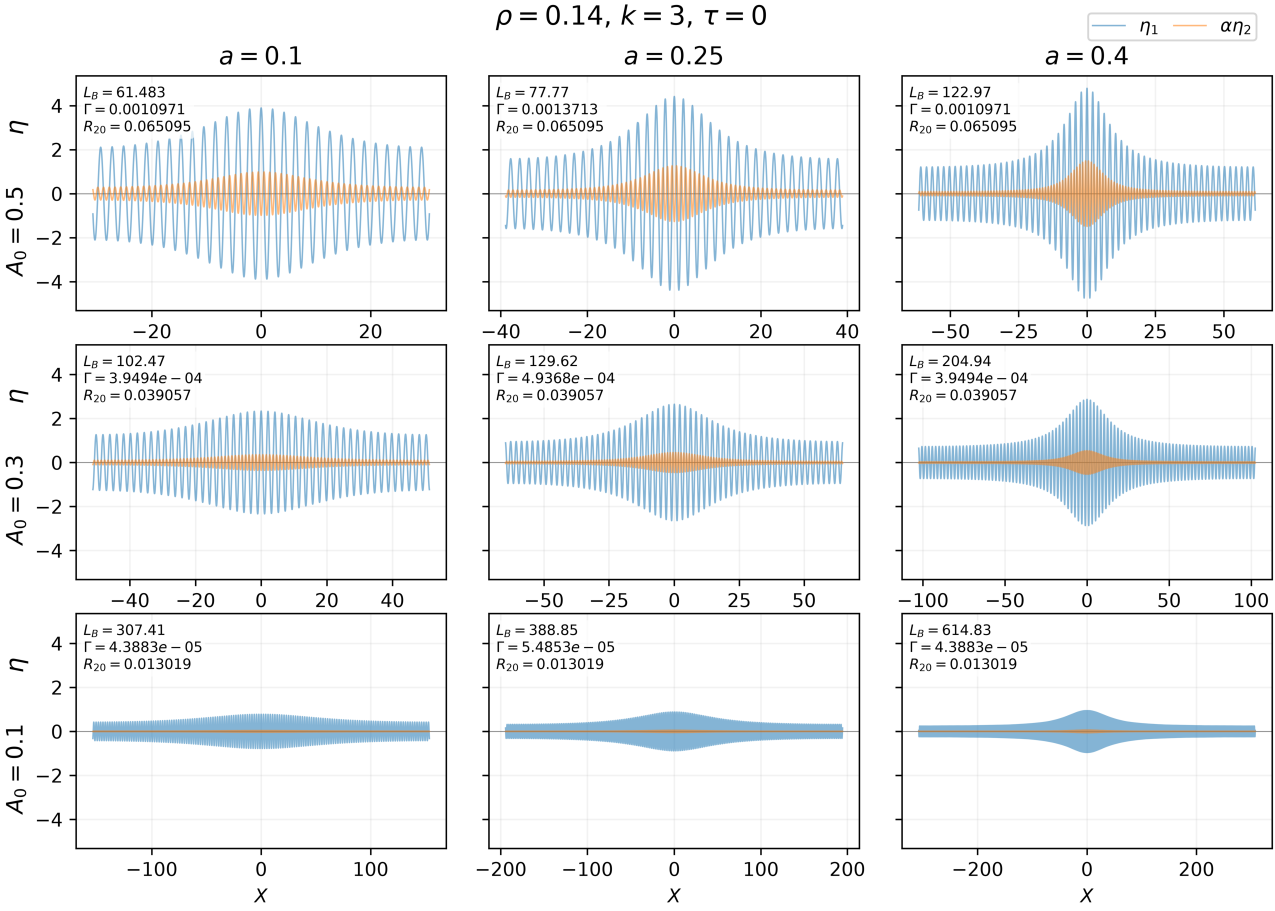}
\caption{\(k=3\)}
\label{fig:rho014_upper_branch_k3}
\end{subfigure}

\vspace{-2mm}

\caption{
Reconstructed Akhmediev breather profiles in the upper
MI region for \(\rho=0.14\).
The left panel of each pair shows the reconstructed
interfacial profile
\(\eta=\eta_1+\alpha\eta_2\).
The right panel shows separately the contributions of
the fundamental harmonic \(\eta_1\) and the bound
second harmonic \(\alpha\eta_2\).
}
\label{fig:rho014_upper_branch}
\end{figure}

\clearpage
\endgroup

Thus, in the vicinity of the lower resonance boundary
of the upper MI region, the
reconstructed interfacial profile undergoes its first
qualitative transformation owing to the substantial
increase in the contribution of the bound second
harmonic.
This region therefore represents a transition between
the classical weakly nonlinear breathers
characteristic of the lower MI
region and regimes in which the second-order nonlinear
corrections exert a significant influence on the
reconstructed interfacial profile.

\textbf{Near the turning point of the upper boundary.}

The analysis next focuses on regimes located near the
turning point of the upper boundary of the upper
MI region.
In this part of the parameter plane, the curve \(J=0\),
which defines the upper boundary of the MI
 region, changes direction and, after the
turning point, approaches the vertical asymptote.
Figure~\ref{fig:rho014_upper_branch} shows the
reconstructed interfacial profiles for \(\rho=0.14\)
and three representative wavenumbers,
\(k=1\), \(2\), and \(3\), corresponding to points
\(B_2\), \(B_3\), and \(B_4\) in
Fig.~\ref{fig:selected_regimes}.

A comparison of the three pairs of panels
(Figs.~\ref{fig:rho014_upper_branch}\subref{fig:rho014_upper_branch_k1}--
\subref{fig:rho014_upper_branch_k3})
shows that the evolution of the reconstructed
interfacial profiles for \(k=1\), \(2\), and \(3\) is
non-monotonic.
For \(\rho=0.14\), increasing the wavenumber initially
reduces the distance to the upper \(J=0\) boundary,
reaching a minimum at \(k=2\), after which the distance
increases again.
As the nonlinearity coefficient \(J\) approaches zero,
the focusing effect weakens: the breather modulation
period \(L_B\) increases substantially, whereas the
MI growth rate \(\Gamma\)
decreases.
This behavior is manifested primarily by an increase in
the spatial extent of the breather, while its
qualitative structure remains essentially unchanged.
In the reconstructed interfacial profile, this
corresponds to a broader envelope containing a larger
number of carrier-wave oscillations.

In Fig.~\ref{fig:rho014_upper_branch}\subref{fig:rho014_upper_branch_k1},
corresponding to \(k=1\), the reconstructed interfacial
profile still retains the characteristic resonance-induced
distortions discussed in the previous subsection.
Near the turning point
(Fig.~\ref{fig:rho014_upper_branch}\subref{fig:rho014_upper_branch_k2},
\(k=2\)), these distortions become much less pronounced.
The dominant effect is now the weakening of the
focusing nonlinearity, which results in a substantial
increase in the breather modulation period.
Beyond the turning point
(Fig.~\ref{fig:rho014_upper_branch}\subref{fig:rho014_upper_branch_k3},
\(k=3\)), the breather becomes more compact again, and
the reconstructed profile approaches the classical
Akhmediev breather with a smooth localized envelope.
Thus, the complex geometry of the upper boundary in the
parameter map does not give rise to new types of
breather structures but is instead accompanied by the
gradual disappearance of the resonance-induced
distortions.

When the wavenumber increases from \(k=1\) to \(k=2\),
the parameter \(R_{20}\) changes only slightly
(\(0.061\rightarrow0.051\)).
The observed evolution of the reconstructed
interfacial profile is therefore governed primarily by
the weakening of the focusing nonlinearity and the
corresponding increase in the breather modulation
period \(L_B\), whereas the contribution of the bound
second harmonic remains nearly unchanged.
Only with a further increase in the wavenumber to
\(k=3\) does the parameter \(R_{20}\) decrease
appreciably, leading to a gradual reduction in the
influence of the bound second harmonic on the
reconstructed interfacial profile.

As in the previous examples, increasing the breather
parameter \(a\) leads to a narrower central region and
stronger localization of the breather.
At the same time, decreasing the background-wave
amplitude \(A_0\) rapidly reduces the contribution of
the bound second harmonic.
For \(A_0=0.1\), this contribution becomes sufficiently
small that, for all three values of \(k\), the
reconstructed interfacial profile is determined almost
entirely by the fundamental harmonic.

Thus, passage through the vicinity of the turning point
of the upper \(J=0\) boundary shows that the complex
geometry of the upper boundary of the MI
 region primarily affects the breather
modulation period and the MI
growth rate, while leaving the qualitative structure of
the breather essentially unchanged.
As the distance from the resonance region increases,
the resonance-induced distortions gradually disappear,
and the influence of the bound second harmonic on the
reconstructed interfacial profile progressively
decreases.

\textbf{Capillary regime.}

Beyond the vicinity of the turning point of the upper
\(J=0\) boundary, a further increase in the wavenumber
moves the system into the capillary part of the upper
MI region.
This region exists only for small density ratios and
terminates before the first vertical asymptote at
\(
\rho\simeq0.1716.
\)
For larger density ratios, the high-wavenumber part of
the parameter plane corresponds to the stable regime up
to the second vertical asymptote,
\(
\rho\simeq5.8284,
\)
and is therefore not considered in the present study.

\begin{figure}
\centering

\begin{subfigure}{\textwidth}
\centering
\includegraphics[width=0.49\linewidth]{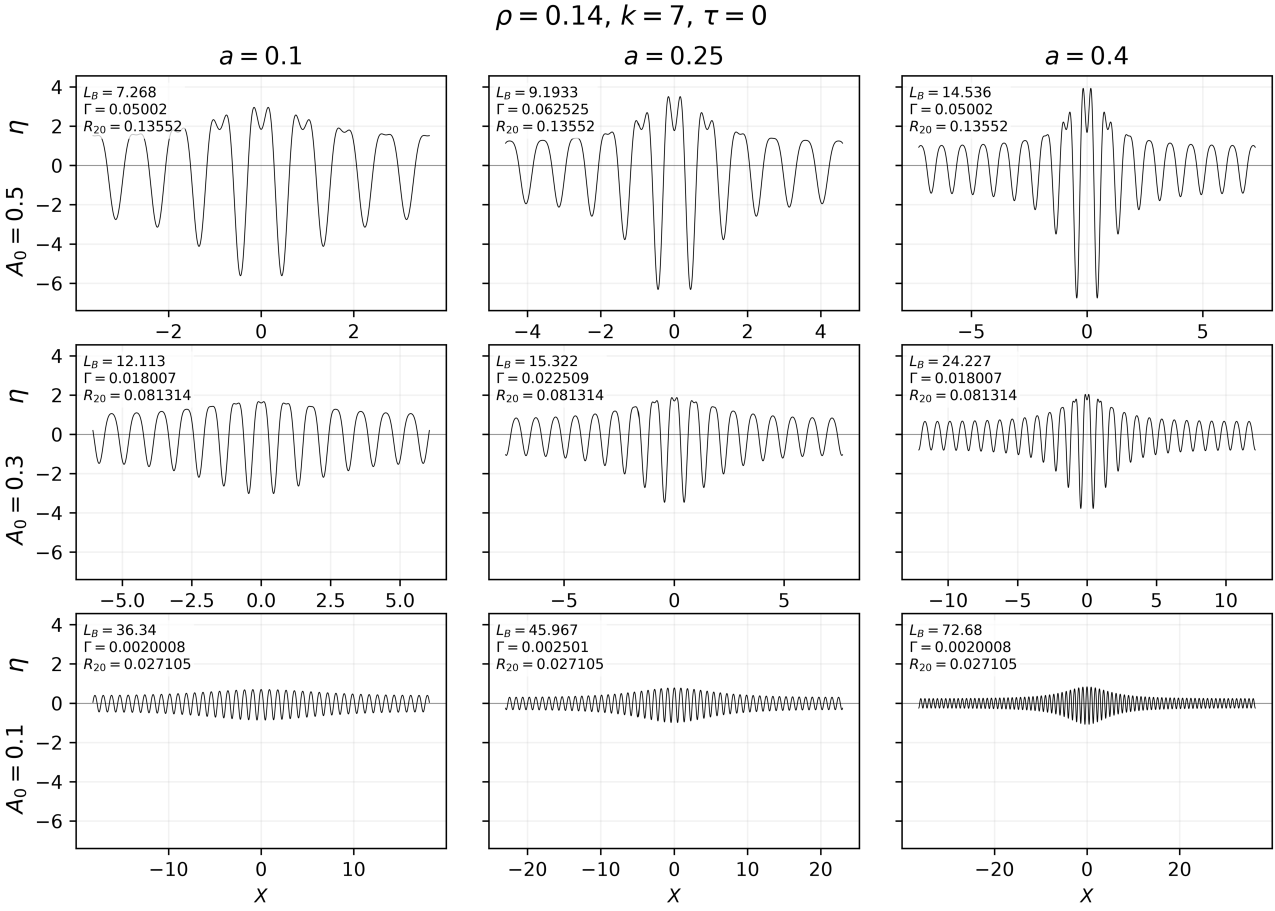}\hfill
\includegraphics[width=0.49\linewidth]{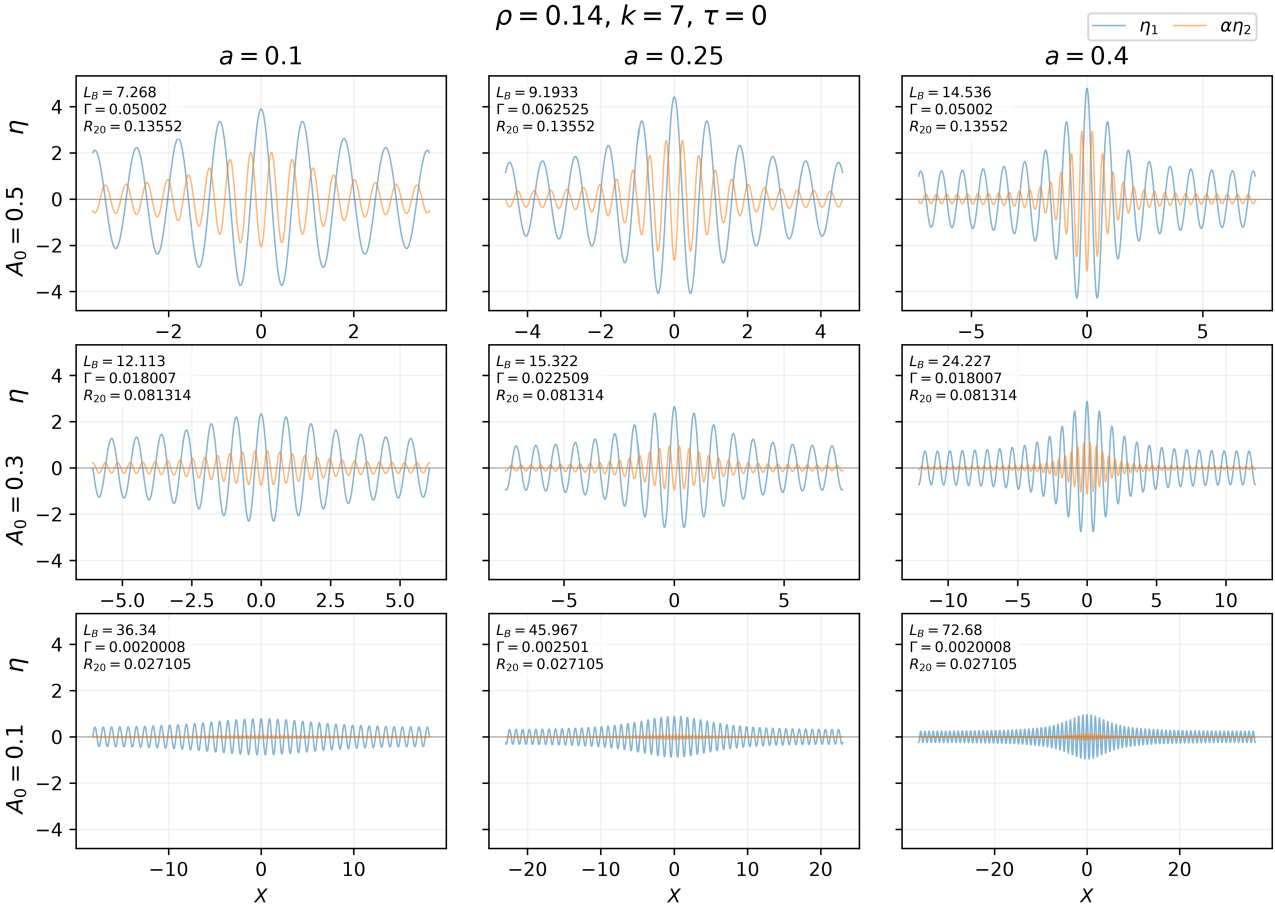}
\caption{$k=7$}
\label{fig:rho014_capillary_k7}
\end{subfigure}

\vspace{2mm}

\begin{subfigure}{\textwidth}
\centering
\includegraphics[width=0.49\linewidth]{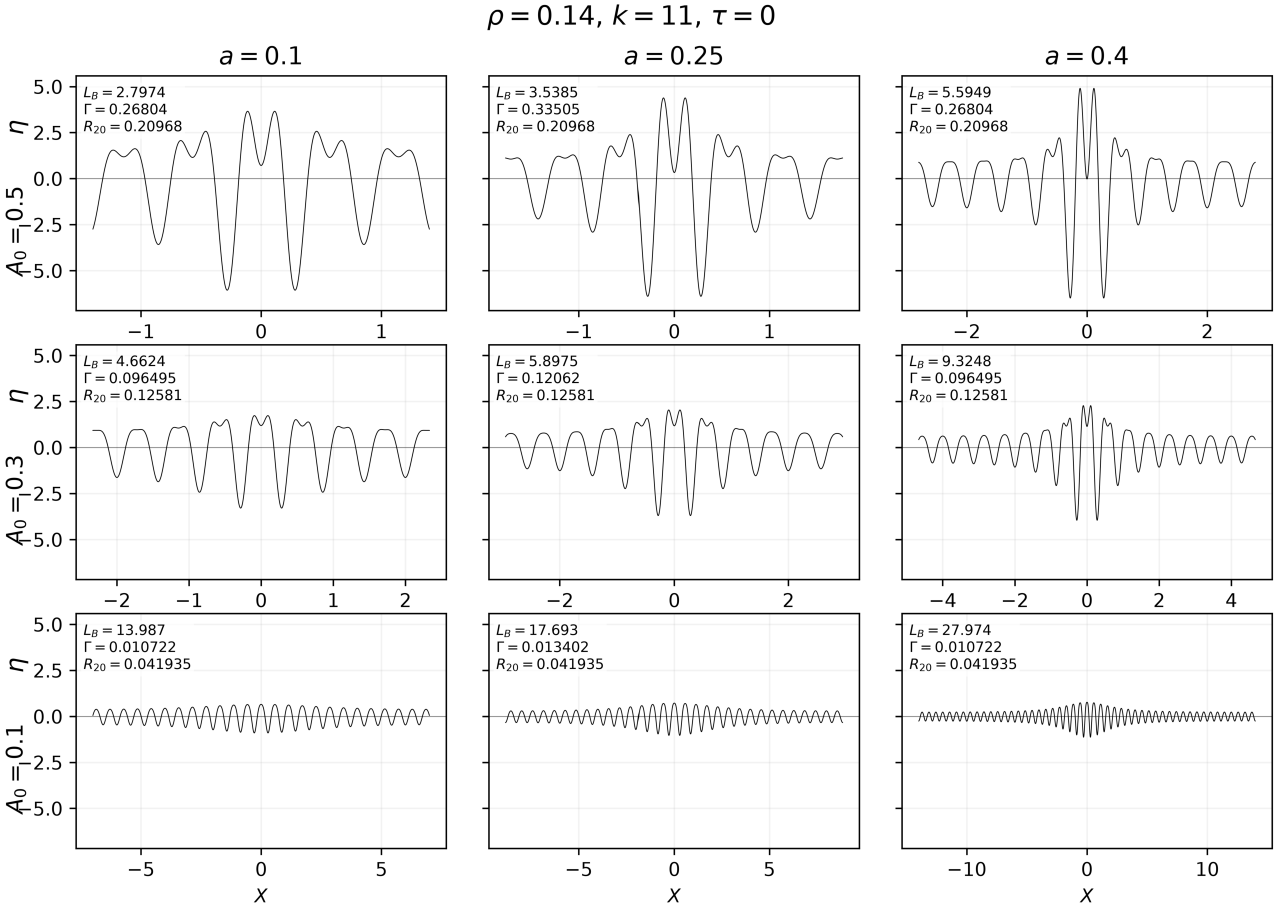}\hfill
\includegraphics[width=0.49\linewidth]{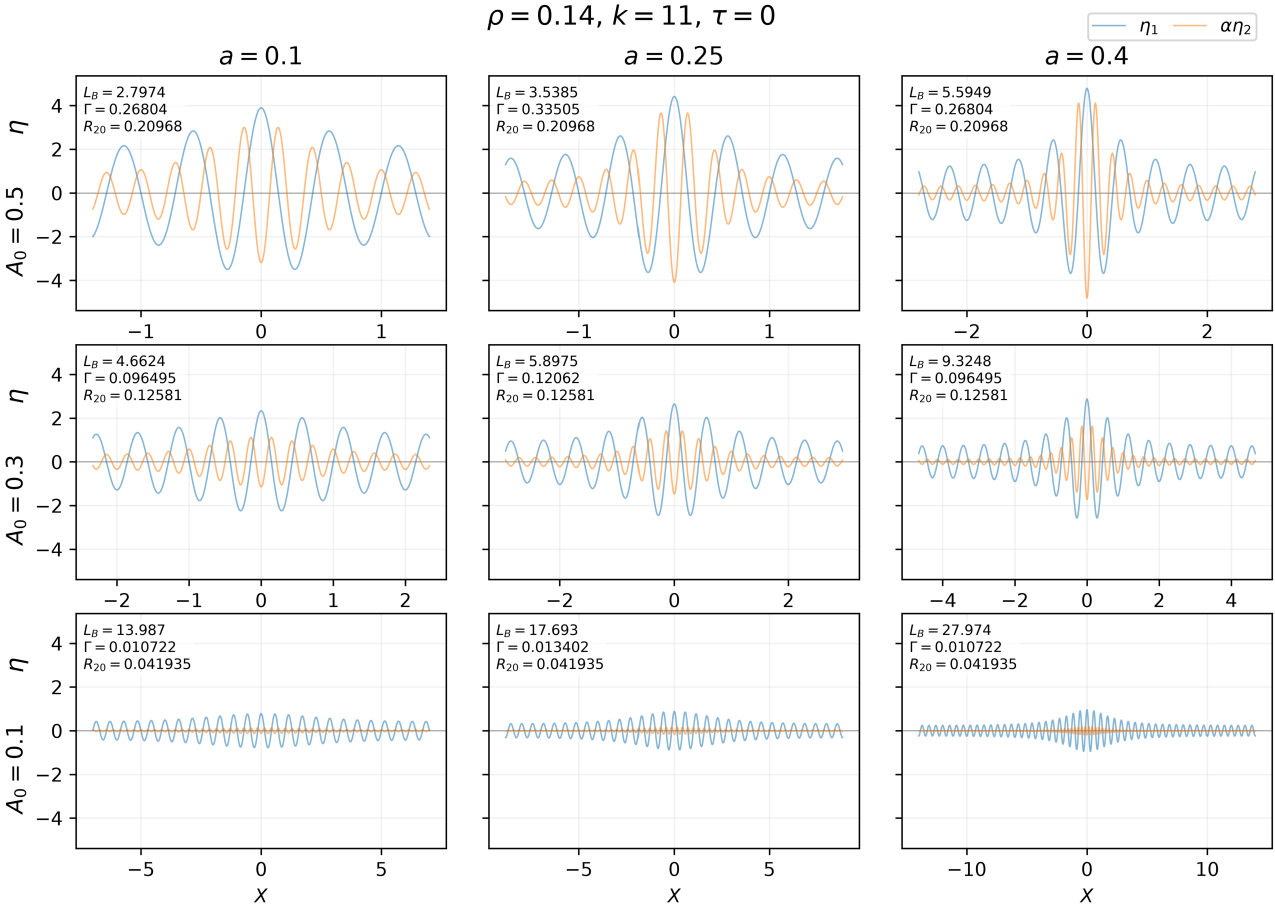}
\caption{$k=11$}
\label{fig:rho014_capillary_k11}
\end{subfigure}

\caption{
Reconstructed Akhmediev breather profiles in the
capillary region for \(\rho=0.14\).
The left panel of each pair shows the reconstructed
interfacial profile
\(\eta=\eta_1+\alpha\eta_2\).
The right panel shows separately the contributions of
the fundamental harmonic \(\eta_1\) and the bound
second harmonic \(\alpha\eta_2\).
}
\label{fig:rho014_capillary}
\end{figure}

A comparison of
Fig.~\ref{fig:rho014_capillary}\subref{fig:rho014_capillary_k7}
and
Fig.~\ref{fig:rho014_capillary}\subref{fig:rho014_capillary_k11}
shows that, as the wavenumber increases from \(k=7\) to
\(k=11\) (points \(B_5\) and \(B_6\) in
Fig.~\ref{fig:selected_regimes}), the breather retains
the characteristic structure of the Akhmediev breather,
whereas its breather modulation period changes
substantially.
The breather modulation period \(L_B\) decreases
markedly, while the MI growth
rate \(\Gamma\) increases significantly.
As a consequence of the shorter modulation period, the
breather becomes more strongly localized: the envelope
becomes more compact, and the number of carrier-wave
oscillations within the wave packet decreases
slightly.

Unlike the region near the turning point, where the
dominant effect was the weakening of the focusing
nonlinearity, the capillary regime is characterized by
a gradual increase in the contribution of the bound
second harmonic.
This trend is reflected in the increase of the
parameter \(R_{20}\) from approximately \(0.14\) at
\(k=7\) to \(0.21\) at \(k=11\).
The right panels of
Fig.~\ref{fig:rho014_capillary}\subref{fig:rho014_capillary_k7}
and
Fig.~\ref{fig:rho014_capillary}\subref{fig:rho014_capillary_k11}
show that, for \(A_0=0.5\), the amplitude of the bound
second harmonic in the central part of the wave packet
becomes comparable to that of the fundamental
harmonic.
As a result, the reconstructed interfacial profile
undergoes pronounced nonlinear deformation.

Despite the substantial increase in the contribution of
the bound second harmonic, the breather undergoes no
qualitative restructuring.
Instead, the stronger nonlinear correction manifests
itself in deeper central troughs, increased asymmetry
of the reconstructed interfacial profile with respect
to the mean level, and a more pronounced deformation of
the interface, while the characteristic breather
localization is preserved.

Thus, along the considered trajectory within the
capillary region, both the focusing nonlinearity and
the contribution of the bound second harmonic increase.
These changes primarily affect the breather modulation
period and the degree of nonlinear deformation of the
reconstructed interfacial profile, while leaving the
characteristic structure of the Akhmediev breather
essentially unchanged.

\subsubsection{Moderate and large density ratios}
\label{subsubsec:upper_largerho}

For moderate and large density ratios, the geometry of
the upper MI region differs
substantially from that observed at small values of
\(\rho\).
Whereas, for small density ratios, the upper
MI region extends into an
unbounded capillary regime beyond the turning point of
the upper boundary, for
\(\rho>\rho_{\mathrm{as}}\approx0.1716\)
it is bounded below by the resonance boundary and above
by the \(J=0\) curve.
As a consequence, throughout the entire existence
region of the breathers, their properties are governed
simultaneously by the proximity to the resonance
boundary and by the distance from the upper \(J=0\)
boundary.

For the subsequent analysis of the reconstructed
interfacial profiles, it is convenient to distinguish
two characteristic subregions.
The first corresponds to moderate density ratios, for
which the upper MI region spans a
relatively wide range of wavenumbers.
The second comprises large density ratios close to
\(\rho=1\), where the breather existence region becomes
extremely narrow in the wavenumber direction.

Within each subregion, a fixed density ratio is
selected, and three representative regimes are
considered: one located near the resonance boundary,
one in the central part of the upper MI region, and one near the upper boundary
defined by the condition \(J=0\).
This choice makes it possible to follow the evolution
of the reconstructed interfacial profiles across the
entire MI region at a fixed
density ratio.

\paragraph{Central part of the upper MI region (moderate density ratios).}

Figure~\ref{fig:rho05_upper_profiles} presents the
reconstructed interfacial profiles for \(\rho=0.5\) at
three representative wavenumbers,
\(k=0.51\), \(0.53\), and \(0.55\),
corresponding to points \(B_7\), \(B_8\), and \(B_9\)
in Fig.~\ref{fig:selected_regimes}, respectively.
These points are located near the resonance boundary,
in the central part of the upper MI region, and near the upper \(J=0\)
boundary.

Figure~\ref{fig:rho05_upper_profiles}\subref{fig:rho05_upper_profiles:a},
corresponding to the regime near the resonance
boundary (\(k=0.51\)), exhibits the strongest nonlinear
deformation of the reconstructed interfacial profile.
For \(A_0=0.5\), the parameter
\(R_{20}\approx0.32\), indicating that the amplitude of
the bound second harmonic in the central part of the
wave packet becomes comparable to that of the
fundamental harmonic.
As a result, the reconstructed profile develops deep
troughs, uneven crests, and pronounced asymmetry with
respect to the mean interface level.
This behavior is fully consistent with the \(R_{20}\)
parameter map, which predicts the largest contribution
of the bound second harmonic in the vicinity of the
resonance boundary.

In
Fig.~\ref{fig:rho05_upper_profiles}\subref{fig:rho05_upper_profiles:b}
(\(k=0.53\)), corresponding to the central part of the
upper MI region, the selected
regime is located farther from the resonance boundary.
Consequently, for \(A_0=0.5\), the parameter
\(R_{20}\) decreases to approximately \(0.11\), while
the breather modulation period \(L_B\) (for
\(a=0.25\)) more than doubles, increasing from
approximately \(104\) to \(225\).
At the same time, the MI growth
rate \(\Gamma\) decreases by nearly a factor of five.
As a result, the reconstructed interfacial profile
becomes more extended, the characteristic nonlinear
deformations in the central part of the wave packet are
substantially smoothed, and the bound second harmonic
no longer dominates the profile shape, although its
contribution remains clearly visible.

Figure~\ref{fig:rho05_upper_profiles}\subref{fig:rho05_upper_profiles:c},
corresponding to the regime near the upper \(J=0\)
boundary (\(k=0.55\)), illustrates a substantial
weakening of the focusing nonlinearity.
This is reflected in a sharp increase in the breather
modulation period \(L_B\) and a further decrease in the
MI growth rate \(\Gamma\).
For \(a=0.25\), the breather modulation period reaches
approximately \(486\), whereas \(\Gamma\) becomes
nearly two orders of magnitude smaller than in the
vicinity of the resonance boundary.
Consequently, the reconstructed interfacial profiles
extend over a much larger spatial scale, while the
contribution of the bound second harmonic decreases:
even for \(A_0=0.5\), the parameter \(R_{20}\)
drops to approximately \(0.07\).
As a result, the reconstructed profile becomes
smoother, and its shape is increasingly governed by the
fundamental harmonic.

As in all the previous examples, decreasing the
background-wave amplitude \(A_0\) rapidly reduces the
contribution of the bound second harmonic.
Already at \(A_0=0.3\), its influence is substantially
weaker than that observed in the top row, whereas at
\(A_0=0.1\) the reconstructed interfacial profile is
determined almost entirely by the fundamental harmonic,
even in the vicinity of the resonance boundary, where
the values of \(R_{20}\) are the largest.

The influence of the breather parameter \(a\) is of a
different nature.
For each of the considered wavenumbers, increasing
\(a\) nearly doubles the breather modulation period
\(L_B\), so that the localized wave packet contains a
larger number of carrier-wave oscillations.
In contrast, the contribution of the bound second
harmonic changes only weakly: it remains confined to
the central part of the wave packet, and its role is
governed primarily by the position of the selected
regime within the upper MI
region rather than by the value of \(a\).

\clearpage

\begingroup
\renewcommand{\floatpagefraction}{0.95}
\renewcommand{\topfraction}{0.99}
\renewcommand{\textfraction}{0.01}

\begin{figure}[p]
\centering

\begin{subfigure}{\textwidth}
\centering
\includegraphics[width=0.49\linewidth]{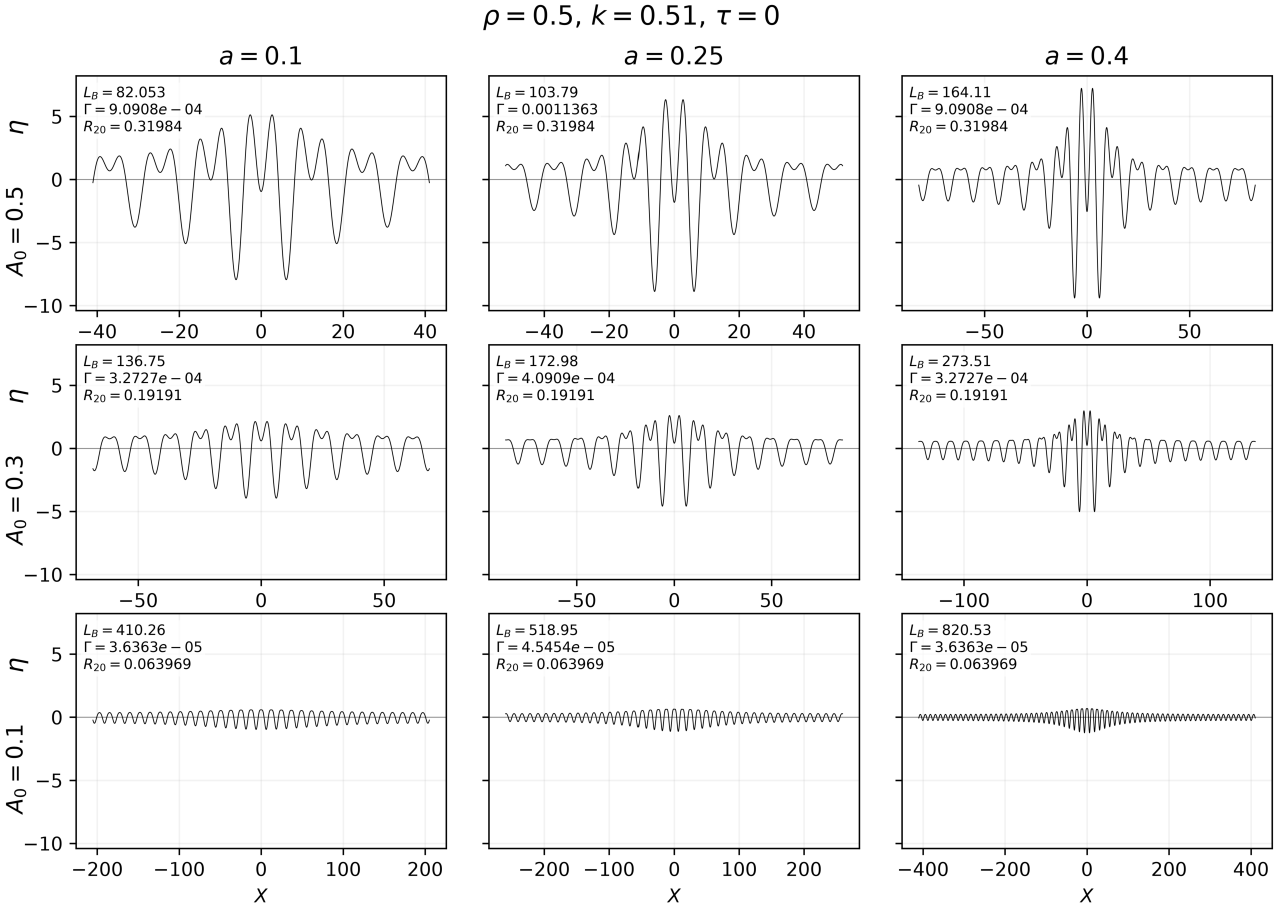}\hfill
\includegraphics[width=0.49\linewidth]{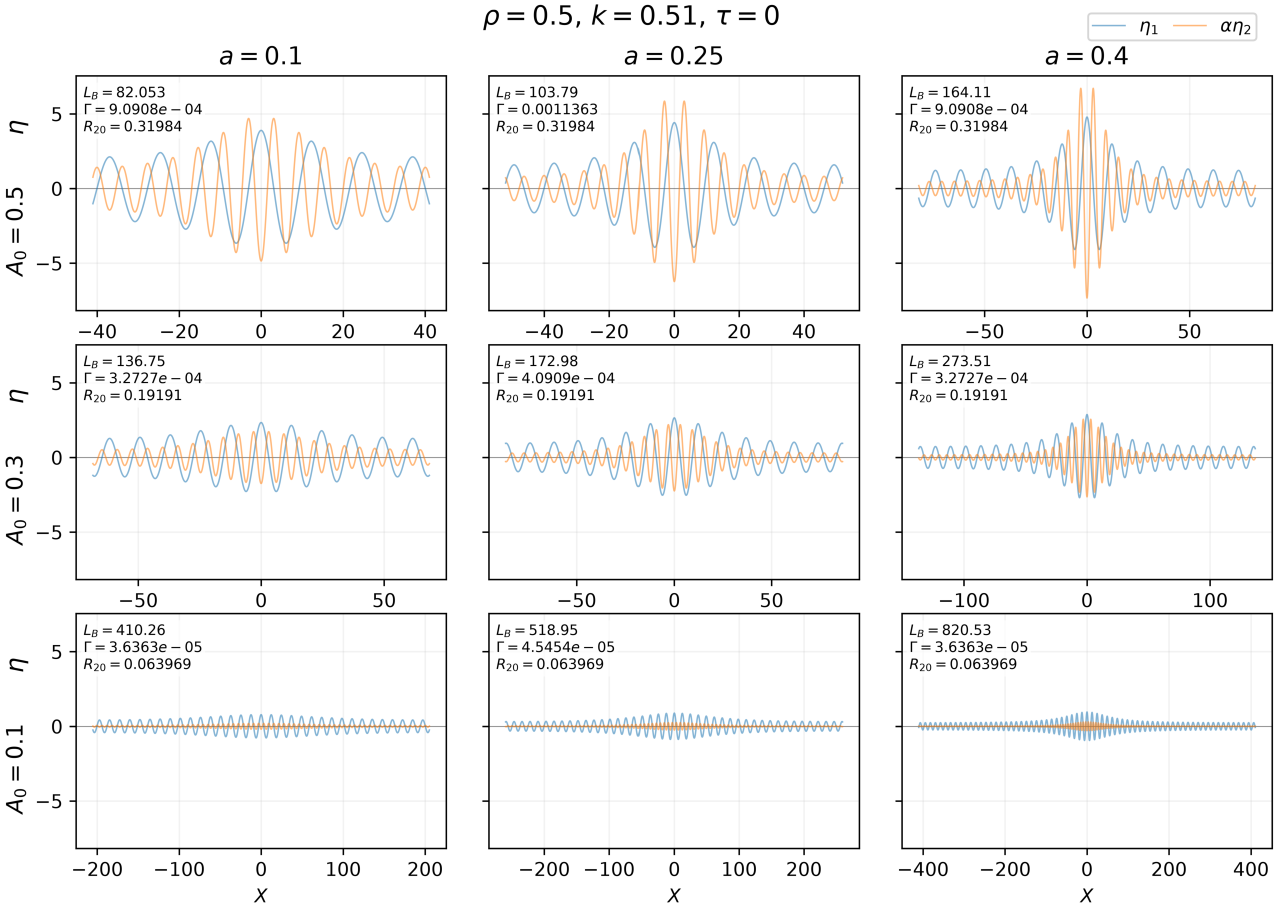}
\caption{$k=0.51$}
\label{fig:rho05_upper_profiles:a}
\end{subfigure}

\vspace{-2mm}

\begin{subfigure}{\textwidth}
\centering
\includegraphics[width=0.49\linewidth]{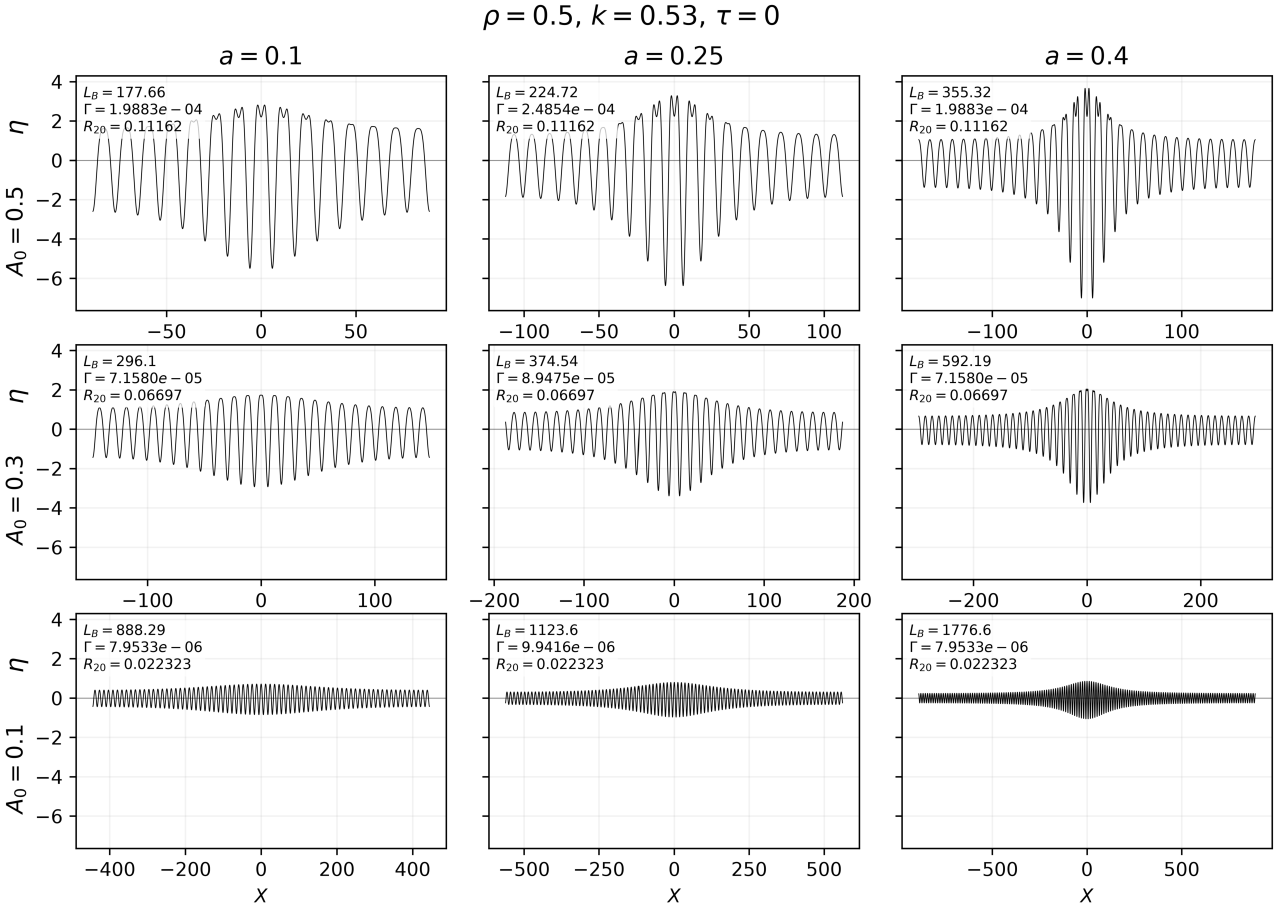}\hfill
\includegraphics[width=0.49\linewidth]{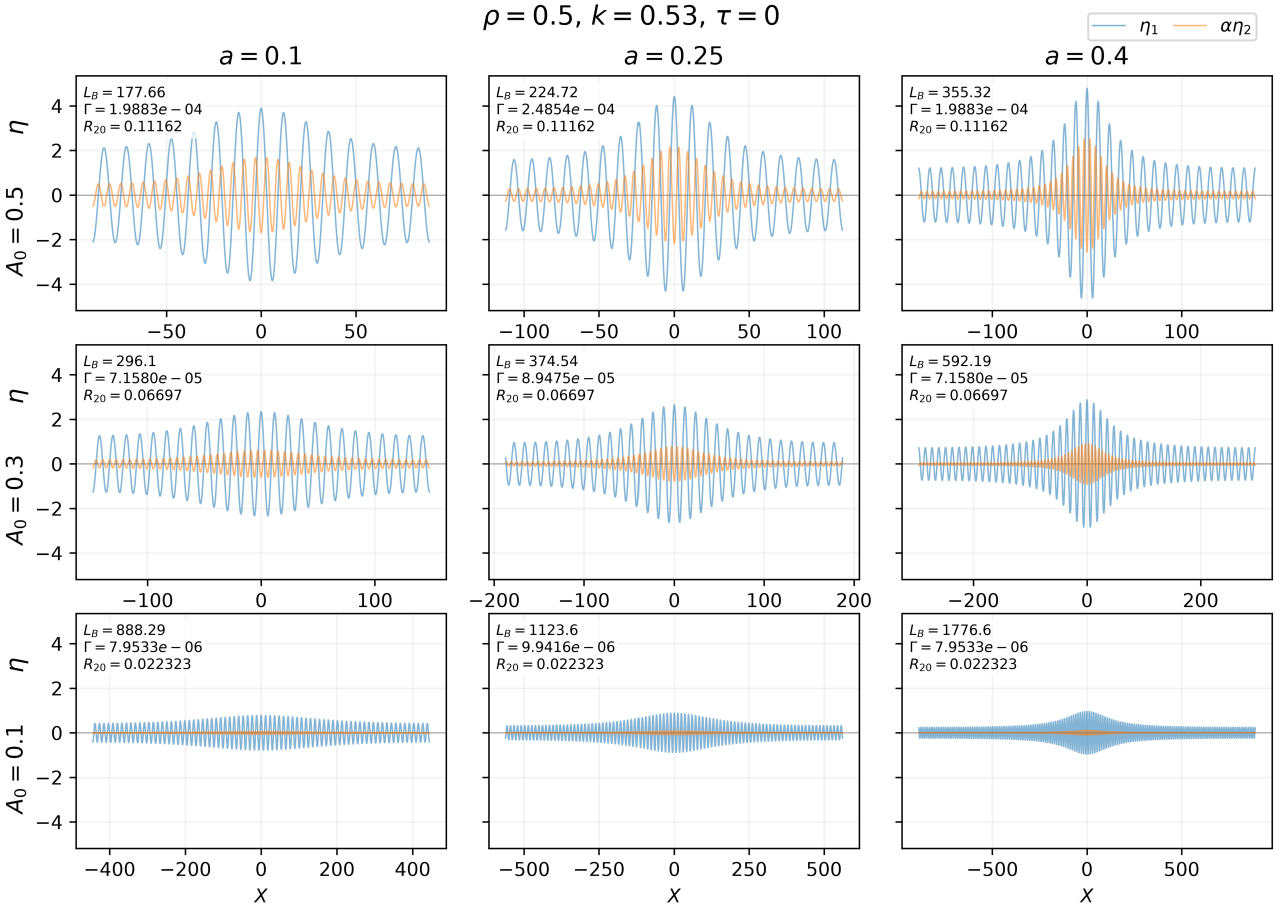}
\caption{$k=0.53$}
\label{fig:rho05_upper_profiles:b}
\end{subfigure}

\vspace{-2mm}

\begin{subfigure}{\textwidth}
\centering
\includegraphics[width=0.49\linewidth]{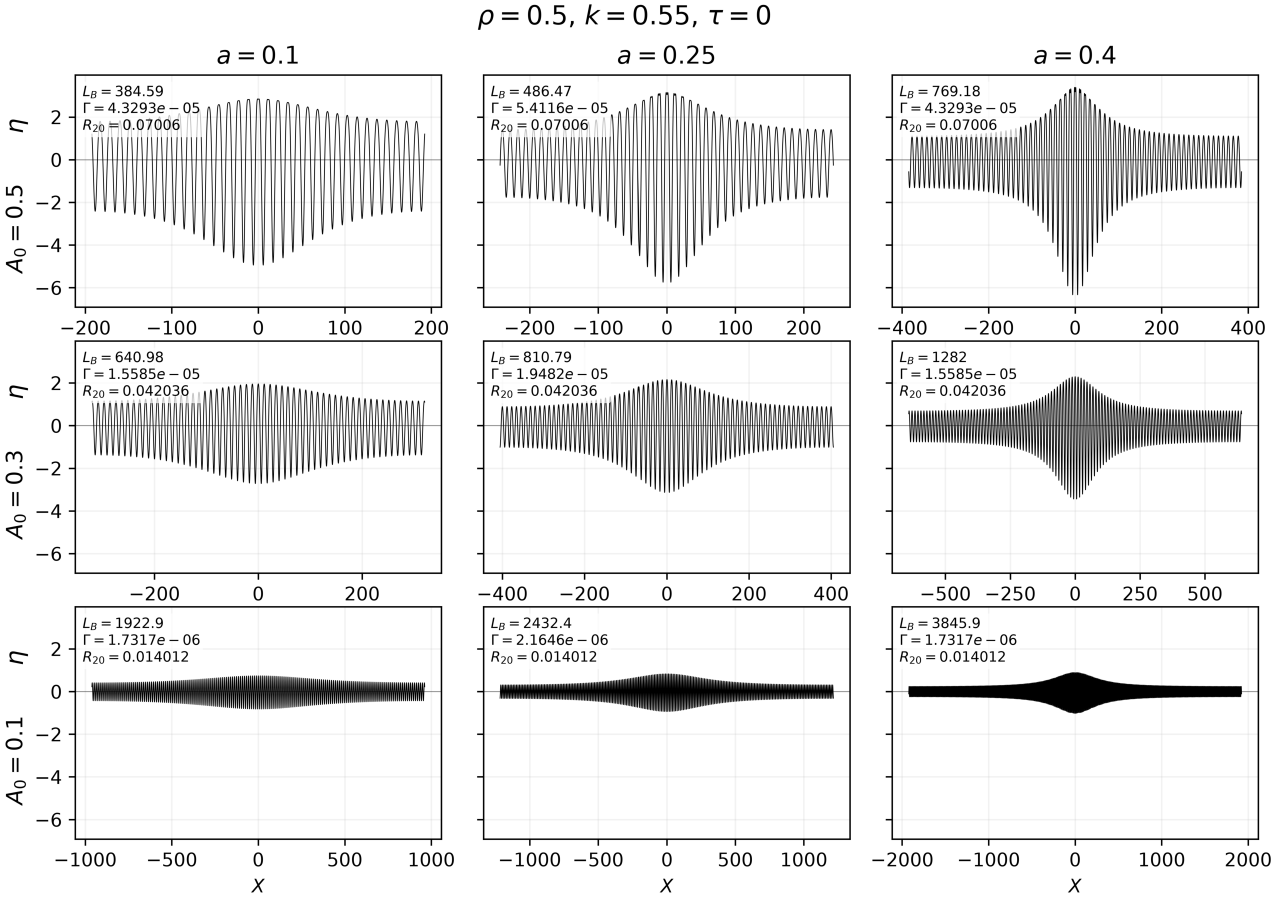}\hfill
\includegraphics[width=0.49\linewidth]{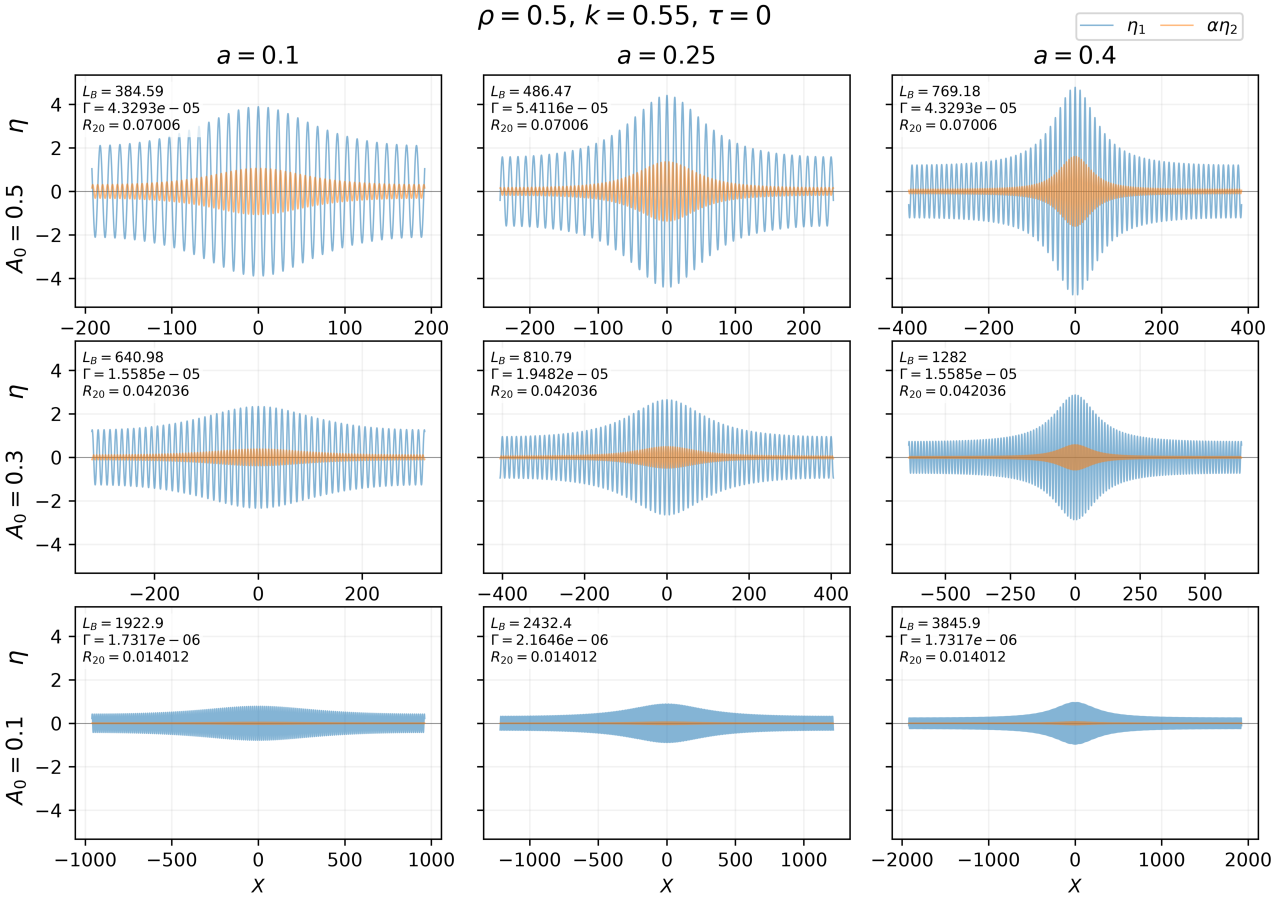}
\caption{$k=0.55$}
\label{fig:rho05_upper_profiles:c}
\end{subfigure}

\vspace{-2mm}

\caption{
Reconstructed Akhmediev breather profiles in the upper
MI region for \(\rho=0.5\).
In each subfigure, the left panel shows the
reconstructed interfacial profile
\(\eta=\eta_1+\alpha\eta_2\),
whereas the right panel shows separately the
contributions of the fundamental harmonic
\(\eta_1\) and the bound second harmonic
\(\alpha\eta_2\).
(a) \(k=0.51\);
(b) \(k=0.53\);
(c) \(k=0.55\).
}
\label{fig:rho05_upper_profiles}

\end{figure}

\clearpage
\endgroup

Thus, for the moderate density ratio \(\rho=0.5\),
moving across the upper MI region
from the resonance boundary to the upper \(J=0\)
boundary is accompanied by a gradual reduction of the
nonlinear deformation of the reconstructed interfacial
profile, an increase in the breather modulation period,
and a decrease in the contribution of the bound second
harmonic.
Only in the vicinity of the resonance boundary does the
bound second harmonic exert a substantial influence on
the profile shape, whereas near the upper \(J=0\)
boundary its contribution becomes minor.
These results are in full agreement with the
distribution of the parameter \(R_{20}\) in the maps
presented in Section~\ref{sec:maps} and further confirm
the validity of the weakly nonlinear reconstruction of
the interfacial profile throughout this parameter
region.

\paragraph{Narrow near-gravity region (large density ratios)}

As the density ratio increases, the geometry of the
upper MI region remains
qualitatively unchanged, whereas its width in the
wavenumber direction decreases rapidly.
For moderate density ratios, this region occupies a
relatively broad interval of wavenumbers, whereas in
the vicinity of \(\rho=1\) it contracts to an
extremely narrow wavenumber interval.
This narrow near-gravity region is analyzed below.

For the representative density ratio \(\rho=0.9\), the
lower MI region at \(k=0.1\) was
already examined in
Section~\ref{subsec:lower_instability}
(Fig.~\ref{fig:rho09_lower_branch}).
In that regime, the reconstructed interfacial profile
was determined almost entirely by the fundamental
harmonic, while the contribution of the bound second
harmonic remained small.

The most distinctive feature of the large-density-ratio
regime is the upper near-gravity MI region, which exists only within an
extremely narrow wavenumber interval of approximately
\(2\times10^{-4}\).
From a physical point of view, it represents the
near-gravity part of the upper MI
region previously identified for smaller density
ratios.
However, as \(\rho\) approaches unity, this region
contracts rapidly and eventually appears as a thin
strip in the parameter plane.

\begin{center}
\vspace{1mm}

\includegraphics[
  width=0.35\textwidth
]{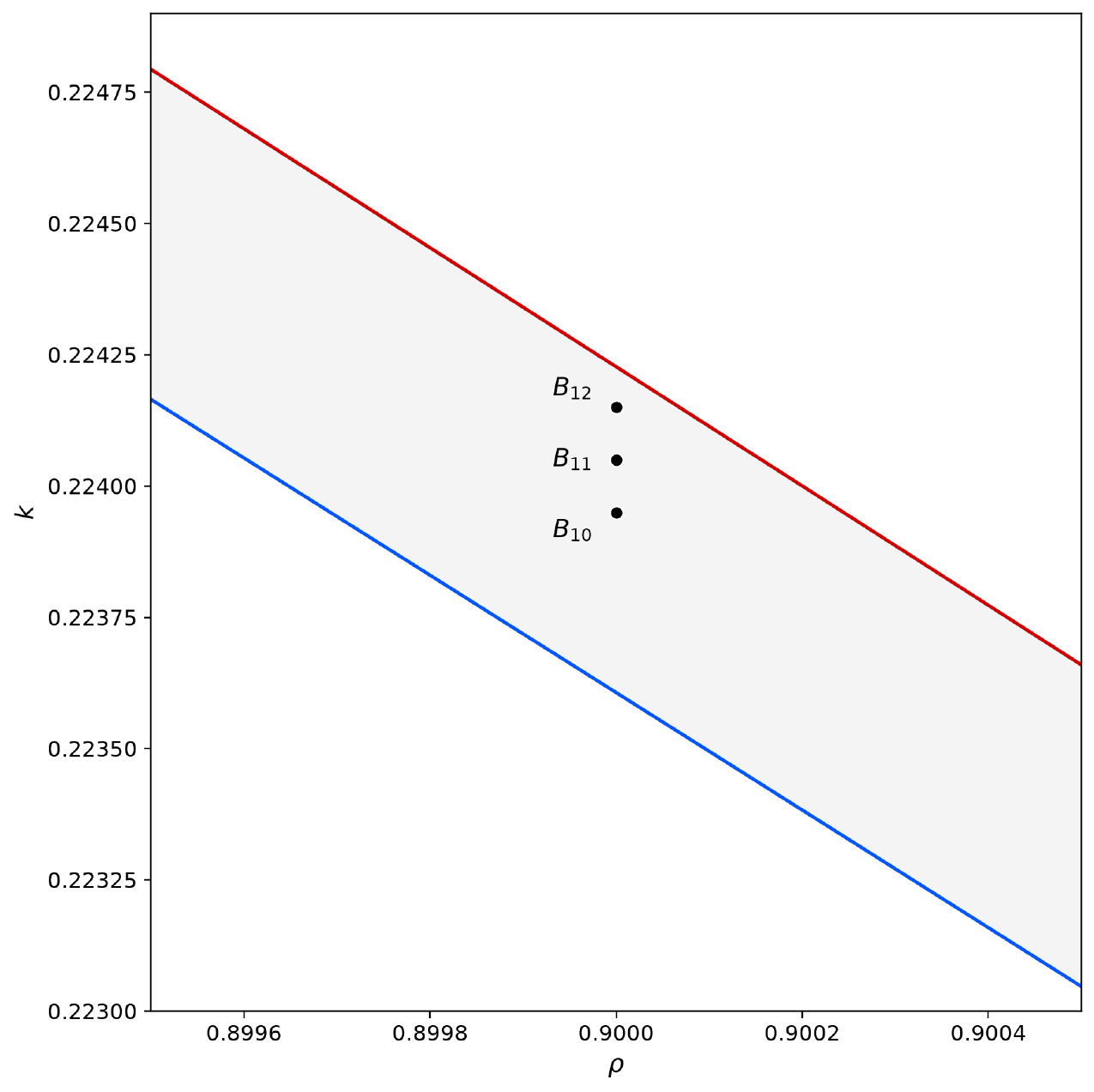}

\vspace{-2mm}

\captionof{figure}{
Narrow near-gravity region of the upper
MI region for
\(\rho\approx0.9\), showing the representative
points \(B_{10}\), \(B_{11}\), and \(B_{12}\)
selected for the reconstruction of the physical
breather profiles.
}
\label{fig:ultranarrow_region}

\vspace{1mm}
\end{center}

Points \(B_{10}\), \(B_{11}\), and \(B_{12}\)
correspond to the wavenumbers
\(k=0.22395\), \(0.22405\), and \(0.22415\),
respectively, and represent regimes located near the
left boundary, in the central part, and near the right
boundary of the narrow upper MI
region.
Because of its extremely small width, these points
cannot be distinguished in the standard parameter map
(Fig.~\ref{fig:selected_regimes}).
An enlarged view of this region is therefore provided
in Fig.~\ref{fig:ultranarrow_region}.
The corresponding reconstructed interfacial profiles
are presented in
Fig.~\ref{fig:rho09_ultranarrow}\subref{fig:rho09_ultranarrow_a}--
\subref{fig:rho09_ultranarrow_c}.

Throughout the entire wavenumber interval considered,
the contribution of the bound second harmonic remains
substantial.
For \(A_0=0.5\), the parameter \(R_{20}\) decreases
only from approximately \(0.29\) to \(0.18\).
Consequently, although the condition \(R_{20}<1\) is
satisfied throughout the region, the second-order
nonlinear correction continues to exert a noticeable
influence on the reconstructed interfacial profile.

In Fig.~\ref{fig:rho09_ultranarrow}\subref{fig:rho09_ultranarrow_a},
corresponding to the left part of the upper
MI region, the contribution of
the bound second harmonic is the largest among the
three regimes considered.
For \(A_0=0.5\), the parameter
\(R_{20}\approx0.29\), so the bound second harmonic
can no longer be regarded as a small correction to the
fundamental harmonic.
As a result, in the central part of the wave packet
\(\alpha\eta_2\) becomes comparable in amplitude to
the fundamental harmonic, leading to a pronounced
asymmetry of the reconstructed interfacial profile and
deeper central troughs.
At the same time, the breather modulation period
remains relatively short
(\(L_B\approx1260\) for \(a=0.25\)), while the
MI growth rate is the largest
among the three regimes considered.

\clearpage

\begingroup
\renewcommand{\floatpagefraction}{0.95}
\renewcommand{\topfraction}{0.99}
\renewcommand{\textfraction}{0.01}

\begin{figure}[p]
\centering

\begin{subfigure}{\textwidth}
\centering
\includegraphics[width=0.49\linewidth]{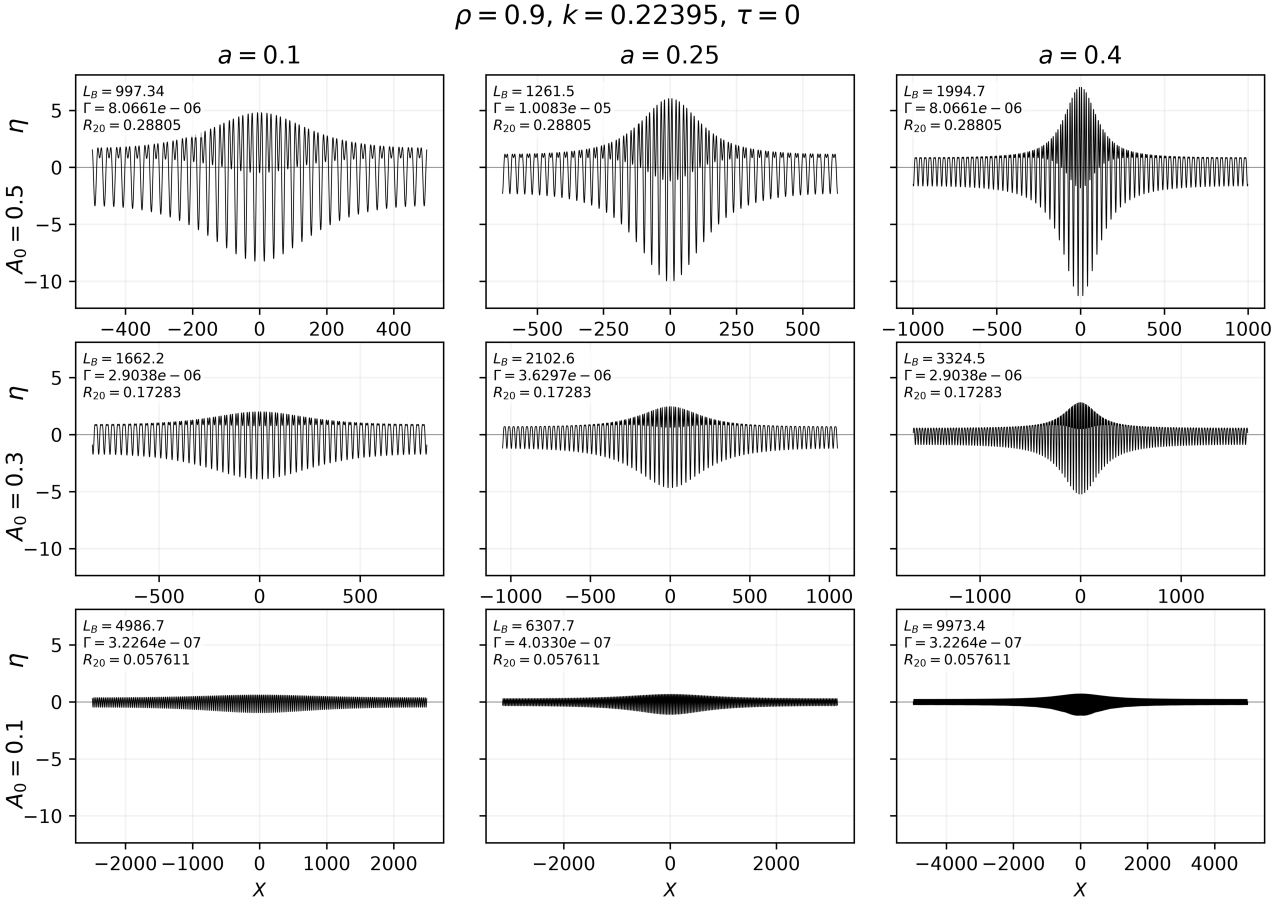}\hfill
\includegraphics[width=0.49\linewidth]{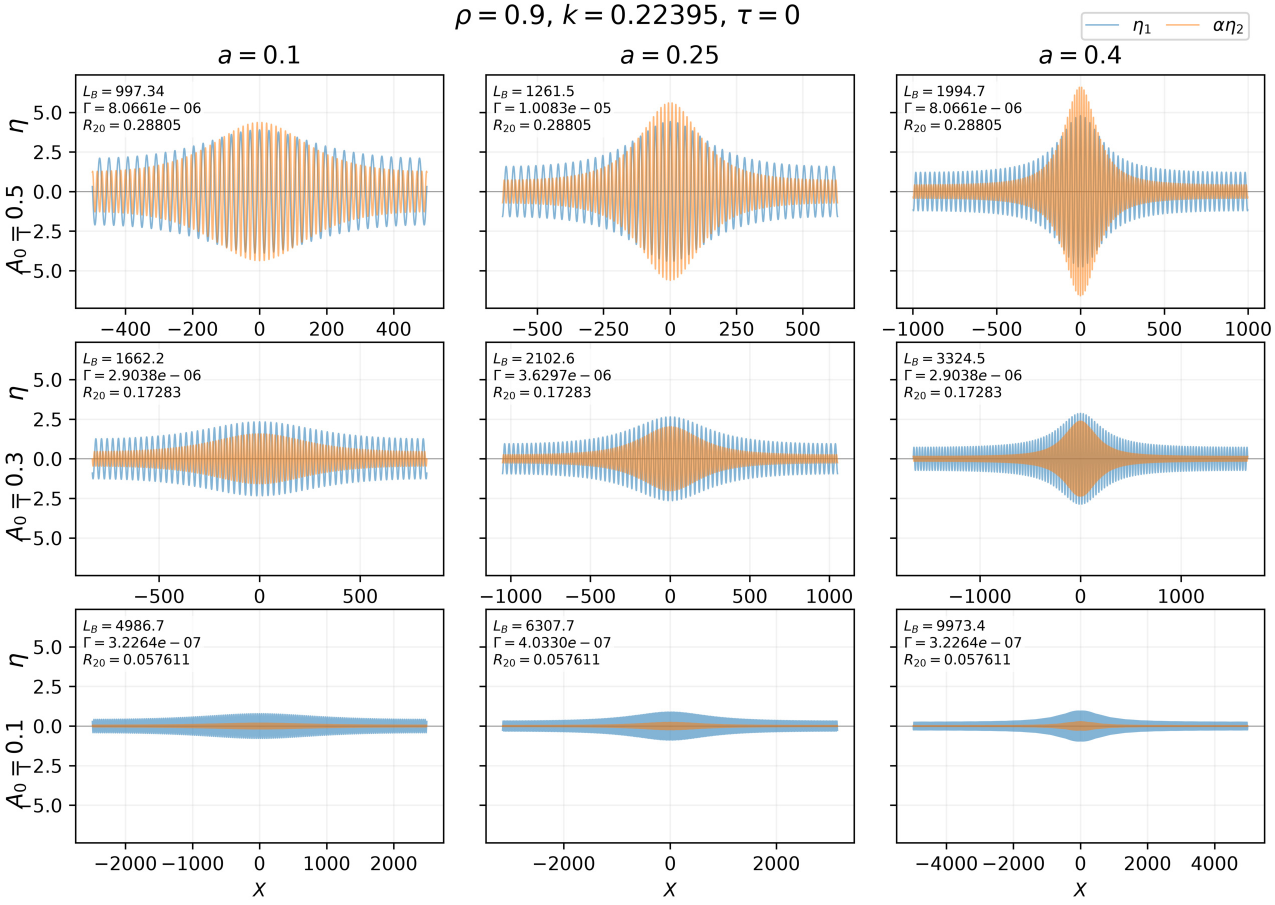}
\caption{$k=0.22395$.}
\label{fig:rho09_ultranarrow_a}
\end{subfigure}

\vspace{-2mm}

\begin{subfigure}{\textwidth}
\centering
\includegraphics[width=0.49\linewidth]{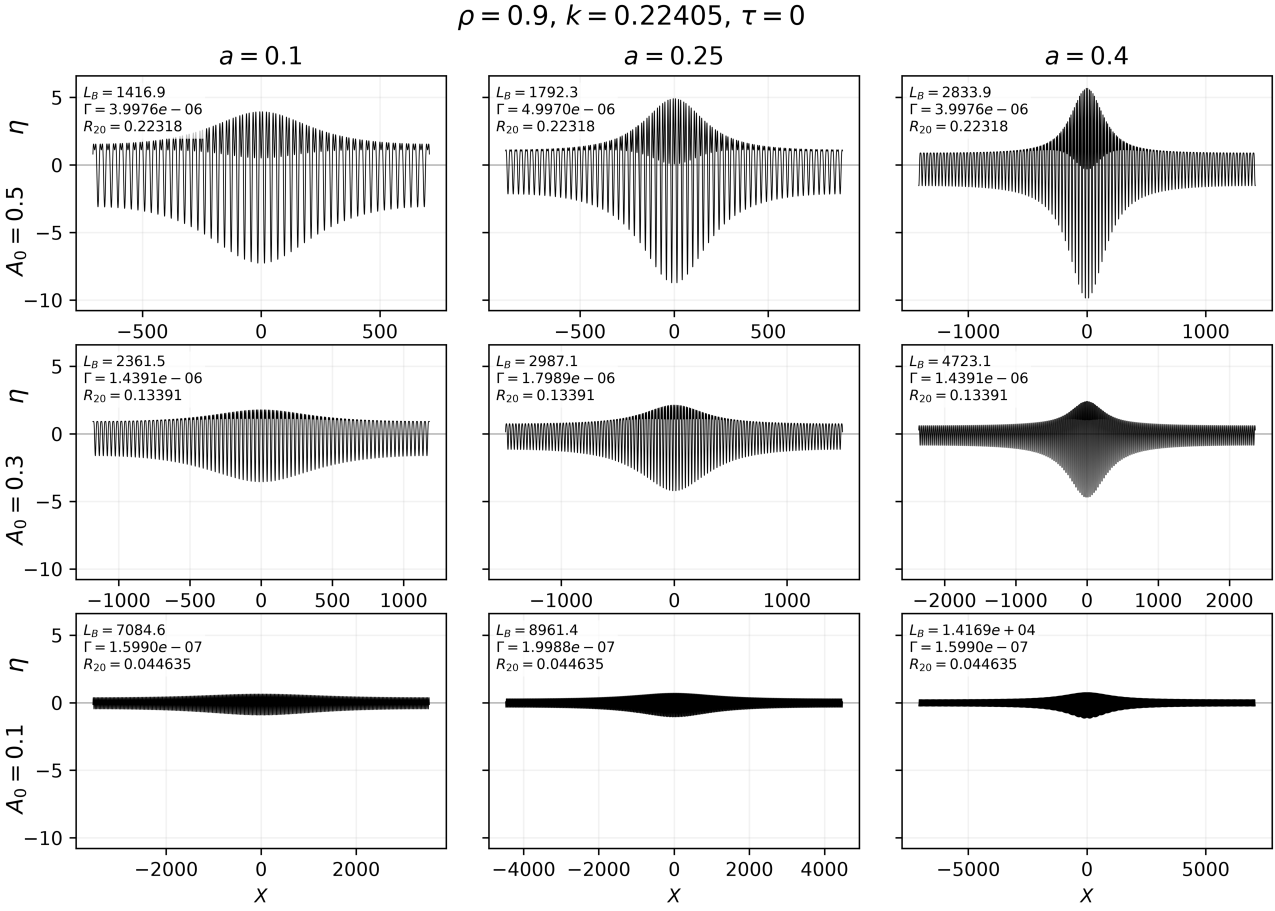}\hfill
\includegraphics[width=0.49\linewidth]{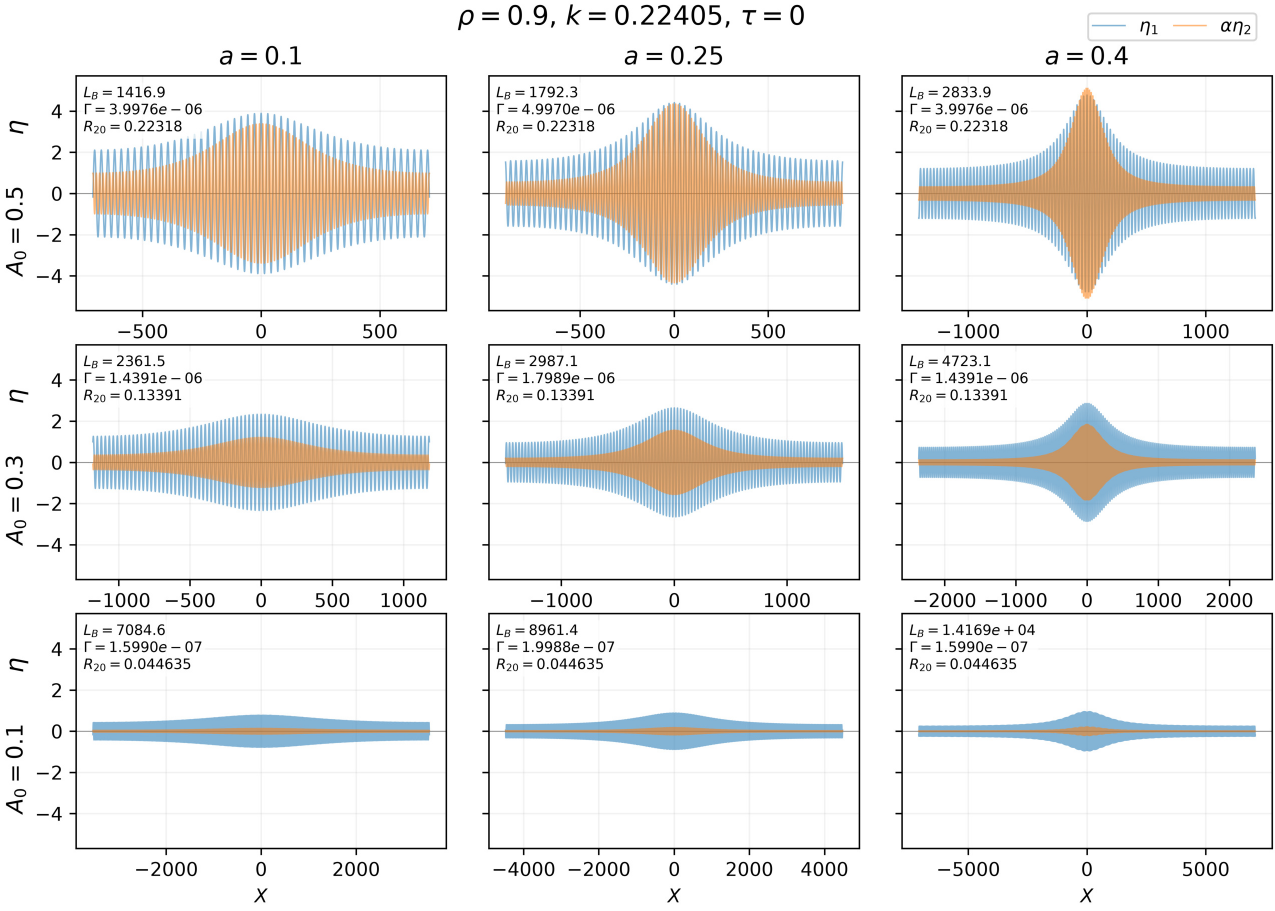}
\caption{$k=0.22405$.}
\label{fig:rho09_ultranarrow_b}
\end{subfigure}

\vspace{-2mm}

\begin{subfigure}{\textwidth}
\centering
\includegraphics[width=0.49\linewidth]{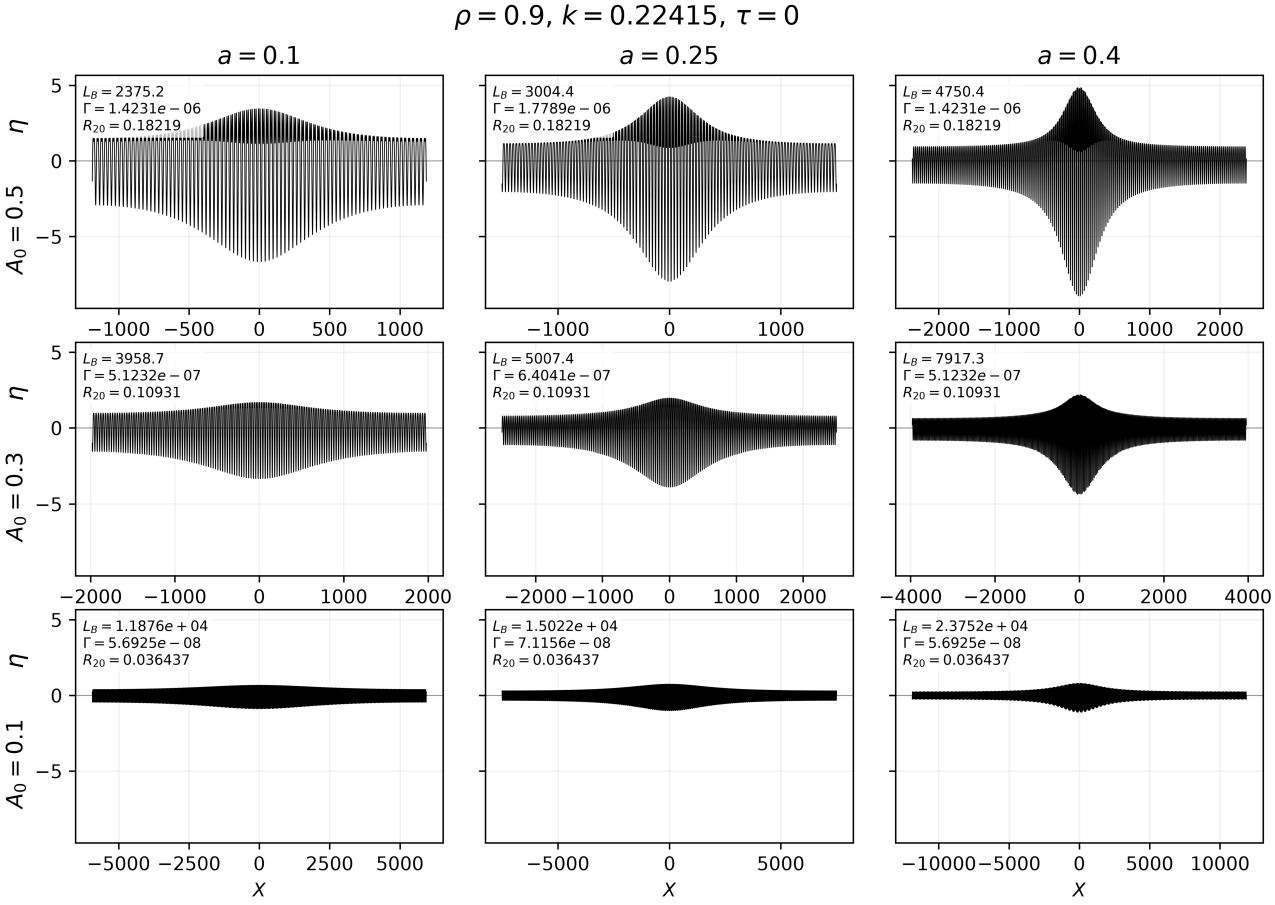}\hfill
\includegraphics[width=0.49\linewidth]{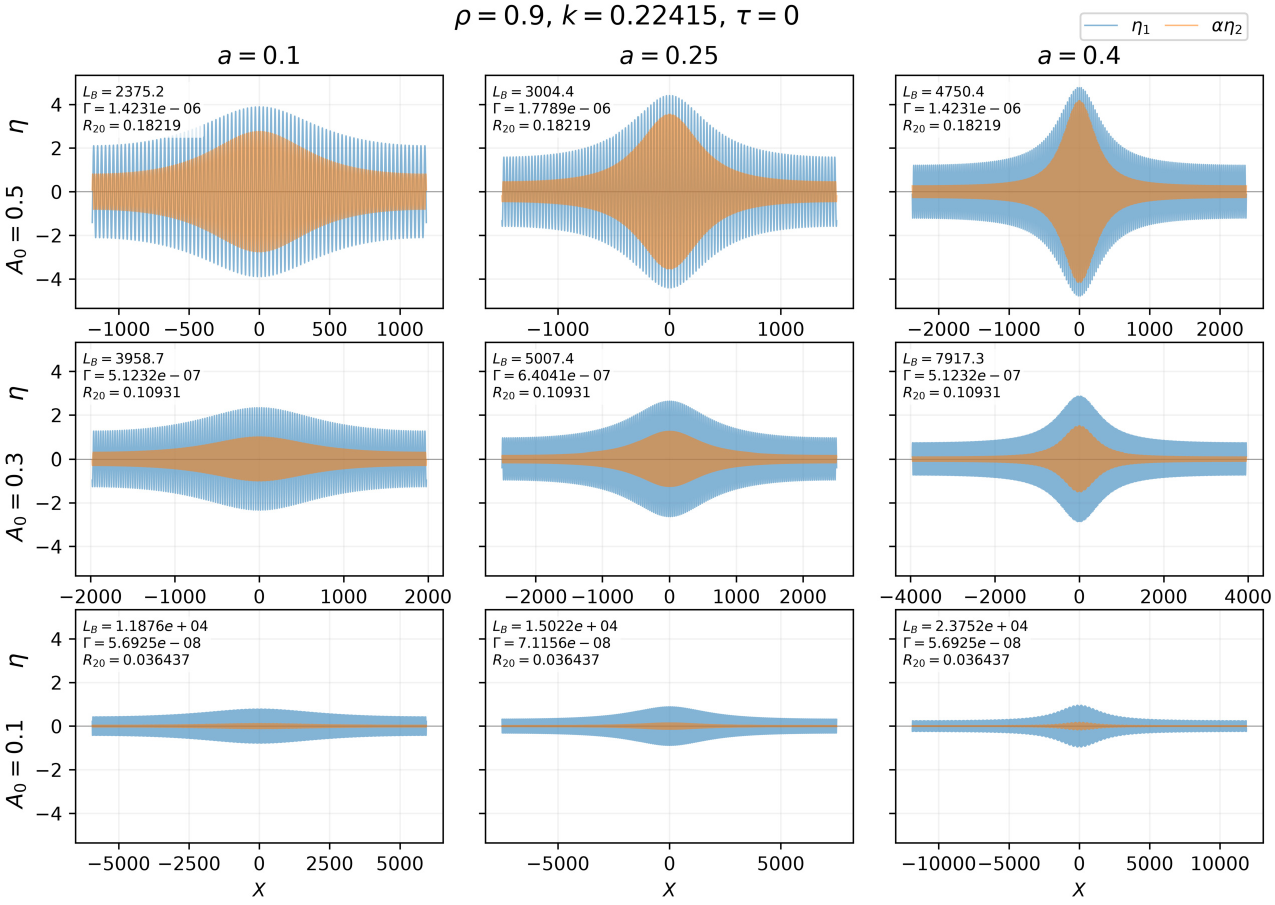}
\caption{$k=0.22415$.}
\label{fig:rho09_ultranarrow_c}
\end{subfigure}

\vspace{-2mm}

\caption{
Reconstructed Akhmediev breather profiles in the
narrow near-gravity region of the upper
MI region for \(\rho=0.9\).
In each subfigure, the left panel shows the
reconstructed interfacial profile
\(\eta=\eta_1+\alpha\eta_2\), while the right panel
shows separately the contributions of the
fundamental harmonic \(\eta_1\) and the bound
second harmonic \(\alpha\eta_2\).
(a) \(k=0.22395\);
(b) \(k=0.22405\);
(c) \(k=0.22415\).
}
\label{fig:rho09_ultranarrow}

\end{figure}

\clearpage
\endgroup

Moving to the central point of the narrow
near-gravity region
(Fig.~\ref{fig:rho09_ultranarrow}\subref{fig:rho09_ultranarrow_b}),
the breather modulation period increases
substantially.
A change in the wavenumber of only \(10^{-4}\)
increases the breather modulation period to
approximately \(L_B\approx1790\) for \(a=0.25\),
whereas the MI growth rate
decreases by nearly a factor of two.
At the same time, the parameter \(R_{20}\)
decreases to approximately \(0.22\).
As a consequence, the central part of the
reconstructed interfacial profile becomes smoother,
and the asymmetry of the profile is noticeably
reduced, although the contribution of the bound
second harmonic remains clearly visible.

A further shift toward the right boundary of the
narrow near-gravity region
(Fig.~\ref{fig:rho09_ultranarrow}\subref{fig:rho09_ultranarrow_c})
is accompanied by a rapid approach to the upper
\(J=0\) boundary.
The breather modulation period increases to
approximately \(L_B\approx3000\) for \(a=0.25\),
whereas the MI growth rate
becomes nearly six times smaller than in the left
part of the region.
The parameter \(R_{20}\) decreases to
approximately \(0.18\), so that the bound second
harmonic no longer governs the shape of the
reconstructed interfacial profile.
The profile becomes smoother and is increasingly
determined by the fundamental harmonic, although a
localized nonlinear deformation of the central part
of the wave packet is still preserved.

As in the previous examples, decreasing the
background-wave amplitude from \(A_0=0.5\) to
\(A_0=0.1\) leads to a substantial reduction in the
parameter \(R_{20}\), a weaker contribution of the
bound second harmonic, and an increasing dominance
of the fundamental harmonic in determining the
reconstructed interfacial profile.
At the same time, increasing the breather parameter
\(a\) nearly doubles the breather modulation period,
resulting in stronger localization of the breather
and a greater concentration of its amplitude in the
central part of the wave packet, whereas the
influence of the bound second harmonic is governed
primarily by the position of the selected regime
within the upper MI region.

Thus, even within this extremely narrow wavenumber
interval, changing \(k\) by only a few
ten-thousandths is accompanied by a pronounced
modification of the breather characteristics.
As the upper \(J=0\) boundary is approached, the
breather modulation period increases rapidly, the
MI growth rate decreases, and
the contribution of the bound second harmonic
gradually weakens.
This behavior is in full agreement with the
distributions of \(L_B\), \(\Gamma\), and
\(R_{20}\) presented in the parameter maps of
Section~\ref{sec:maps} and confirms that even minute variations
of the wavenumber within the narrow near-gravity
region lead to substantial changes in the properties
of the breather regime.

Overall, the results of this section fully confirm
the trends established from the parameter maps
presented in
Section~\ref{sec:maps}.
As the density ratio increases from moderate to
large values, the physical picture remains
qualitatively unchanged: the principal features of
breather evolution are preserved.
At the same time, the upper MI region contracts rapidly, making the
breather characteristics increasingly sensitive to
small variations in the wavenumber.

\section{Discussion}
\label{sec:Discussion}

The parameter maps constructed in this work
demonstrate that the focusing condition defined by
Eq.~(\ref{eq:focusing}) specifies only the
mathematical domain in which Akhmediev breathers
exist within the framework of the NLS model.
By itself, however, this condition is not sufficient
to characterize their physical properties.
The breather modulation period, the MI growth rate, and the magnitude of the
nonlinear corrections vary largely independently and
therefore cannot be described by a single criterion.
Consequently, an analysis based solely on the
focusing condition of
Eq.~(\ref{eq:focusing}) is insufficient to identify
those regimes for which the weakly nonlinear
reconstruction of the physical interfacial profile is
most reliable.

The approach proposed in this work is based on the
combined analysis of three complementary
characteristics: the breather modulation period
\(L_B\), the MI growth rate
\(\Gamma\), and the dimensionless parameter
\(R_{20}\).
The parameter \(L_B\) determines the characteristic
spatial scale of the breather structure,
\(\Gamma\) characterizes its temporal development,
whereas \(R_{20}\) quantifies the magnitude of the
nonlinear corrections involved in the reconstruction
of the physical interfacial profile.
Their combined analysis makes it possible not only
to identify the regions where breather solutions
exist, but also to distinguish the regimes for which
the weakly nonlinear description is most reliable.
Thus, the proposed parameter maps serve not only as
a means of visualization, but also as a practical
tool for classifying breather regimes and selecting
physically meaningful regions of the parameter
space.

An equally important outcome of this study is the
reconstruction of the physical interfacial profile.
The results demonstrate that the parameter
\(R_{20}\) is not only a measure of the contribution
of the bound second harmonic, but also an indicator
of the degree of nonlinear deformation of the
reconstructed interface.
For \(R_{20}\ll1\), the interfacial shape is
determined predominantly by the fundamental
harmonic.
When \(R_{20}=O(0.1)\), the bound second harmonic
already exerts a noticeable influence on the wave
geometry, while its contribution remains moderate
relative to the fundamental harmonic.
Finally, for \(R_{20}=O(1)\), the contribution of
the bound second harmonic becomes comparable to
that of the fundamental harmonic, leading to
pronounced deformation of both crests and troughs
and a substantial modification of the reconstructed
interfacial profile.
Thus, combining the parameter maps with the
reconstruction of the physical interfacial profile
provides a direct physical interpretation of the
parameters of the Nayfeh model and bridges the gap
between the mathematical description of the complex
envelope and the actual shape of the fluid
interface.

The present analysis is based on the Nayfeh model
and the NLS and is
therefore limited to the second-order weakly
nonlinear approximation.
The effects of viscosity, finite fluid depth,
three-dimensional dynamics, wave breaking, and
higher-order nonlinear corrections have not been
taken into account.
Accordingly, the parameter maps and reconstructed
interfacial profiles presented here define the range
of applicability of the weakly nonlinear model
rather than the full spectrum of possible
interfacial-wave dynamics.

The results obtained for small density ratios are of
particular interest because of their close connection
with the classical problem of gravity--capillary
waves on a free fluid surface.
In the limit \(\rho\to0\), the two-fluid system
reduces continuously to the case of surface waves
beneath a light gas phase, for which breather
dynamics have been observed experimentally in wave
tanks.
In particular, \cite{Chabchoub2011} reported the
experimental realization of the Peregrine breather,
which represents the limiting case of the Akhmediev
breather corresponding to an infinite modulation
period.
The experiment exhibited the characteristic
development of MI: growth of a
localized perturbation on a finite background,
attainment of a maximum amplitude, and subsequent
return to an almost undisturbed wave field.
The interfacial breather solutions obtained here for
small values of \(\rho\) display qualitatively
similar localization and delocalization dynamics,
indicating good qualitative agreement with these
experimental observations for free-surface waves.
From this perspective, the interfacial breathers
investigated in the present work may be viewed as a
natural extension of this scenario to the interface
between two fluids with an arbitrary density ratio.

The proposed approach is not restricted to the
Akhmediev breather and can be extended to other
localized solutions of the NLS and its generalizations.
It also provides a general framework for
constructing physically reconstructed wave profiles
and parameter maps in a broad class of weakly
nonlinear wave systems.
A particularly promising direction for future work
is the application of this methodology to other
stratified hydrodynamic systems, including models of
internal waves in two-layer fluids of finite depth.

\section{Conclusions}

A system of parameter maps for Akhmediev breathers
at the interface between two fluid half-spaces has
been constructed on the basis of the Nayfeh model.
The resulting maps describe the breather modulation
period \(L_B\), the MI growth
rate \(\Gamma\), and the dimensionless parameter
\(R_{20}\), which characterizes the relative
contribution of the bound second harmonic to the
reconstructed physical interfacial profile.

It has been shown that the combined analysis of
\(L_B\), \(\Gamma\), and \(R_{20}\) makes it
possible not only to determine the existence domain
of breather regimes, but also to assess their
physical properties.
The proposed approach provides a unified framework
for analyzing the spatial, temporal, and nonlinear
characteristics of Akhmediev breathers and enables
the identification of parameter regions in which the
weakly nonlinear reconstruction of the physical
interfacial profile remains reliable.

The physical interfacial profile has been
reconstructed by incorporating the bound second
harmonic, and the relationship between the
dimensionless parameter \(R_{20}\) and the
nonlinear deformation of the interface has been
established.
As \(R_{20}\) increases, the influence of the bound
second harmonic on the interfacial shape becomes
more pronounced, whereas small values of this
parameter correspond to regimes in which the weakly
nonlinear approximation remains quantitatively
reliable.

The proposed methodology can be extended to other
localized solutions of the NLS and its generalizations, as well as to
other stratified hydrodynamic systems, including
models of internal waves in two-layer fluids of
finite depth.
Overall, the proposed approach provides a direct
link between the mathematical description of
Akhmediev breathers and their physical
interpretation in terms of interfacial-wave
dynamics.

\section*{Acknowledgments}

Olga Avramenko acknowledges the support of the
Research Council of Lithuania in the preparation of
this article.


\printcredits

\bibliographystyle{cas-model2-names}

\bibliography{Article_Achmed_SH_SH_fixed}



\end{document}